 \documentclass[twocolumn,showpacs,prb,amsfonts,amsmath,amssymb,floatfix]{revtex4}
\usepackage{amsfonts}
\usepackage{amsmath}
\usepackage{amssymb}
\usepackage{graphicx}%
\begin{document}

\preprint{ }
\title{Interpolation across a muffin-tin interstitial using localized linear
combinations of spherical waves}
\author{Yoshiro Nohara}
\altaffiliation{Current address: ASMS Co., Ltd., 1-7-11 Higashi-gotanda Shinagawa-ku, Tokyo 141-0022, Japan}
\author{O. K. Andersen}
\affiliation{Max Planck Institute for Solid State Research, Heisenbergstrasse 1, D-70569, Stuttgart, Germany}
\date{\today}


\begin{abstract}
A method for 3D interpolation between hard spheres is described. The function
to be interpolated could be the charge density between atoms in condensed
matter. Its electrostatic potential is found analytically, and so are various
integrals. Periodicity is not required. The interpolation functions are
localized structure-adapted linear combinations of spherical waves, socalled
unitary spherical waves (USWs), $\psi_{RL}\left(  \varepsilon,\mathbf{r}%
\right)  ,$ centered at the spheres, $R,$ where they have cubic-harmonic
character, $L.$ Input to the interpolation are the coefficients in the
cubic-harmonic expansions of the target function at and slightly outside the
spheres; specifically, the values and the 3 first radial derivatives labelled
by $d=0$ (value), and 1-3 (derivatives). To fit this, we use USWs with 4
negative energies, $\varepsilon=\epsilon_{1},\epsilon_{2},\epsilon_{3}$ and
$\epsilon_{4}.$ Each interpolation function, $\varrho_{dRL}\left(
\mathbf{r}\right)  ,$ is actually a linear combination of these 4 sets of USWs
with the following properties: (1) It is centered at a specific sphere where
it has a specific cubic-harmonic character and radial derivative. (2) Its
value and first 3 radial derivatives vanish at all other spheres and for all
other cubic-harmonic characters, and is therefore highly localized,
essentially inside its Voronoi cell. Value-and-derivative (v\&d) functions
were originally introduced and used by Methfessel [Phys. Rev. B \textbf{38},
1537 (1988)], but only for the first radial derivative. Explicit expressions
are given for the v\&d functions and their Coulomb potentials in terms of the
USWs at the 4 energies, plus $\epsilon_{0}\equiv0$ for the potentials. The
coefficients, as well as integrals over the interstitial such as the
electrostatic energy, are given entirely in terms of the structure matrix,
$S_{RL,R^{\prime}L^{\prime}}\left(  \epsilon_{n}\right)  $, describing the
slopes of the USWs at the 5 energies and their expansions in Hankel functions.
For open structures, additional constraints are installed to pinpoint the
interpolated function deep in the interstitial. The strong localization of the
v\&d functions makes the method uniquely suited for complicated structures.
Use of point- and space-group symmetries can significantly reduce matrix sizes
and the number of v\&d functions. As simple examples, we consider a constant
density and the valence-electron densities in zinc-blende structured Si, ZnSe,
and CuBr.

\end{abstract}
\maketitle

\section{Introduction}

An often-met problem in computational physics, chemistry, and biology, and a
key one in electronic density-functional calculations, is to express a smooth,
global function, $\rho\left(  \mathbf{r}\right)  ,$ say the charge density, in
the region between the atoms in a form suitable for finding its electrostatic
potential, $V\left(  \mathbf{r}\right)  ,$ and for evaluating integrals such
as an electrostatic energy $\int\rho\left(  \mathbf{r}\right)  V\left(
\mathbf{r}\right)  d^{3}r.$

It has been found useful to expand
\[
\rho\left(  \mathbf{r}\right)  =\sum_{mn}\psi_{m}\left(  \epsilon
_{n},\mathbf{r}\right)  c_{mn},
\]
in solutions of the wave equation,%
\begin{equation}
\left(  \Delta+\varepsilon\right)  \psi\left(  \varepsilon,\mathbf{r}\right)
=0, \label{wave eq}%
\end{equation}
because then, the solution of Poisson's equation:%
\begin{equation}
-\Delta V\left(  \mathbf{r}\right)  =8\pi\rho\left(  \mathbf{r}\right)
\label{Poisson}%
\end{equation}
(in atomic Ry units), is simply:%
\begin{equation}
V\left(  \mathbf{r}\right)  =8\pi\left[  \psi\left(  0,\mathbf{r}\right)
+\sum_{nm}\psi_{m}\left(  \epsilon_{n},\mathbf{r}\right)  c_{mn}/\epsilon
_{n}\right]  . \label{Vr}%
\end{equation}
The expansion functions, $\psi_{m}\left(  \epsilon_{n},\mathbf{r}\right)  ,$
need only be defined in the region of interest, while the particular solution,
$\psi\left(  0,\mathbf{r}\right)  ,$ of the Laplace equation extends in all space.

A common choice of expansion functions is wave-equation solutions with
transform according to an irreducible representation $\left(  \mathbf{k}%
\right)  $ of a crystal or supercell, i.e. plane waves, $e^{i\left(
\mathbf{k+G}_{n}\right)  \cdot\mathbf{r}},$ with $\left\vert \mathbf{k+G}%
_{n}\right\vert ^{2}=\varepsilon\geq0$ and $\mathbf{G}_{n}$ running over the
vectors of the reciprocal lattice. For expanding a periodic function like the
charge density, $\mathbf{k=0.}$ The plane-wave set is complete and orthonormal
in the primitive cell, but is overcomplete and non-orthogonal in the
interstitial, say between muffin-tin (MT) spheres surrounding the atoms and
voids. Electronic-structure methods give the charge density as the sum of
products of the electronic basis functions, and using plane waves for the
latter, yields also the charge density as a sum of plane waves: $\sum
e^{i\left(  \mathbf{G}_{m}-\mathbf{G}_{n}\right)  \cdot\mathbf{r}}c_{m}%
c_{n}^{\ast}$. With pseudopotential methods, this density is a smooth part of
the true density and extends in all space, whereas with augmented plane-wave
methods, it is the true density in the MT interstitial
\cite{Martin,12MartinandTerakuras}. Even in cases where the electronic basis
functions are not plane waves, plane-wave expansion of the charge density is
often used. With a small basis set of MT orbitals, for instance, a smooth part
of the orbitals is either Fourier transformed and then multiplied together
\cite{85Weyrich, 96Savrasov, 00Wills}, or multiplied together directly on a
mesh and then Fourier transformed \cite{00Methfessel, 10Kotani}.

In this paper, we shall \emph{not} expand in plane waves because they are
extended, but in MT-centered, decaying spherical waves, $h_{l}^{\left(
1\right)  }\left(  \kappa r_{R}\right)  Y_{L}\left(  \mathbf{\hat{r}}%
_{R}\right)  ,$ the natural choice for dealing with local point symmetries.
Here and in the following, $r_{R}\equiv\left\vert \mathbf{r-R}\right\vert ,$
$\mathbf{\hat{r}}_{R}\equiv\widehat{\mathbf{r-R}},$ $L\equiv lm,$ and
$\kappa^{2}=\varepsilon\lesssim0.$ Actually, we shall combine linearly the
$RL$-set of spherical waves for a given energy and structure, as specified by
its centers, $\mathbf{R,}$ and radii, $a_{R},$ into a set of even more
localized, structure-adapted unitary spherical waves (USWs) \cite{94Trieste}%
$,$ each of which is a cubic harmonic, $Y_{L}\left(  \mathbf{\hat{r}}%
_{R}\right)  ,$ on the own sphere and vanishes on all other spheres. Because
of this requirement, the spheres cannot overlap. Moreover, the USWs are
defined to vanish inside all spheres, which we shall therefore call hard-
rather than MT spheres \cite{hardvsMT}. A set of USWs is thus a set of
localized, structure-adapted spherical waves with a given energy. Localization
is essential if the computational effort is to increase merely proportional to
the size of the system (or the number of inequivalent sites) \cite{12orderN}.
The USW set, $\psi_{RL}\left(  \varepsilon,\mathbf{r}\right)  ,$ is specified
by a structure- or slope matrix whose element, $S_{RL,R^{\prime}L^{\prime}%
}\left(  \varepsilon\right)  ,$ equals the radial derivative (slope) of the
$R^{\prime}L^{\prime}$-projection of $\psi_{RL}\left(  \varepsilon
,\mathbf{r}\right)  $, and also gives the coefficient to $h_{l^{\prime}%
}^{\left(  1\right)  }\left(  \kappa r_{R^{\prime}}\right)  Y_{L^{\prime}%
}\left(  \mathbf{\hat{r}}_{R^{\prime}}\right)  $ in the expansion of
$\psi_{RL}\left(  \varepsilon,\mathbf{r}\right)  $.

When $\rho\left(  \mathbf{r}\right)  $ is better known -- or simpler to
evaluate -- near the surfaces of the spheres than throughout the topologically
complicated interstitial, it is advantageous to interpolate $\rho\left(
\mathbf{r}\right)  $ across the interstitial rather than to project it onto
the interstitial. This is the case for the electronic density: Near the
surface of any sphere surrounding an atom, this density is essentially the sum
of products of occupied atomic orbitals and therefore has a cubic-harmonic
expansion with $l_{\max}$ about twice the highest $l$ of an occupied atomic
orbital. Near each sphere, one can therefore easily project $\rho\left(
\mathbf{r}\right)  $ onto cubic harmonics obtaining the radial functions,
$\mathcal{\hat{P}}_{RL}\left(  r\right)  \rho\left(  \mathbf{r}\right)  ,$ and
then search an expansion:%
\begin{equation}
\rho\left(  \mathbf{r}\right)  =\sum_{n=1}^{d_{\max}+1}\sum_{RL}\,\psi
_{RL}\left(  \epsilon_{n},\mathbf{r}\right)  c_{nRL}, \label{rho}%
\end{equation}
in USWs with $d_{\max}+1$ different energies, which fits the values and first
$d_{\max}$ radial derivatives at $a_{R}$ for all $RL.$ It is obvious that if
we fit only values $\left(  d_{\max}\mathrm{=}0\right)  ,$ the unitary
property of, $\psi_{RL}\left(  \epsilon_{1},\mathbf{r}\right)  ,$ leads to the
simple result: $c_{0RL}=\mathcal{\hat{P}}_{RL}\left(  a_{R}\right)
\rho\left(  \mathbf{r}\right)  .$ In order to fit also slopes $\left(
d_{\max}\mathrm{=}1\right)  $, we must solve $N_{R}N_{L}$ linear equations
where $N_{R}$ the number of sites within the range of a USW and $N_{L}$ is the
number of $L$ values. In case we need to interpolate the charge density many
times for a given structure, as is the case in charge-selfconsistent
electronic-structure calculations, we would invert the corresponding matrix
for the linear equations once and for all. This matrix is the 1st
energy-divided difference,
\begin{equation}
\frac{S\left(  \epsilon_{1}\right)  -S\left(  \epsilon_{2}\right)  }%
{\epsilon_{1}-\epsilon_{2}}\equiv S_{12}, \label{S12}%
\end{equation}
of the slope matrix \cite{94Trieste}$.$ In fact, $aS_{12}$ equals the integral
$\left\langle \psi\left(  \epsilon_{1}\right)  \mid\psi\left(  \epsilon
_{2}\right)  \right\rangle $ over the interstitial.

For general $d_{\max},$ the dimension of the matrix to be inverted would be
$d_{\max}+1$ times as large$.$ Considering the fact that changing the energy
or the structure requires another inversion, this could be computationally
demanding. In the present paper, we shall therefore derive explicit
expressions involving merely the first $d_{\max}+1$ energy-divided differences
of the slope matrix. This is achieved by exploiting the radial wave equation,%
\begin{equation}
\left[  r\psi_{l}\left(  \varepsilon,r\right)  \right]  ^{\prime\prime
}=-\left[  \varepsilon-l\left(  l+1\right)  /r^{2}\right]  r\psi_{l}\left(
\varepsilon,r\right)  , \label{rad}%
\end{equation}
with the two unitary boundary conditions: $r\psi_{l}\left(  \varepsilon
,r\right)  |_{a}=1$ or $0.$ The simplest way to think about this approach is
that for a given structure and symmetry, but independently of the function to
be interpolated, we find those linear combinations, $\varrho_{dRL}\left(
\mathbf{r}\right)  ,$ of the USW sets for the energies $\epsilon
_{1},\,\epsilon_{2},\,..,\,\epsilon_{d_{\max}}$ which have the "super-unitary"
property that the $d^{\prime}$th radial derivative, $\left[  r\mathcal{\hat
{P}}_{R^{\prime}L^{\prime}}\left(  r\right)  \varrho_{dRL}\left(
\mathbf{r}\right)  \right]  ^{\left(  d^{\prime}\right)  }$ at $a_{R^{\prime}%
},$ for all $R^{\prime},L^{\prime}$, and $d^{\prime}=0,...,d_{\max},$ vanish,
except the own derivative $\left(  d\right)  $ of the own cubic-harmonics
projection $\left(  L\right)  $ at the own sphere $\left(  R\right)  $. In
terms of these value-and-derivative\emph{ }(v\&d) functions,\emph{ }which are
even more localized than the USWs, the density interpolated from its radial
derivatives,
\begin{equation}
\mathcal{R}_{RL}^{\left(  d\right)  }\left(  a_{R}\right)  \equiv\left[
r\mathcal{\hat{P}}_{RL}\left(  r\right)  \rho\left(  \mathbf{r}\right)
\right]  _{a_{R}}^{\left(  d\right)  }, \label{Input}%
\end{equation}
at the spheres is then:%
\begin{equation}
\rho\left(  \mathbf{r}\right)  =\sum_{d=0}^{d_{\max}}\sum_{RL}\varrho
_{dRL}\left(  \mathbf{r}\right)  \mathcal{R}_{RL}^{\left(  d\right)  }\left(
a_{R}\right)  . \label{interpolation}%
\end{equation}
The v\&d functions are localized essentially inside the Voronoi (Wigner-Seitz)
cells and the expansion (\ref{interpolation}) is therefore similar to, but
more general and efficient, than the one-center, cubic-harmonic expansion of
the cell-truncated density \cite{73Mol} used in KKR \cite{91KKRcell} and
LMTO\cite{00Kollar} Green-function methods to treat molecules, crystals,
impurities, random alloys, amorphous systems, surfaces, interfaces, etc., when
going beyond the atomic-spheres approximation (ASA)
\cite{73Simple,75PRB,80Segall,83imp,84Skriver,LamG,Green,Schilf,Pessoa,Nowak,Bose,Vargas,Mook,Abrikosov,Ruban,Turek,Weinberger,Stuttgart}%
.

For each v\&d function, $\varrho_{dRL}\left(  \mathbf{r}\right)  ,$ we can
solve Poisson's equation (\ref{Poisson}) and find the \emph{localized}
potential, $\varphi_{dRL}^{loc}\left(  \mathbf{r}\right)  ,$ and the
multipoles which have been subtracted in order to make it localized. This
potential and its multipole moments are expressed in terms of energy-divided
differences \cite{00DivDif} of the USWs and the slope matrix over the energy
mesh, $\epsilon_{0},\,\epsilon_{1},...,\,\epsilon_{d_{\max}+1},$ to which the
energy $\epsilon_{0}\equiv0$ has been added. The latter takes care of the
particular solution in eq.$\,$(\ref{Vr}) which picks the localized part of the
potential. In terms of these potentials from the v\&d charge densities, the
localized Coulomb potential from $\rho\left(  \mathbf{r}\right)  $ is then:
\begin{equation}
V^{loc}\left(  \mathbf{r}\right)  =\sum_{d=0}^{d_{\max}}\sum_{RL}\varphi
_{dRL}^{loc}\left(  \mathbf{r}\right)  \mathcal{R}_{RL}^{\left(  d\right)
}\left(  a_{R}\right)  . \label{Potential}%
\end{equation}
At the end of a calculation, the localizing multipoles are added to those from
the remaining charge density in the system and the resulting Laplace potential
is expanded in zero-energy USWs, $\psi_{RL}\left(  0,\mathbf{r}\right)  .$

With the charge density and the Coulomb potential in the interstitial
expressed in terms of USWs and their slope matrix, and with the integral of a
product of USWs over the interstitial expressed in terms of the slope matrix
(\ref{S12}), so is the electrostatic energy of the interstitial charge
density. Also one-center cubic-harmonic expansions, such as:%
\begin{equation}
\varphi_{dRL}^{loc}\left(  \mathbf{r}\right)  \approx\sum_{L^{\prime}%
}Y_{L^{\prime}}\left(  \mathbf{\hat{r}}_{R^{\prime}}\right)  \mathcal{\hat{P}%
}_{R^{\prime}L^{\prime}}\left(  r\right)  \varphi_{dRL}^{loc}\left(
\mathbf{r}\right)  , \label{one-centre}%
\end{equation}
are given in terms of the slope matrix and two radial wave-equation solutions
(\ref{rad}). The spherically symmetric averages $\left(  L^{\prime}%
\mathrm{=}0\right)  $ are for instance used to generate the potential in the
overlapping MT approximation (OMTA) \cite{98MRS, 09OMTA} which defines the 3rd
generation LMTO \cite{94Trieste, 98MRS,00Rashmi} and NMTO \cite{00NMTO,
00Odile, 12Juelich,15FPNMTO} basis sets.

The present paper reformulates and extends beyond 1st radial derivatives an
approach proposed nearly 30 years ago by Methfessel \cite{88Methfessel} for
use in charge-selfconsistent electronic-structure calculations in which the
smooth part of the electronic wave functions are expanded in relatively few
LMTOs \cite{89Methfessel}, rather than in many plane waves or many Gaussians
\cite{Martin,12MartinandTerakuras}. In the latter methods, also the charge
density --being wave-function products-- is a sum of plane waves or Gaussians,
for which Poisson's equation has an analytical solution. This is a main reason
for the popularity of those methods. Unfortunately, products of spherical
waves (LMTO envelopes) are not sums of spherical waves, except in the (warped
\cite{86WASA}) ASA \cite{73Simple,75PRB,80Segall,83imp,84Skriver}, but
Methfessel noted that this product is easily formed near the surfaces of the
spheres, and then interpolated across the interstitial using spherical waves.
Hence, he saw interpolation across the interstitial as an approximate way to
reduce the product to a sum: $\psi\psi\approx\sum c\psi$. Moreover, since the
integral over the interstitial of a product of spherical waves is a surface
integral over the spheres, and thus analytical, Methfessel's interpolation
approach also serves to compute multi-center integrals over the interstitial,
a task which had been solved analytically \cite{87Springborg}, but with an
impractically complicated result.

Systems of current interest often have interstitials so complex that insertion
of interstitial, so-called "empty" (E) spheres in the voids \cite{80Segall}
can be insufficient for achieving the accuracy needed for interpolating the
charge density across the interstitial. Moreover, in molecular-dynamics
calculations empty spheres are useless because they are not "conserved". These
are reasons why fitting to higher radial derivatives has become necessary.
Whereas Andersen et al. \cite{94Trieste,98MRS} fitted values and 1st radial
derivatives exactly with USWs and their first energy derivatives, $\psi\left(
\epsilon_{1},\mathbf{r}\right)  $ and $\dot{\psi}\left(  \epsilon
_{1},\mathbf{r}\right)  ,$ Tank and Arcangeli \cite{00Rashmi} added
$\ddot{\psi}\left(  \epsilon_{1},\mathbf{r}\right)  $ and could then
least-squares fit also at selected points in the interstitial. Since forming
high-order energy derivatives is numerically troublesome, energy-divided
differences were used in ref. \cite{00Odile}.

In this paper we give the details of the v\&d formalism for $d_{\max
}\mathrm{=}3$ and test it on the charge density in some diamond-structured
$sp^{3}$-bonded and ionic semiconductors. This technique has been developed
for solving Poisson's equation in our newly developed full-potential NMTO
electronic-structure method \cite{15FPNMTO} used in Ref. \cite{15LiPB}.
Obviously, the technique could be useful for any electronic-structure method
which does not use a plane-wave or Gaussian basis set, and --actually-- for
interpolating any 3D function across a hard-sphere interstitial from the cubic
harmonic projections at and closely outside the spheres. The purpose could be
decomposition into atom-centered, strongly localized functions, evaluation of
integrals over the interstitial, or solving Poisson's equation; but \emph{not}
evaluation of differential properties like the kinetic energy. The v\&d
technique should be particularly useful for treating Coulomb effects beyond
the ASA in systems without translational symmetry, such as liquids, amorphous
and disordered systems, systems with impurities, interfaces, surfaces, and
biological molecules. In the latter cases, it will be necessary to constrain
the charge density as described towards the end of the paper.

Although uniquely suited for interpolating functions without symmetry, point
symmetry can significantly reduce the number of cubic harmonics needed when
generating the slope matrix by inversion, and space-group symmetry can reduce
the number of sites needed when generating the v\&d functions. For the charge
density in diamond-structured Si, for example, we need $4$ rather than 25
cubic harmonics, and a cluster of $\sim150$ sites to generate the slope matrix
in real space. To subsequently form the v\&d functions, we need only 2 sites
after the slope matrix has been Bloch summed with $\mathbf{k=0,}$ and merely 1
site using the space group-symmetry.

The paper is organized as follows: Sect.$\,$\ref{Prel} gives preliminaries for
the derivation of the v\&d functions. \ref{SectInput} specifies the input for
the interpolation and the boundary conditions for the v\&d functions.
\ref{USW} reviews the transformation from Hankel functions to USWs, in fair
detail because we shall use it in a following paper \cite{15FPNMTO}.
\ref{Taylor} expands the two radial wave functions in energy-dependent Taylor
series in $r-a,$ and \ref{Sectdivdif} forms their energy-divided differences.
In Sect.$\,$\ref{v&dfunctions} we derive the v\&d functions with $d_{\max
}\mathrm{=}3$ as linear combinations of USWs. In Sect.$\,$\ref{Poissons eq} we
solve Poisson's equation for the v\&d functions, obtaining potentials which
are either localized or regular and long-ranged. Analytical expressions for
the integral over the interstitial of a single USW, a product of USWs, or of
their energy-divided differences -- and herewith of the electrostatic energy
-- are given in Sect.$\,$\ref{integrals}. Sect.$\,$\ref{ParamValues} discusses
how to set the parameters: \ref{N} the size of the cluster used to generate
the slope matrix, \ref{Symmetry} how to use symmetry to reduce matrix sizes,
and \ref{Emesh} how to choose the energy mesh. Here we use the examples of bcc
and diamond-structured interstitials, first with a constant density, and then
with the valence densities in $sp^{3}$-bonded and ionic semiconductors
obtained from FP NMTO calculations. Sect.$\,$\ref{Constraints} deals with
extra constraints needed in open structures. Finally, in Sect.$\,$\ref{Concl}
we conclude. One-centre expansions of the v\&d functions and their localized
potentials are derived in the Appendix.

\section{Preliminaries\label{Prel}}

\subsection{Input to the interpolation\label{SectInput}}

The method derived in this paper interpolates a 3D function, $\rho\left(
\mathbf{r}\right)  ,$ across a hard-sphere interstitial from the value and
first $d_{\max}=3$ radial derivatives of each $L$-projection at and outside
each sphere, $R$:%
\begin{align}
&  r\mathcal{\hat{P}}_{RL}\left(  r\right)  \rho\left(  \mathbf{r}\right)
\equiv r\int d^{3}r\,\delta\left(  r_{R}-r\right)  Y_{L}^{\ast}\left(
\mathbf{\hat{r}}_{R}\right)  \rho\left(  \mathbf{r}\right) \label{proj}\\
&  \equiv\mathcal{R}_{RL}\left(  r\right)  =\mathcal{R}_{RL}\left(
a_{R}\right)  +\frac{r-a_{R}}{1!}\mathcal{R}_{RL}^{\prime}\left(  a_{R}\right)
\nonumber\\
&  +\frac{\left(  r-a_{R}\right)  ^{2}}{2!}\mathcal{R}_{RL}^{\prime\prime
}\left(  a_{R}\right)  +\frac{\left(  r-a_{R}\right)  ^{3}}{3!}\mathcal{R}%
_{RL}^{\prime\prime\prime}\left(  a_{R}\right)  +o\nonumber\\
&  \equiv\sum_{d=0}^{3}\frac{\left(  r-a_{R}\right)  ^{d}}{d!}\mathcal{R}%
_{RL}^{\left(  d\right)  }\left(  a_{R}\right)  +o. \label{input}%
\end{align}
Here and in the following, $Y_{L}\left(  \mathbf{\hat{r}}\right)  $ in denotes
a real, cubic harmonic \cite{Y}, and a global coordinate system is assumed for
simplicity. Moreover, terms of order higher than 3rd in $r-a,$ i.e. smaller
than $\left(  r-a\right)  ^{3},$ are denoted:%
\begin{equation}
o\equiv o\left(  \left(  r-a\right)  ^{3}\right)  . \label{o}%
\end{equation}

Input to the interpolation is thus the vector $\mathcal{R}_{RL}^{\left(
d\right)  }\left(  a_{R}\right)  $ with components $dRL.$ It could be output
from an electronic-structure calculation.

We shall construct a set of v\&d functions, $\varrho_{dRL}\left(
\mathbf{r}\right)  ,$ which satisfies the following super-unitary boundary
condition on the hard spheres:
\begin{equation}
r\mathcal{\hat{P}}_{R^{\prime}L^{\prime}}\left(  r\right)  \varrho
_{dRL}\left(  \mathbf{r}\right)  =\delta_{R^{\prime}R}\delta_{L^{\prime}%
L}\frac{\left(  r-a_{R}\right)  ^{d}}{d!}+o, \label{v&d}%
\end{equation}
for $l^{\prime}\leq l_{\max},$ in terms of which, the interpolation is given
by eq.$\,$(\ref{interpolation}). We shall also find the localized Coulomb
potential, $\varphi_{dRL}^{loc}\left(  \mathbf{r}\right)  ,$ from
$\varrho_{dRL}\left(  \mathbf{r}\right)  $ in terms of which the localized
potential from $\varrho\left(  \mathbf{r}\right)  $ is as given by
eq.$\,$(\ref{Potential}). Similarly for the regular potential.

Note that we have defined the value and derivatives as those of $r$ times the
$L$-projection. This has been done in order to simplify the derivation of the
v\&d functions through use of the radial wave equation (\ref{rad}).

The v\&d functions will be constructed from 4 sets of USWs with 4 different
energies, $\varepsilon=\epsilon_{1},\,\epsilon_{2},\,\epsilon_{3},$ and
$\epsilon_{4}.$ But first, we consider a single energy.

\subsection{USWs and their slope matrix\label{USW}}

A unitary spherical wave (USW), $\psi_{RL}\left(  \varepsilon,\mathbf{r}%
\right)  ,$ is a wave-equation solution (\ref{wave eq}) in the interstitial
and satisfies the boundary condition on the spheres that, for $l^{\prime}\leq
l_{\max},$%
\begin{equation}
\mathcal{\hat{P}}_{R^{\prime}L^{\prime}}\left(  a\right)  \psi_{RL}\left(
\varepsilon,\mathbf{r}\right)  =\delta_{R^{\prime}R}\delta_{L^{\prime}L}%
Y_{L}\left(  \mathbf{\hat{r}}_{R}\right)  . \label{USWdef}%
\end{equation}
That is, the projection onto the cubic harmonic, $Y_{L^{\prime}}\left(
\mathbf{\hat{r}}_{R^{\prime}}\right)  ,$ on the sphere centered at
$\mathbf{R}^{\prime}$ with radius $a_{R^{\prime}}$ vanishes, unless
$\mathbf{R}^{\prime}=\mathbf{R}$ and $L^{\prime}=L,$ in which case the
projection is unity. Since this holds for any $\mathbf{R}^{\prime}$ and
$L^{\prime},$ $\psi_{RL}\left(  \varepsilon,\mathbf{r}\right)  $ has
cubic-harmonic character, $L$, on its "own" sphere, $R,$ while on all other
spheres, it has vanishing $L^{\prime}$-projections for all $l^{\prime}\leq
l_{\max}$. As a consequence, the USW is localized in the interstitial close to
its own sphere (but its analytical continuation diverges at the sphere centers).

While the USW is defined to vanish inside all spheres, its projection at and
outside any sphere is \cite{94Trieste,98MRS}:%
\begin{align}
&  \mathcal{\hat{P}}_{R^{\prime}L^{\prime}}\left(  r\right)  \psi_{RL}\left(
\varepsilon,\mathbf{r}\right) \label{r9}\\
&  =f_{R^{\prime}l^{\prime}}(\varepsilon,r)\delta_{R^{\prime}R}\delta
_{L^{\prime}L}+g_{R^{\prime}l^{\prime}}(\varepsilon,r)S_{R^{\prime}L^{\prime
},RL}\left(  \varepsilon\right)  ,\nonumber
\end{align}
where $f$ and $g$ are the two linearly independent, dimensionless solutions of
the radial wave equation (\ref{rad}), defined by the boundary conditions:%
\begin{equation}
f_{Rl}\left(  \varepsilon,a_{R}\right)  =1,\;\;f_{Rl}^{\prime}\left(
\varepsilon,a_{R}\right)  =0, \label{f}%
\end{equation}
and%
\begin{equation}
g_{Rl}\left(  \varepsilon,a_{R}\right)  =0,\;\;g_{Rl}^{\prime}\left(
\varepsilon,a_{R}\right)  =1/a_{R}. \label{g}%
\end{equation}
$S\left(  \varepsilon\right)  $ is the dimensionless \emph{slope matrix} for
the USW set. Its on-site diagonal element, $S_{RL,RL}\left(  \varepsilon
\right)  ,$ is the radial logarithmic derivative, $a_{R}\left.  \partial
/\partial r\right\vert _{a_{R}}$, of the $L$-projection of $\psi_{RL}\left(
\varepsilon,\mathbf{r}\right)  $ at its own sphere, while the off-site
element, $S_{RL,R^{\prime}L^{\prime}}\left(  \varepsilon\right)  ,$ is the
dimensionless slope, $a_{R^{\prime}}\left.  \partial/\partial r_{R^{\prime}%
}\right\vert _{a_{R^{\prime}}},$ of the $L^{\prime}$-projection at the
$R^{\prime}$-sphere. The on-site off-diagonal element, $S_{RL,RL^{\prime}%
}\left(  \varepsilon\right)  ,$ gives the dimensionless slope at the own
sphere of another $L^{\prime}$-projection.

We need to generate the slope matrix from analytically known functions. For
this purpose, we first express the set of USWs as superpositions of the
decaying solutions of the wave equation:%
\begin{equation}
\psi_{RL}\left(  \varepsilon,\mathbf{r}\right)  =\sum_{R^{\prime}L^{\prime}%
}h_{l^{\prime}}\left(  \varepsilon,r_{R^{\prime}}\right)  Y_{L^{\prime}%
}\left(  \mathbf{\hat{r}}_{R^{\prime}}\right)  M_{R^{\prime}L^{\prime}%
,RL}\left(  \varepsilon\right)  . \label{USWinHank}%
\end{equation}
valid in the interstitial. Here, the radial function,%
\begin{align}
h_{l}\left(  \varepsilon,r\right)   &  \equiv-i\kappa^{l+1}h_{l}^{\left(
1\right)  }\left(  \kappa r\right)  =\kappa^{l+1}\left[  n_{l}\left(  \kappa
r\right)  -ij_{l}\left(  \kappa r\right)  \right] \nonumber\\
&  \equiv n_{l}\left(  \varepsilon,r\right)  -i\kappa\varepsilon^{l}%
j_{l}\left(  \varepsilon,r\right)  , \label{h}%
\end{align}
is the spherical Hankel function of the 1st kind, renormalized so that it is
an analytical function of $\varepsilon\equiv\kappa^{2}$ (for $0\leq
\angle\varepsilon<2\pi)$ and a decaying function of $r$ (when $0<\angle
\varepsilon<2\pi)$. It is real for real, non-positive energy and $r>0$. In the
second line of eq.$\,$(\ref{h}), we have expressed the Hankel function in
terms of spherical Neumann and Bessel functions$,$ renormalized such that they
are real for \emph{all} real $\varepsilon:$%
\begin{equation}
n_{l}\left(  \varepsilon,r\right)  \equiv\kappa^{l+1}n_{l}\left(  \kappa
r\right)  \;\;\mathrm{and\;\;}j_{l}\left(  \varepsilon,r\right)  \equiv
\kappa^{-l}j_{l}\left(  \kappa r\right)  . \label{nandj}%
\end{equation}
The Bessel function is regular and the Neumann function irregular at the
origin. As examples, for $\varepsilon=0$:%
\begin{align}
h_{l}\left(  0,r\right)   &  =n_{l}\left(  0,r\right)  =-\frac{\left(
2l-1\right)  !!}{r^{l+1}}\nonumber\\
\mathrm{and}\;\;j_{l}\left(  0,r\right)   &  =\frac{r^{l}}{\left(
2l+1\right)  !!}, \label{eps0}%
\end{align}
where $\left(  2l+1\right)  !!\equiv\left(  2l+1\right)  \left(  2l-1\right)
..1$ and $\left(  -1\right)  !!\equiv1.$ For $l=0$:
\begin{align*}
&  h_{0}\left(  \varepsilon,r\right) \\
&  =-\frac{\exp\left(  -\sqrt{-\varepsilon}r\right)  }{r}\,=-\frac
{\cosh\left(  \sqrt{-\varepsilon}r\right)  }{r}+\frac{\sinh\left(
\sqrt{-\varepsilon}r\right)  }{r}\\
&  =n_{0}\left(  \varepsilon,r\right)  +\sqrt{-\varepsilon}j_{0}\left(
\varepsilon,r\right)  =-\frac{\exp\left(  i\kappa r\right)  }{r}\\
&  =-\frac{\cos\left(  \kappa r\right)  }{r}-\frac{i\sin\left(  \kappa
r\right)  }{r}=n_{0}\left(  \varepsilon,r\right)  -i\kappa j_{0}\left(
\varepsilon,r\right)  .
\end{align*}

In analogy with eq.$\,$(\ref{r9}), the $L^{\prime}$-projection around site
$R^{\prime}$ of a Hankel function times a cubic harmonic is:%
\begin{align}
&  \mathcal{\hat{P}}_{R^{\prime}L^{\prime}}\left(  r\right)  h_{l}\left(
\varepsilon,r_{R}\right)  Y_{L}\left(  \mathbf{\hat{r}}_{R}\right)
\label{Ph}\\
&  =n_{l^{\prime}}(\varepsilon,r)\delta_{R^{\prime}R}\delta_{L^{\prime}%
L}+j_{l^{\prime}}(\varepsilon,r)B_{R^{\prime}L^{\prime},RL}\left(
\varepsilon\right)  ,\nonumber
\end{align}
where $B\left(  \varepsilon\right)  $ is the \emph{bare structure matrix} with
the analytically known elements \cite{KKR,HamSegall,kappa}:
\begin{equation}
B_{RL^{\prime},RL}\left(  \varepsilon\right)  =\sqrt{-\varepsilon}%
\varepsilon^{l}\delta_{LL^{\prime}}=-i\kappa\varepsilon^{l}\delta_{LL^{\prime
}} \label{bareonsite}%
\end{equation}
and, for $\mathbf{R}^{\prime}\mathbf{\neq R:}$%
\begin{align}
B_{R^{\prime}L^{\prime},RL}\left(  \varepsilon\right)   &  \equiv
\sum_{L^{\prime\prime}}4\pi i^{-l+l^{\prime}-l^{\prime\prime}}\kappa
^{l+l^{\prime}-l^{\prime\prime}}\mathcal{C}_{LL^{\prime}L^{\prime\prime}%
}\nonumber\\
&  \times h_{l^{\prime\prime}}\left(  \varepsilon,\left\vert \mathbf{R}%
^{\prime}\mathbf{-R}\right\vert \right)  Y_{L^{\prime\prime}}^{\ast}\left(
\widehat{\mathbf{R}^{\prime}\mathbf{-R}}\right)  . \label{bare}%
\end{align}
The Gaunt coefficients,%
\[
\mathcal{C}_{LL^{\prime}L^{\prime\prime}}\equiv\int Y_{L}\left(
\mathbf{\hat{r}}\right)  Y_{L^{\prime}}^{\ast}\left(  \mathbf{\hat{r}}\right)
Y_{L^{\prime\prime}}\left(  \mathbf{\hat{r}}\right)  d\mathbf{\hat{r},}%
\]
for the cubic harmonics are real and the $L^{\prime\prime}$-sum includes only
the terms with $l^{\prime\prime}=\left\vert l^{\prime}-l\right\vert ,$
$\left\vert l^{\prime}-l\right\vert +2,...,$ and $l^{\prime}+l,$ for which the
factor $i^{-l+l^{\prime}-l^{\prime\prime}}\kappa^{l+l^{\prime}-l^{\prime
\prime}}$ is $\left(  -\right)  ^{l}$ times respectively $\left(
-\varepsilon\right)  ^{\min\left\{  l,l^{\prime}\right\}  },$ $\left(
-\varepsilon\right)  ^{\min\left\{  l,l^{\prime}\right\}  -1},$ $...,$ and
$1$. Hence, the bare structure matrix is real and symmetric for $\varepsilon
\leq0.$ For $\varepsilon=0,$ it reduces to:%
\begin{align}
&  B_{R^{\prime}L^{\prime},RL}\left(  0\right) \label{bare0}\\
&  \equiv4\pi\left(  -\right)  ^{l+1}\sum_{m^{\prime\prime}}\mathcal{C}%
_{LL^{\prime}L^{\prime\prime}}\frac{\left(  2l^{\prime\prime}-1\right)
!!}{\left\vert \mathbf{R}^{\prime}\mathbf{-R}\right\vert ^{l^{\prime\prime}%
+1}}Y_{L^{\prime\prime}}^{\ast}\left(  \widehat{\mathbf{R}^{\prime}%
\mathbf{-R}}\right)  ,\nonumber
\end{align}
with $l^{\prime\prime}=l+l^{\prime},$ and the projection (\ref{Ph}) becomes
that of the potential from an electrostatic multipole. Apart from
normalizations, $B\left(  0\right)  $ is also the structure matrix used in
canonical band theory \cite{75PRB}. For $\varepsilon>0,$ the real and
imaginary parts of $B\left(  \varepsilon\right)  $ are symmetric, i.e.
$B\left(  \varepsilon\right)  $ is \emph{not} Hermitian for $\varepsilon>0$.

The imaginary part of the Hankel function is according to (\ref{h}) the
free-electron solution in all space with angular-momentum $L$ and energy
$\varepsilon$ of Schr\"{o}dinger's equation while the real part is the
solution irregular at the origin and decaying. Applied to the projection
(\ref{Ph}), this means that only when the bare structure matrix has an
imaginary part, do free-electron solutions exist, otherwise the solutions are localized.

We now relate to the hard-sphere solutions, the USWs. First, we express the
Bessel-Neumann set of linearly independent solutions of the radial wave
equation in terms of the value-slope set:%
\begin{equation}
\left\{
\begin{array}
[c]{c}%
n_{l}\left(  \varepsilon,r\right) \\
j_{l}\left(  \varepsilon,r\right)
\end{array}
\right\}  =\left[
\begin{array}
[c]{cc}%
n_{l}\left(  \varepsilon,a_{R}\right)  & a_{R}n_{l}^{\prime}\left(
\varepsilon,a_{R}\right) \\
j_{l}\left(  \varepsilon,a_{R}\right)  & a_{R}j_{l}^{\prime}\left(
\varepsilon,a_{R}\right)
\end{array}
\right]  \left\{
\begin{array}
[c]{c}%
f_{Rl}\left(  \varepsilon,r\right) \\
g_{Rl}\left(  \varepsilon,r\right)
\end{array}
\right\}  , \label{njinfg}%
\end{equation}
where we have used eq.s$\,$(\ref{f})-(\ref{g}). Here, the values and radial
derivatives are related by the Wronskian:
\[
r^{2}\left[  j_{l}\left(  \varepsilon,r\right)  n_{l}^{\prime}\left(
\varepsilon,r\right)  -n_{l}\left(  \varepsilon,r\right)  j_{l}^{\prime
}\left(  \varepsilon,r\right)  \right]  =1.
\]
The inverse transformation is seen to be:%
\begin{equation}
\left\{
\begin{array}
[c]{c}%
f\left(  \varepsilon,r\right) \\
g\left(  \varepsilon,r\right)
\end{array}
\right\}  =a\left[
\begin{array}
[c]{cc}%
-aj^{\prime}\left(  \varepsilon,a\right)  & an^{\prime}\left(  \varepsilon
,a\right) \\
j\left(  \varepsilon,a\right)  & -n\left(  \varepsilon,a\right)
\end{array}
\right]  \left\{
\begin{array}
[c]{c}%
n\left(  \varepsilon,r\right) \\
j\left(  \varepsilon,r\right)
\end{array}
\right\}  , \label{fginnj}%
\end{equation}
where we have used the Wronskian and have dropped the subscripts $R$ and $l$.

Next, we proceed with expanding the set of USWs in terms of the set of
decaying Hankel functions (\ref{USWinHank}). It is, however, simpler to derive
the inverse expansion:%
\begin{equation}
\sum_{R^{\prime}L^{\prime}}\psi_{R^{\prime}L^{\prime}}\left(  \varepsilon
,\mathbf{r}\right)  \left[  M\left(  \varepsilon\right)  \right]  _{R^{\prime
}L^{\prime},RL}^{-1}=h_{l}\left(  \varepsilon,r_{R}\right)  Y_{L}\left(
\mathbf{\hat{r}}_{R}\right)  , \label{HankinUSW}%
\end{equation}
because for this, we can exploit the unitary properties (\ref{r9})-(\ref{g})
of the USWs together with the projections (\ref{Ph}) of the Hankel function.
Projection onto values, $\mathcal{\hat{P}}_{R^{\prime}L^{\prime}}\left(
a_{R^{\prime}}\right)  ,$ immediately yields:%
\begin{align*}
&  \left[  M\left(  \varepsilon\right)  \right]  _{R^{\prime}L^{\prime}%
,RL}^{-1}\\
&  =n_{l^{\prime}}(\varepsilon,a_{R^{\prime}})\delta_{R^{\prime}R}%
\delta_{L^{\prime}L}+j_{l^{\prime}}(\varepsilon,a_{R^{\prime}})B_{R^{\prime
}L^{\prime},RL}\left(  \varepsilon\right)  ,
\end{align*}
so that the solution is:%
\begin{align}
M\left(  \varepsilon\right)   &  =\left[  n(\varepsilon,a)+j(\varepsilon
,a)B\left(  \varepsilon\right)  \right]  ^{-1}\nonumber\\
&  =\left[  \frac{n(\varepsilon,a)}{j(\varepsilon,a)}+B\left(  \varepsilon
\right)  \right]  ^{-1}\frac{1}{j(\varepsilon,a)} \label{M}%
\end{align}
\qquad Here, and often in the following, we use matrix notation where
$n(\varepsilon,a)/j(\varepsilon,a)$ and $1/j(\varepsilon,a)$ are diagonal
matrices with the respective elements $n_{l}(\varepsilon,a_{R})/j_{l}%
(\varepsilon,a_{R})$ and $1/j_{l}(\varepsilon,a_{R}).$ The matrix in the
square parenthesis in (\ref{M}) is symmetric, with $n_{l}(\varepsilon
,a_{R})/j_{l}(\varepsilon,a_{R})=\kappa^{2l+1}\left[  \cot\eta_{Rl}\left(
\varepsilon\right)  -i\right]  $ and $\eta_{Rl}\left(  \varepsilon\right)  $
the hard-sphere phase shifts.

The slope matrix is derived by projecting the multi-center expansion
(\ref{HankinUSW}) onto slopes. Application of $\mathcal{\hat{P}}_{R^{\prime
}L^{\prime}}\left(  r\right)  $ first yields:
\[
\left[  f(\varepsilon,r)+g(\varepsilon,r)S\left(  \varepsilon\right)  \right]
M\left(  \varepsilon\right)  ^{-1}=n(\varepsilon,r)+j(\varepsilon,r)B\left(
\varepsilon\right)
\]
in matrix notation. Its right-hand side becomes after transformation
(\ref{njinfg}) to the $\left\{  f,g\right\}  $ set:%
\begin{align*}
&  f\left(  \varepsilon,r\right)  n\left(  \varepsilon,a\right)  +g\left(
\varepsilon,r\right)  an^{\prime}\left(  \varepsilon,a\right) \\
&  +\left[  f\left(  \varepsilon,r\right)  j\left(  \varepsilon,r\right)
+g\left(  \varepsilon,r\right)  aj^{\prime}\left(  \varepsilon,a\right)
\right]  B\left(  \varepsilon\right)  .
\end{align*}
Equating now the coefficients to $f\left(  \varepsilon,r\right)  $ of course
yields expression (\ref{M}), while equating those to $g\left(  \varepsilon
,r\right)  $ yields:%
\begin{align*}
S\left(  \varepsilon\right)  M\left(  \varepsilon\right)  ^{-1}  &
=an^{\prime}\left(  \varepsilon,a\right)  +aj^{\prime}\left(  \varepsilon
,a\right)  B\left(  \varepsilon\right) \\
&  =\frac{a}{j\left(  \varepsilon,a\right)  }\left[  n^{\prime}\left(
\varepsilon,a\right)  j\left(  \varepsilon,a\right)  -j^{\prime}\left(
\varepsilon,a\right)  n\left(  \varepsilon,a\right)  \right] \\
&  +aj^{\prime}\left(  \varepsilon,a\right)  \left[  \frac{n\left(
\varepsilon,a\right)  }{j\left(  \varepsilon,a\right)  }+B\left(
\varepsilon\right)  \right] \\
&  =\frac{1}{aj\left(  \varepsilon,a\right)  }+aj^{\prime}\left(
\varepsilon,a\right)  \left[  \frac{n\left(  \varepsilon,a\right)  }{j\left(
\varepsilon,a\right)  }+B\left(  \varepsilon\right)  \right] \\
&  =\frac{1}{aj\left(  \varepsilon,a\right)  }+\frac{aj^{\prime}\left(
\varepsilon,a\right)  }{j\left(  \varepsilon,a\right)  }M\left(
\varepsilon\right)  ^{-1}.
\end{align*}
In order to simplify the solution for $S\left(  \varepsilon\right)  ,$ we have
on the right-hand side separated a term proportional to $M\left(
\varepsilon\right)  ^{-1}$ and used the Wronskian. As a result, we obtain the
most important relation:%
\begin{align}
S\left(  \varepsilon\right)   &  =\frac{1}{aj\left(  \varepsilon,a\right)
}M\left(  \varepsilon\right)  +\frac{aj^{\prime}\left(  \varepsilon,a\right)
}{j\left(  \varepsilon,a\right)  }\label{aS}\\
&  =\frac{aj^{\prime}\left(  \varepsilon,a\right)  }{j\left(  \varepsilon
,a\right)  }+\frac{1}{aj\left(  \varepsilon,a\right)  }\left[  \frac
{n(\varepsilon,a)}{j(\varepsilon,a)}+B\left(  \varepsilon\right)  \right]
^{-1}\frac{1}{j\left(  \varepsilon,a\right)  }.\nonumber
\end{align}
between the dimensionless slope matrix, $S\left(  \varepsilon\right)  ,$ and
the bare structure matrix, $B\left(  \varepsilon\right)  $. In (\ref{aS}) all
quantities \emph{other than} $S,$ $M,$ and $B$ are diagonal matrices.
Specifically the elements $a_{R}j_{l}^{\prime}\left(  \varepsilon
,a_{R}\right)  /j_{l}\left(  \varepsilon,a_{R}\right)  $ are the radial
logarithmic derivatives of the Bessel functions. The dimensionless slope
matrix is not symmetric, but $aS\left(  \varepsilon\right)  $ with the
elements $a_{R}S_{RL,R^{\prime}L^{\prime}}\left(  \varepsilon\right)  ,$ the
so-called \emph{screened} structure matrix
\cite{84TB-LMTO,92MRS,94Trieste,95Zeller,98MRS,00Rashmi,00NMTO,Martin,12MartinandTerakuras}%
, is seen to be symmetric and real. This holds not only for $\varepsilon\leq
0$, but for all energies where no solution exists of Schr\"{o}dinger's
equation for the hard-sphere interstitial. That is, where no wave-equation
solution exists which satisfies the homogeneous boundary condition that the
solution vanishes at \emph{all} spheres for \emph{all} $l\leq l_{\max}$. As
seen from expansion (\ref{r9}), such solutions are given by the imaginary part
of the slope matrix. It is by forbidding the space region inside spheres, i.e.
by insertion of hard spheres, that the lowest energy, $\varepsilon_{\hom},$
for which solutions of the homogeneous problem exist, is pushed above zero.
$\varepsilon_{\hom}$ is the highest energy for which the USW set is localized.
With $\varepsilon<\varepsilon_{\hom}$, the matrix inversion in eq.$\,$%
(\ref{aS}) can be done in real space for a local cluster with the range of the
USWs (rather than that of the Hankel functions), which contains at the order
of 100 sites \cite{84TB-LMTO,92MRS,95Zeller} (see Sect.\thinspace\ref{N}).

For application to charge densities in condensed matter \cite{SSW}, we need
$l_{\max}\sim4$ and $\varepsilon\lesssim0$.

It may be noted that the spheres are hard only for the\emph{ low} angular
momenta, $l\leq l_{\max},$ but transparent for the high ones, $l>l_{\max}.$
This means that the USWs have high-$l$ tails from the Hankel functions
surviving inside the spheres. Specifically: The set of USWs, $\psi_{RL}\left(
\varepsilon,\mathbf{r}\right)  ,$ with $R$ being any site and $L$ low$,$ is
given by the superpositions (\ref{USWinHank}) of Hankel functions times cubic
harmonics with $R^{\prime}$ running over all sites and $L^{\prime}$ over all
low angular momenta. The low-$l$ components of the Hankel functions (\ref{Ph})
are truncated inside all spheres while the high-$l$ ones remain. Those
high-$l$ parts of the Hankel-function tails contribute
\begin{align}
\psi_{RL}\left(  \varepsilon,\mathbf{r}\right)   &  =\sum_{l^{\prime\prime
}>l\max}^{\infty}\sum_{m^{\prime\prime}=-l^{\prime\prime}}^{l^{\prime\prime}%
}j_{l^{\prime\prime}}(\varepsilon,r)Y_{R^{\prime\prime}L^{\prime\prime}%
}\left(  \mathbf{\hat{r}}_{R^{\prime\prime}}\right) \nonumber\\
&  \times\sum_{R^{\prime}\neq R^{\prime\prime}}\sum_{L^{\prime}}%
B_{R^{\prime\prime}L^{\prime\prime},R^{\prime}L^{\prime}}\left(
\varepsilon\right)  M_{R^{\prime}L^{\prime},RL}\left(  \varepsilon\right)  ,
\label{high}%
\end{align}
to the USW \emph{inside} the $R^{\prime\prime}$-sphere. We usually avoid
evaluating this contribution. Rather, we use the multi-center expansion
(\ref{USWinHank}) in all space and subtract the low-$l$ components inside the
spheres. When finally adding to the interpolation the proper function inside
the spheres, we only add its low-$l$ components and let the high-$l$ ones be
those of the interpolation. This makes the final function smooth, but
approximate as regards the high-$l$ components inside the spheres.

USWs look like the ones shown in Fig. 1. Here we have chosen $L\mathrm{=}0,$
which is the most appropriate for expanding charge densities. In the two first
panels$,$ we show the $s$-USW for six different USW sets, specifically sets
with 3 different energies and for 2 different hard-sphere structures. The
latter are body-centered cubic (bcc), which is closely packed, and diamond
(dia), which is open and can be viewed as bcc with every second sphere removed
to be part of the interstitial. In both structures, all spheres are
equivalent. We see that the USW for the higher energies spread into the voids
but, nevertheless, stays essentially inside its Voronoi (Wigner-Seitz) cell.

Because they are solutions of the wave equation (\ref{wave eq}), the USWs are
invariant to a uniform scaling $\left(  t\right)  $ of the structure, provided
that they are considered as functions of a dimensionless space variable
$\mathbf{r/}t$ and the dimensionless energy variable $\varepsilon t^{2}.$

\bigskip

\begin{figure}[htbp]
  \vspace{0cm}
  \begin{center}
  \includegraphics[width=0.45\textwidth]{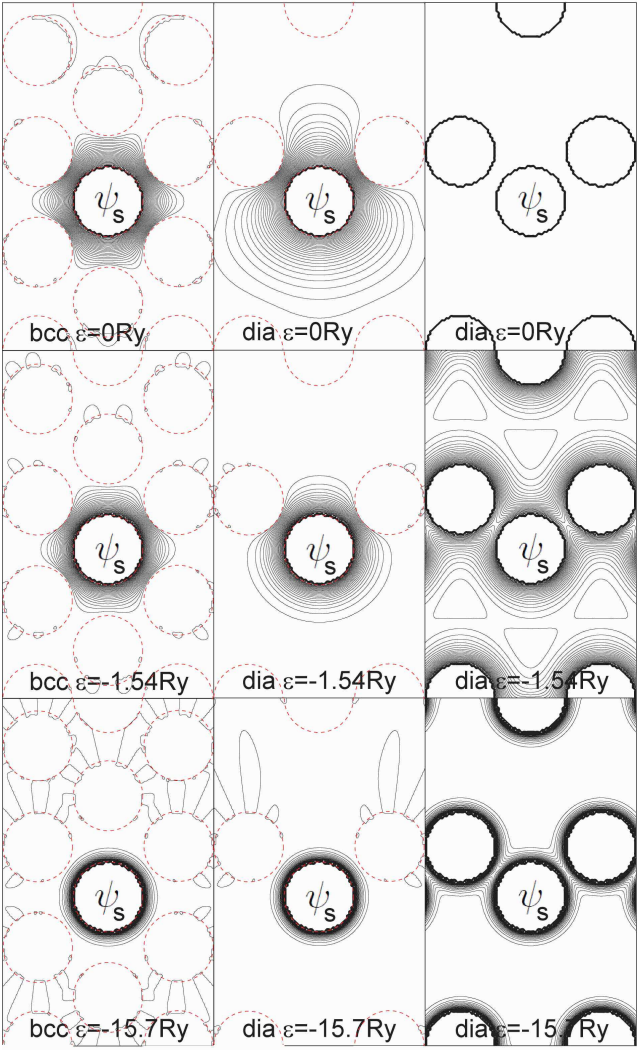}
  \caption{(Color online)
USW with $L=0$ from six different sets: The ones with $\varepsilon
t^{2}=0$ (top), $-7.6$ (middle), or $-77$ (bottom) and the bcc structure
(left) or the diamond structure (middle and right). Here, $t$ is the radius of
touching spheres, i.e. half the nearest-neighbor distance. For
diamond-structured silicon, $t=2.22$ Bohr radii and, for this case, the
energies are therefore $\varepsilon=0$ (top)$,$ $-1.54$ (middle) or $-15.7$ Ry
(bottom). The contours of the USW are in the (110) diagonal plane and range
from $0$ to $Y_{00}=1/\sqrt{4\pi}\approx0.28$ in steps of 0.01. The (red) dots
indicate the hard spheres whose radius was taken to be $a=0.8t$. In the bcc
structure, the spheres are at the corners and body-centers of cubes. The
diamond structure follows from bcc structure by deleting every second sphere,
i.e. by including it in the interstitial. In both structures, all sites are
equivalent (translationally in bcc, and with every second site inverted in
dia), so that for a given structure and energy there is only one $s$-USW
shape, $\psi_{s}$. It was generated from eq.s (\ref{USWinHank}) and (\ref{M})
by screening the bare spherical $s$-wave, $e^{-r\sqrt{-\varepsilon}}/r,$ with
all waves centered at the $51$ (bcc) or $87$ (dia) nearest sites and having
$l\leq l_{\max}=9$. This high value of $l_{\max}$ was chosen in order that the
lowest contour display the hard spheres. The last panel shows the $s$-USW
symmetrized with respect to the identity representation of the diamond
space-group (Fd3m). The symmetrized USW with the lowest energy decays rapidly
into the interstitial, the one with intermediate energy decays moderately
fast, and the one with zero energy stays constant (see eq.$\,$(\ref{1=LCUSW})
and Sect.$\,$\ref{flat}). For the symmetrized $s$-USW, $l_{\max}=3$ suffices.
See Sect.s \ref{USW} and \ref{Symmetry}.
}
  \end{center}
\end{figure}

\bigskip

\subsection{Taylor series in $r-a$ of the radial functions\label{Taylor}}

In order to combine USW-sets with different energies linearly into a set of
v\&d functions, $\varrho_{dRL}\left(  \mathbf{r}\right)  ,$ with the
super-unitary property (\ref{v&d}), we need to expand the radial functions $f$
and $g$ defined in (\ref{f}) and (\ref{g}) in $\varepsilon$-dependent Taylor
series in $r-a.$ Since the radial wave equation (\ref{rad}) is simplest when
expressed in terms of $r$ times the radial function, it is convenient instead
of $f$ and $g$ to use%
\begin{equation}
U\left(  \varepsilon,r\right)  \equiv ru\left(  \varepsilon,r\right)  \equiv
r\left[  f\left(  \varepsilon,r\right)  -g\left(  \varepsilon,r\right)
\right]  \label{UU}%
\end{equation}
and
\begin{equation}
G\left(  \varepsilon,r\right)  \equiv rg\left(  \varepsilon,r\right)  ,
\label{GG}%
\end{equation}
because they satisfy the boundary conditions:%
\begin{equation}
U\left(  \varepsilon,a\right)  =a,\;\;U^{\prime}\left(  \varepsilon,a\right)
=0, \label{U}%
\end{equation}
and%
\begin{equation}
G\left(  \varepsilon,a\right)  =0,\;\;G^{\prime}\left(  \varepsilon,a\right)
=1. \label{V}%
\end{equation}
Here again, the subscripts $R$ and $l$ have been dropped for simplicity. The
projection (\ref{r9}) of the USWs, expressed in the form needed for the
definition of the v\&d functions, is then:%
\begin{align}
&  r\mathcal{\hat{P}}_{R^{\prime}L^{\prime}}\left(  r\right)  \psi_{RL}\left(
\varepsilon,\mathbf{r}\right) \nonumber\\
&  =ru_{R^{\prime}l^{\prime}}(\varepsilon,r)\delta_{R^{\prime}R}%
\delta_{L^{\prime}L}+rg_{R^{\prime}l^{\prime}}(\varepsilon,r)\mathcal{S}%
_{R^{\prime}L^{\prime},RL}\left(  \varepsilon\right) \label{r02}\\
&  =U_{R^{\prime}l^{\prime}}(\varepsilon,r)\delta_{R^{\prime}R}\delta
_{L^{\prime}L}+G_{R^{\prime}l^{\prime}}(\varepsilon,r)\mathcal{S}_{R^{\prime
}L^{\prime},RL}\left(  \varepsilon\right)  , \label{r12}%
\end{align}
where the script slope matrix is the one appropriate for $r\mathcal{\hat{P}%
}_{R^{\prime}L^{\prime}}\left(  r\right)  \psi_{RL}\left(  \varepsilon
,\mathbf{r}\right)  :$
\begin{equation}
\mathcal{S}_{R^{\prime}L^{\prime},RL}\left(  \varepsilon\right)  \equiv
S_{R^{\prime}L^{\prime},RL}\left(  \varepsilon\right)  +\delta_{R^{\prime}%
R}\delta_{L^{\prime}L} \label{r13}%
\end{equation}

We now expand the $r$-dependence of the radial functions entering the
projection (\ref{r12}) in an $\varepsilon$-dependent Taylor series in $r-a$
using the radial wave equation (\ref{rad}). For a radial solution,
$\Psi\left(  \varepsilon,r\right)  \equiv r\psi\left(  \varepsilon,r\right)
,$ with boundary conditions $\Psi\left(  a\right)  $ and $\Psi^{^{\prime}%
}\left(  a\right)  $ chosen to be independent of energy, the 2nd and 3rd
radial derivatives are simply:
\begin{align*}
\Psi^{\prime\prime}\left(  \varepsilon,r\right)   &  =\left[  w\left(
r\right)  -\varepsilon\right]  \Psi\left(  \varepsilon,r\right)
,\;\mathrm{and}\\
\Psi^{\prime\prime\prime}\left(  \varepsilon,r\right)   &  =w^{\prime}\left(
r\right)  \Psi\left(  \varepsilon,r\right)  +\left[  w\left(  r\right)
-\varepsilon\right]  \Psi^{\prime}\left(  \varepsilon,r\right)
\end{align*}
where $w\left(  r\right)  \equiv l\left(  l+1\right)  /r^{2}$ is the
centrifugal potential. Using these derivatives for the functions with the
boundary conditions (\ref{U}) and (\ref{V}), yield the following Taylor
series:%
\begin{equation}
U\left(  \varepsilon,r\right)  =a+a\left(  w-\varepsilon\right)  \frac{\left(
r-a\right)  ^{2}}{2!}+aw^{\prime}\frac{\left(  r-a\right)  ^{3}}{3!}+o
\label{p1}%
\end{equation}
and%
\begin{equation}
G\left(  \varepsilon,r\right)  =\left(  r-a\right)  +\left(  w-\varepsilon
\right)  \frac{\left(  r-a\right)  ^{3}}{3!}+o, \label{p2}%
\end{equation}
with $o$ as defined in (\ref{o}). Moreover,%
\begin{equation}
w\equiv\frac{l\left(  l+1\right)  }{a^{2}}\;\;\mathrm{and\;\;}aw^{\prime
}\equiv-\frac{2l\left(  l+1\right)  }{a^{2}} \label{centrif}%
\end{equation}
are respectively the value- and $a$ times the first derivative of the
centrifugal potential at the hard sphere.

As an example, for $l=0,$ $w=0,$ so that $U_{s}\left(  \varepsilon,r\right)
/a$ is a\emph{\ }function\emph{\ }of $\left(  r-a\right)  ^{2}\varepsilon$ and
$G_{s}\left(  \varepsilon,r\right)  $ is an\emph{\ }odd function\emph{\ }of
$\left(  r-a\right)  \sqrt{-\varepsilon}.$ In fact, $U_{s}\left(
\varepsilon,r\right)  =a\cosh\left\{  \left(  r-a\right)  \sqrt{-\varepsilon
}\right\}  $ and $G_{s}\left(  \varepsilon,r\right)  =\sqrt{-\varepsilon}%
^{-1}\sinh\left\{  \left(  r-a\right)  \sqrt{-\varepsilon}\right\}  .$

\subsection{Energy-divided differences\label{Sectdivdif}}

Next, we must form linear combinations with \emph{zero }value, 1st, 2nd, and
3rd radial derivatives in all \emph{non}-eigenchannels of the USW-sets with 4
different energies.\emph{\ }This means: the projections formed from
(\ref{r12}) of those linear combinations must have all terms with $R^{\prime
}\neq R$ or $L^{\prime}\neq L$ smaller than $\left(  r-a\right)  ^{3}.$ From
eq.s$\,$(\ref{p1}) and (\ref{p2}) we see that the second and higher
\emph{energy} derivatives of $U\left(  \varepsilon,r\right)  $ and $G\left(
\varepsilon,r\right)  $ are smaller than $\left(  r-a\right)  ^{3},$ and
differentiation of the USWs in eq.$\,$(\ref{r12}) with respect to energy can
therefore be used to single out functions which satisfy equations (\ref{v&d}).

Rather than using derivatives at one energy, it is far more flexible and
accurate to use energy-divided differences for \emph{discrete sets} of
energies. From the theory of polynomial approximation (Newton-Lagrange),
remember that if we approximate a function of energy, $\psi\left(
\varepsilon\right)  ,$ by the polynomial of $d$th order which coincides with
$\psi\left(  \varepsilon\right)  $ at the energies $\epsilon_{1}%
,..,\epsilon_{d+1},$ then the highest non-vanishing energy derivative --the
$d$th-- of this polynomial is $d!$ times the $d$th \emph{divided difference.
}The latter can be written in many ways, but the most general and compact is
\cite{00DivDif}:%
\begin{equation}
\sum_{n=1}^{d+1}\frac{\psi\left(  \epsilon_{n}\right)  }{\Pi_{m=1,\neq
n}^{d+1}\left(  \epsilon_{n}-\epsilon_{m}\right)  }\equiv\psi_{1..d+1}\,.
\label{q0}%
\end{equation}
On the right-hand side we have introduced a \emph{notation} according to which
the value (the 0th divided difference), $\psi\left(  \epsilon_{n}\right)  ,$
at $\epsilon_{n}$ is denoted $\psi_{n}.$ The first divided difference,%
\[
\frac{\psi\left(  \epsilon_{m}\right)  -\psi\left(  \epsilon_{n}\right)
}{\epsilon_{m}-\epsilon_{n}}=\frac{\psi_{m}-\psi_{n}}{\epsilon_{m}%
-\epsilon_{n}},
\]
taken at the two energy points, $\epsilon_{m}$ and $\epsilon_{n},$ is denoted
$\psi_{mn}$ like in eq.$\,$(\ref{S12}). The second divided difference,%
\[
\frac{\frac{\psi\left(  \epsilon_{l}\right)  -\psi\left(  \epsilon_{m}\right)
}{\epsilon_{l}-\epsilon_{m}}-\frac{\psi\left(  \epsilon_{m}\right)
-\psi\left(  \epsilon_{n}\right)  }{\epsilon_{m}-\epsilon_{n}}}{\epsilon
_{l}-\epsilon_{n}}=\frac{\psi_{lm}-\psi_{mn}}{\epsilon_{l}-\epsilon_{n}},
\]
taken at the three energy points, $\epsilon_{l},$ $\epsilon_{m},$ and
$\epsilon_{n},$ is denoted $\psi_{lmn}.$ Hence, the general notation is:%
\[
\frac{\psi_{m,m+1,...,n}-\psi_{m+1,...,n,n+1}}{\epsilon_{m}-\epsilon_{n+1}%
}\equiv\psi_{m,m+1,..,n,n+1}.
\]

Note that a divided difference (\ref{q0}) depends on the energy points at
which it is formed, but \emph{not} on their \emph{order, }e.g. $\psi
_{lmn}=\psi_{nlm}.$

If $\psi\left(  \varepsilon\right)  $ itself is a polynomial of order $d,$
then all divided differences formed for more than $d+1$ energies, i.e. of
order higher than $d,$ vanish. From eq.s$\,$(\ref{p1}) and (\ref{p2}),
therefore, all the second and higher energy-divided differences of the
functions $U\left(  \varepsilon,r\right)  $ and $G\left(  \varepsilon
,r\right)  $ are smaller than $\left(  r-a\right)  ^{3}.$ The energy-divided
differences of increasing order are seen to be:
\begin{align}
U_{n}\left(  r\right)   &  =a-a\left(  \epsilon_{n}-w\right)  \frac{\left(
r-a\right)  ^{2}}{2!}\nonumber\\
&  +aw^{\prime}\frac{\left(  r-a\right)  ^{3}}{3!}+o,\label{q14}\\
U_{mn}\left(  r\right)   &  =-a\frac{\left(  r-a\right)  ^{2}}{2!}%
+o,\label{q24}\\
U_{lmn..}\left(  r\right)   &  =o, \label{q4}%
\end{align}
with $o$ defined in (\ref{o}), and:%
\begin{align}
G_{n}\left(  r\right)   &  =\left(  r-a\right)  -\left(  \epsilon
_{n}-w\right)  \frac{\left(  r-a\right)  ^{3}}{3!}+o,\label{q15}\\
G_{mn}\left(  r\right)   &  =-\frac{\left(  r-a\right)  ^{3}}{3!}%
+o,\label{q25}\\
G_{lmn}..\left(  r\right)   &  =o. \label{q5}%
\end{align}

In the following, e.g. in eq.$\,$(\ref{(r-a)3}) below, we shall be using that
the $d$th-order divided difference of a \emph{product} is given by the
binomial rule \cite{00DivDif}:%
\begin{equation}
\left(  G\mathcal{S}\right)  _{1...d+1}=\sum_{m=1}^{d+1}G_{1..m}%
\,\mathcal{S}_{m.d+1}. \label{q1}%
\end{equation}

\section{V\&d functions\label{v&dfunctions}}

After these preliminaries, we are finally in a position form the set of v\&d
functions, $\varrho_{dRL}\left(  \mathbf{r}\right)  ,$ with the super-unitary
property (\ref{v&d}) from the four sets of USWs, $\psi_{RL}\left(
\epsilon_{n},\mathbf{r}\right)  $ with $n=1,2,3,4,$ or -- more conveniently --
from the set of four energy-divided differences: $\psi_{1;RL}\left(
\mathbf{r}\right)  ,$ $\psi_{12;RL}\left(  \mathbf{r}\right)  ,$
$\psi_{123;RL}\left(  \mathbf{r}\right)  ,$ and $\psi_{1234;RL}\left(
\mathbf{r}\right)  .$

Since energy-divided differences are formed for a given element of a vector or
a matrix, we can avoid the $R^{\prime}L^{\prime}$ and $R^{\prime}L^{\prime
},RL$ subscripts by using a matrix notation in which the projection
(\ref{r12}) is written as: $r\mathcal{\hat{P}}\left(  r\right)  \psi\left(
\varepsilon,\mathbf{r}\right)  =U(\varepsilon,r)1+G(\varepsilon,r)\mathcal{S}%
\left(  \varepsilon\right)  .$ Forming energy-divided differences of
increasing order -- from 0th to 3rd -- by use of the binomial formula
(\ref{q1}) then yields:%
\begin{align}
r\mathcal{\hat{P}}\left(  r\right)  \psi_{1}\left(  \mathbf{r}\right)   &
=U_{1}\left(  r\right)  1+G_{1}\left(  r\right)  \mathcal{S}_{1},\nonumber\\
r\mathcal{\hat{P}}\left(  r\right)  \psi_{12}\left(  \mathbf{r}\right)   &
=U_{12}\left(  r\right)  1+G_{1}\left(  r\right)  S_{12}+G_{12}\left(
r\right)  \mathcal{S}_{2},\nonumber\\
r\mathcal{\hat{P}}\left(  r\right)  \psi_{123}\left(  \mathbf{r}\right)   &
=G_{1}\left(  r\right)  S_{123}+G_{12}\left(  r\right)  S_{23}+o,\nonumber\\
r\mathcal{\hat{P}}\left(  r\right)  \psi_{1234}\left(  \mathbf{r}\right)   &
=G_{1}\left(  r\right)  S_{1234}+G_{12}\left(  r\right)  S_{234}+o.
\label{(r-a)3}%
\end{align}
Here, we have chosen to use the energy point with the lower index first, e.g.
$\epsilon_{1}$ before $\epsilon_{2}.$ Moreover, we have used that the
energy-divided differences, except the 0th, of the slope matrices
$\mathcal{S}\left(  \varepsilon\right)  $ and $S\left(  \varepsilon\right)  $
are identical because they differ by merely a constant (see eq.$\,$%
(\ref{r13})). Most importantly, we have made use of eq.s$\,$(\ref{q4}) and
(\ref{q5}).

The set of \emph{3rd-derivative functions,} $\varrho_{3}\left(  \mathbf{r}%
\right)  ,$ is seen from eq.s$\,$(\ref{v&d}) and (\ref{q25}) to have the
projection $-G_{12}\left(  r\right)  1.$ We therefore eliminate $G_{1}\left(
r\right)  $ from the last two equations (\ref{(r-a)3}):%
\begin{align}
&  r\mathcal{\hat{P}}\left(  r\right)  \psi_{1234}\left(  \mathbf{r}\right)
\left(  S_{1234}\right)  ^{-1}-r\mathcal{\hat{P}}\left(  r\right)  \psi
_{123}\left(  \mathbf{r}\right)  \left(  S_{123}\right)  ^{-1}\nonumber\\
&  =G_{12}\left(  r\right)  \left[  S_{234}\left(  S_{1234}\right)
^{-1}-S_{23}\left(  S_{123}\right)  ^{-1}\right]  +o, \label{V01}%
\end{align}
and find that:%
\begin{align}
\varrho_{3}\left(  \mathbf{r}\right)   &  =-\psi_{123}\left(  \mathbf{r}%
\right)  \left[  S_{23}-S_{234}\left(  S_{1234}\right)  ^{-1}S_{123}\right]
^{-1}\nonumber\\
&  -\psi_{1234}\left(  \mathbf{r}\right)  \left[  S_{234}-S_{23}\left(
S_{123}\right)  ^{-1}S_{1234}\right]  ^{-1}\label{DIII}\\
&  \equiv\psi_{123}\left(  \mathbf{r}\right)  D_{3,3}+\psi_{1234}\left(
\mathbf{r}\right)  D_{4,3}.\nonumber
\end{align}
Here we have used that in the matrix notation, the projection of a linear
combination $\sum_{RL}\psi_{1.n;RL}\left(  \mathbf{r}\right)  c_{RL}$ equals
the projection of $\psi_{1.n;RL}\left(  \mathbf{r}\right)  $ right-multiplied
by $c_{RL}.$ In eq.$\,$(\ref{V01}) and in the following, functions like
$\psi_{123}\left(  \mathbf{r}\right)  $ and $\varrho_{3}\left(  \mathbf{r}%
\right)  $ are row vectors with the respective components $\varrho
_{3;R^{\prime}L^{\prime}}\left(  \mathbf{r}\right)  $ and $\psi_{123;RL}%
\left(  \mathbf{r}\right)  ,$ while constants like $\left[  S_{23}%
-S_{234}\left(  S_{1234}\right)  ^{-1}S_{123}\right]  ^{-1}$ are square
matrices with $RL,R^{\prime}L^{\prime}$-elements. Hence, the subscripts $RL$
are summed over, and the square parentheses in eq.$\,$(\ref{DIII}) contain
matrix products and inversions. In the last line of eq.$\,$(\ref{DIII}) and in
the following, a matrix $D_{n,d}$ with elements $D_{nR^{\prime}L^{\prime}%
,dRL}$ is defined to be the coefficient to $\psi_{1.n}\left(  \mathbf{r}%
\right)  $ in the expansion of $\varrho_{d}\left(  \mathbf{r}\right)  .$
Remember, that we have chosen to number the energy points, $n=1,..,4,$ and the
radial derivatives, $d=0,..,\,3.$

In order to find the set of \emph{1st-derivative functions,} we eliminate
$G_{12}\left(  r\right)  $ from the last two equations (\ref{(r-a)3}) and
subsequently insert expression (\ref{q15}) for $G_{1}\left(  r\right)  $:%
\begin{align}
&  r\mathcal{\hat{P}}\left(  r\right)  \left\{  \psi_{1234}\left(
\mathbf{r}\right)  \left(  S_{234}\right)  ^{-1}-\psi_{123}\left(
\mathbf{r}\right)  \left(  S_{23}\right)  ^{-1}\right\} \nonumber\\
&  =G_{1}\left(  r\right)  \left[  S_{1234}\left(  S_{234}\right)
^{-1}-S_{123}\left(  S_{23}\right)  ^{-1}\right]  +o\nonumber\\
&  =\left[  \left(  r-a\right)  -\frac{\left(  r-a\right)  ^{3}}{3!}\left(
\epsilon_{1}-w\right)  \right] \nonumber\\
&  \qquad\times\left[  S_{1234}\left(  S_{234}\right)  ^{-1}-S_{123}\left(
S_{23}\right)  ^{-1}\right]  +o. \label{V0}%
\end{align}
As a result:%
\begin{align}
\varrho_{1}\left(  \mathbf{r}\right)   &  =\varrho_{3}\left(  \mathbf{r}%
\right)  \left(  \epsilon_{1}-w\right) \nonumber\\
&  +\left[  \psi_{1234}\left(  \mathbf{r}\right)  \left(  S_{234}\right)
^{-1}-\psi_{123}\left(  \mathbf{r}\right)  \left(  S_{23}\right)  ^{-1}\right]
\nonumber\\
&  \times\left[  S_{1234}\left(  S_{234}\right)  ^{-1}-S_{123}\left(
S_{23}\right)  ^{-1}\right]  ^{-1}\nonumber\\
&  =\psi_{123}\left(  \mathbf{r}\right)  D_{3,3}\left[  S_{234}\left(
S_{1234}\right)  ^{-1}+\epsilon_{1}-w\right] \nonumber\\
&  +\psi_{1234}\left(  \mathbf{r}\right)  D_{4,3}\left[  S_{23}\left(
S_{123}\right)  ^{-1}+\epsilon_{1}-w\right]  . \label{DI}%
\end{align}
As usual, quantities like $\epsilon_{1}-w$ are diagonal matrices.

The set of \emph{2nd-derivative functions,} $\varrho_{2}\left(  \mathbf{r}%
\right)  ,$ is seen from eq.s$\,$(\ref{v&d}) and (\ref{q24}) to have the
projection $-U_{12}\left(  r\right)  \frac{1}{a}.$ From the second equation
(\ref{(r-a)3}) and from expressions (\ref{V01}) and (\ref{V0}) therefore:%
\begin{align*}
&  \varrho_{2}\left(  \mathbf{r}\right)  a=-\psi_{12}\left(  \mathbf{r}%
\right)  -\psi_{123}\left(  \mathbf{r}\right) \\
&  \times\left\{
\begin{array}
[c]{c}%
\left(  S_{23}\right)  ^{-1}\left[  S_{1234}\left(  S_{234}\right)
^{-1}-S_{123}\left(  S_{23}\right)  ^{-1}\right]  ^{-1}S_{12}\\
+\left(  S_{123}\right)  ^{-1}\left[  S_{234}\left(  S_{1234}\right)
^{-1}-S_{23}\left(  S_{123}\right)  ^{-1}\right]  ^{-1}\mathcal{S}_{2}%
\end{array}
\right\} \\
&  +\psi_{1234}\left(  \mathbf{r}\right) \\
&  \times\left\{
\begin{array}
[c]{c}%
\left(  S_{234}\right)  ^{-1}\left[  S_{1234}\left(  S_{234}\right)
^{-1}-S_{123}\left(  S_{23}\right)  ^{-1}\right]  ^{-1}S_{12}\\
+\left(  S_{1234}\right)  ^{-1}\left[  S_{234}\left(  S_{1234}\right)
^{-1}-S_{23}\left(  S_{123}\right)  ^{-1}\right]  ^{-1}\mathcal{S}_{2}%
\end{array}
\right\}  ,
\end{align*}
which reduces to:%
\begin{align}
\varrho_{2}\left(  \mathbf{r}\right)  a  &  =-\psi_{12}\left(  \mathbf{r}%
\right) \nonumber\\
&  -\psi_{123}\left(  \mathbf{r}\right)  D_{3,3}\left[  \mathcal{S}%
_{2}-S_{234}\left(  S_{1234}\right)  ^{-1}S_{12}\right] \nonumber\\
&  -\psi_{1234}\left(  \mathbf{r}\right)  D_{4,3}\left[  \mathcal{S}%
_{2}-S_{23}\left(  S_{123}\right)  ^{-1}S_{12}\right]  . \label{DII}%
\end{align}

Of the divided-difference functions, $\psi_{1..d+1}\left(  \mathbf{r}\right)
,$ only the 0th does not vanish at all spheres, and it must therefore be
included in the value functions $\varrho_{0}\left(  \mathbf{r}\right)  .$ From
the first eq.$\,$(\ref{(r-a)3}) and eq.$\,$(\ref{q14}):%
\begin{align*}
r\mathcal{\hat{P}}\left(  r\right)  \psi_{1}\left(  \mathbf{r}\right)   &
=a-\frac{\left(  r-a\right)  ^{2}}{2!}a\left(  \epsilon_{1}-w\right) \\
&  +\frac{\left(  r-a\right)  ^{3}}{3!}aw^{\prime}+G_{1}\left(  r\right)
\mathcal{S}_{1}+o,
\end{align*}
so that with the help of the first lines of expressions (\ref{V0}) and
(\ref{DI}), we get:
\begin{align*}
\psi_{1}\left(  \mathbf{r}\right)   &  =\varrho_{0}\left(  \mathbf{r}\right)
a-\varrho_{2}\left(  \mathbf{r}\right)  a\left(  \epsilon_{1}-w\right)
+\varrho_{3}\left(  \mathbf{r}\right)  aw^{\prime}\\
&  +\left[  \varrho_{1}\left(  \mathbf{r}\right)  -\varrho_{3}\left(
\mathbf{r}\right)  \left(  \epsilon_{1}-w\right)  \right]  \mathcal{S}_{1}.
\end{align*}
As a result, the set of \emph{value functions} is given by:%
\begin{align*}
\varrho_{0}\left(  \mathbf{r}\right)  a  &  =\psi_{1}\left(  \mathbf{r}%
\right)  -\psi_{12}\left(  \mathbf{r}\right)  \left(  \epsilon_{1}-w\right)
+\psi_{123}\left(  \mathbf{r}\right) \\
&  \times\left\{
\begin{array}
[c]{c}%
\left[  S_{23}-S_{234}\left(  S_{1234}\right)  ^{-1}S_{123}\right]  ^{-1}\\
\times\left[  \mathcal{S}_{2}-S_{234}\left(  S_{1234}\right)  ^{-1}%
S_{12}\right]  \left(  \epsilon_{1}-w\right) \\
+\left[  S_{23}-S_{234}\left(  S_{1234}\right)  ^{-1}S_{123}\right]
^{-1}aw^{\prime}\\
-\left[  S_{123}-S_{1234}\left(  S_{234}\right)  ^{-1}S_{23}\right]
^{-1}\mathcal{S}_{1}%
\end{array}
\right\} \\
&  +\psi_{1234}\left(  \mathbf{r}\right) \\
&  \times\left\{
\begin{array}
[c]{c}%
\left[  S_{234}-S_{23}\left(  S_{123}\right)  ^{-1}S_{1234}\right]  ^{-1}\\
\times\left[  \mathcal{S}_{2}-S_{23}\left(  S_{123}\right)  ^{-1}%
S_{12}\right]  \left(  \epsilon_{1}-w\right) \\
+\left[  S_{234}-S_{23}\left(  S_{123}\right)  ^{-1}S_{1234}\right]
^{-1}aw^{\prime}\\
-\left[  S_{1234}-S_{123}\left(  S_{23}\right)  ^{-1}S_{234}\right]
^{-1}\mathcal{S}_{1}%
\end{array}
\right\}  ,
\end{align*}
which simplifies to:
\begin{align}
&  \varrho_{0}\left(  \mathbf{r}\right)  a\nonumber\\
&  =\psi_{1}\left(  \mathbf{r}\right)  -\psi_{12}\left(  \mathbf{r}\right)
\left(  \epsilon_{1}-w\right) \nonumber\\
&  -\psi_{123}\left(  \mathbf{r}\right)  D_{3,3}\left\{
\begin{array}
[c]{c}%
aw^{\prime}+\mathcal{S}_{2}\left(  \epsilon_{1}-w\right)  +\\
S_{234}\left(  S_{1234}\right)  ^{-1}\left[  \mathcal{S}_{1}-S_{12}\left(
\epsilon_{1}-w\right)  \right]
\end{array}
\right\} \nonumber\\
&  -\psi_{1234}\left(  \mathbf{r}\right)  D_{4,3}\left\{
\begin{array}
[c]{c}%
aw^{\prime}+\mathcal{S}_{2}\left(  \epsilon_{1}-w\right)  +\\
S_{23}\left(  S_{123}\right)  ^{-1}\left[  \mathcal{S}_{1}-S_{12}\left(
\epsilon_{1}-w\right)  \right]
\end{array}
\right\}  . \label{DO}%
\end{align}

Hence, for use in the interpolation (\ref{interpolation}), we have succeeded
in forming a set of four v\&d functions, $\varrho_{d}\left(  \mathbf{r}%
\right)  $ with $d\mathrm{=}0$ to $3,$ from the USW-sets at four different
energies, $\psi_{n}\left(  \mathbf{r}\right)  \equiv\psi\left(  \epsilon
_{n},\mathbf{r}\right)  \,$with $n\mathrm{=}1$ to $4.$ The result is:%
\begin{align}
\varrho_{dRL}\left(  \mathbf{r}\right)   &  =\sum_{n=1}^{4}\sum_{R^{\prime
}L^{\prime}}\sum_{n^{\prime}=1}^{n}\frac{\psi_{n^{\prime}R^{\prime}L^{\prime}%
}\left(  \mathbf{r}\right)  }{\Pi_{m=1,\neq n^{\prime}}^{n}\left(
\epsilon_{n^{\prime}}-\epsilon_{m}\right)  }\,D_{nR^{\prime}L^{\prime}%
,dRL}\nonumber\\
&  =\sum_{n=1}^{4}\sum_{R^{\prime}L^{\prime}}\psi_{1..n;R^{\prime}L^{\prime}%
}\left(  \mathbf{r}\right)  \,D_{nR^{\prime}L^{\prime},dRL}. \label{vandd}%
\end{align}
The similarity transformation, $D_{n,d},$ from the four energy-divided
differences (\ref{q0}) of USWs, $\psi_{1..n}\left(  \mathbf{r}\right)  ,$ to
the v\&d functions, $\varrho_{d}\left(  \mathbf{r}\right)  ,$ is given by the
coefficients found in eq.s$\,$(\ref{DIII}), (\ref{DI}), (\ref{DII}), and
(\ref{DO}) in terms of energy-divided differences $\mathcal{S}_{1..n}$ of the
screened structure matrix (\ref{r13}). For the odd and even derivatives'
functions they are respectively:%
\begin{align}
D_{1,3}  &  =D_{2,3}=0,\nonumber\\
D_{3,3}  &  =-\left[  S_{23}-S_{234}\left(  S_{1234}\right)  ^{-1}%
S_{123}\right]  ^{-1},\nonumber\\
D_{4,3}  &  =-\left[  S_{234}-S_{23}\left(  S_{123}\right)  ^{-1}%
S_{1234}\right]  ^{-1}\nonumber\\
&  =-\left(  S_{1234}\right)  ^{-1}S_{123}D_{3,3},\nonumber\\
D_{1,1}  &  =D_{2,1}=0,\nonumber\\
D_{3,1}  &  =D_{3,3}\left[  S_{234}\left(  S_{1234}\right)  ^{-1}+\epsilon
_{1}-w\right]  ,\nonumber\\
D_{4,1}  &  =D_{4,3}\left[  S_{23}\left(  S_{123}\right)  ^{-1}+\epsilon
_{1}-w\right] \nonumber\\
&  =A-\left(  S_{1234}\right)  ^{-1}S_{123}D_{3,3}\left(  \epsilon
_{1}-w\right)  , \label{odd}%
\end{align}
and:%
\begin{align}
D_{1,2}a  &  =0,\nonumber\\
D_{2,2}a  &  =-1,\nonumber\\
D_{3,2}a  &  =-D_{3,3}\left[  \mathcal{S}_{2}-S_{234}\left(  S_{1234}\right)
^{-1}S_{12}\right]  ,\nonumber\\
D_{4,2}a  &  =-D_{4,3}\left[  \mathcal{S}_{2}-S_{23}\left(  S_{123}\right)
^{-1}S_{12}\right]  =-D_{4,3}\mathcal{S}_{2}+AS_{12},\nonumber\\
D_{1,0}a  &  =1,\nonumber\\
D_{2,0}a  &  =-\left(  \epsilon_{1}-w\right)  ,\nonumber\\
D_{3,0}a  &  =-D_{3,3}\left\{
\begin{array}
[c]{c}%
aw^{\prime}+\mathcal{S}_{2}\left(  \epsilon_{1}-w\right)  +\\
S_{234}\left(  S_{1234}\right)  ^{-1}\left[  \mathcal{S}_{1}-S_{12}\left(
\epsilon_{1}-w\right)  \right]
\end{array}
\right\}  ,\nonumber\\
D_{4,0}a  &  =-D_{4,3}\left\{
\begin{array}
[c]{c}%
aw^{\prime}+\mathcal{S}_{2}\left(  \epsilon_{1}-w\right)  +\\
S_{23}\left(  S_{123}\right)  ^{-1}\left[  \mathcal{S}_{1}-S_{12}\left(
\epsilon_{1}-w\right)  \right]
\end{array}
\right\} \nonumber\\
&  =-D_{4,3}\left\{  aw^{\prime}+\mathcal{S}_{2}\left(  \epsilon_{1}-w\right)
\right\}  -A\left\{  \mathcal{S}_{1}-S_{12}\left(  \epsilon_{1}-w\right)
\right\}  . \label{even}%
\end{align}
Here, we have defined the matrix:%
\begin{equation}
A\equiv D_{4,3}S_{23}\left(  S_{123}\right)  ^{-1}=\left[  S_{1234}%
-S_{123}\left(  S_{23}\right)  ^{-1}S_{234}\right]  ^{-1}, \label{DD}%
\end{equation}
and $\epsilon_{1}-w$ and $aw^{\prime}$ are diagonal matrices with the
respective components $\epsilon_{1}-l\left(  l+1\right)  \left(  a_{R}\right)
^{-2}$ and $-2l\left(  l+1\right)  \left(  a_{R}\right)  ^{-2}.$ Expressions
involving inversion of the second divided difference, $S_{123}$, which may not
be positive definite, have been rewritten in terms of inverted first and the
third divided differences, $S_{23}$ and $S_{1234}.$ The latter are likely to
be positive-definite because, as seen from eq.$\,$(\ref{ovl}) in
Sect.$\,$\ref{integrals}, their elements are overlap integrals over nearly
identical functions.

It should be noted, that the v\&d functions are invariant to the numbering of
the four energies.

Here, we have chosen to express the v\&d functions in terms of the
dimensionless slope matrix, $\mathcal{S}\left(  \varepsilon\right)  $ given by
eq.s (\ref{r13}) and (\ref{aS}), because it has a simple physical
interpretation. To rewrite eq.s (\ref{odd}) and (\ref{even}) in terms of the
symmetric matrix $a\mathcal{S}$ is a trivial matter.

In order to generate all 16 submatrices (\ref{odd})-(\ref{even}), one thus
needs to invert 4 matrices, e.g. $S_{1234},$ $S_{23}-S_{234}\left(
S_{1234}\right)  ^{-1}S_{123},$ $S_{23},$ and $S_{1234}-S_{123}\left(
S_{23}\right)  ^{-1}S_{234},$ in addition to the 4 matrices $B\left(
\epsilon_{n}\right)  +n(\epsilon_{n},a)/j(\epsilon_{n},a)$ in eq.$\,$%
(\ref{aS}). The remaining matrix operations in eq.s$\,$(\ref{odd}) and
(\ref{even}) are merely products and sums. The dimensions of the matrices will
be discussed in Sect.s \ref{N} and \ref{Symmetry}.

This value-and-first-3-derives' formalism achieves to invert 4 slope matrices
instead of one, four times larger matrix. Apart from this, the 4 v\&d
functions of a given $RL$ are more localized than the 4 USWs of the same $RL$
because any v\&d function has vanishing values and first 3 radial derivatives
at \emph{all} spheres \emph{other} than its own. Moreover, with increasing
derivative order, $d,$ the v\&d function, $\varrho_{dRL}\left(  \mathbf{r}%
\right)  ,$ extends further and further into the interstitial around site $R,$
at the same time as remaining localized to inside the Voronoi cell,
approximately. This can clearly be seen in the top row of Fig.s 2 and 3, where
we show the value-, first-, second-, and third-derivative $s$-functions for
bcc- and dia-structured interstitials.

The interpolation (\ref{interpolation}) is \emph{local,} that is: to the
density at point $\mathbf{r,}$ essentially only the v\&d functions centered at
the cell containing $\mathbf{r}$ contribute. However, the generation of the
v\&d functions is not local. They are multi-centered linear combinations
(\ref{vandd}) of USWs which, themselves, are multi-centered linear
combinations (\ref{USWinHank}) of Hankel functions. This generation of
shorter-ranged functions from longer-ranged ones -- and even in two stages --
was used to produce the contour plots in Fig.s 2 and 3. For the purpose of
simplifying the plotting of $\rho\left(  \mathbf{r}\right)  $ from
eq.$\,$(\ref{interpolation}), one might use v\&d functions tabulated on a mesh
spanning their own cell and its near neighborhood. Faster, but less accurate,
it is to approximate the v\&d functions in the interstitial \emph{near} an
arbitrary site $\mathbf{R}^{\prime}$ by their cubic-harmonic expansion around
that site:%
\begin{equation}
\rho\left(  \mathbf{r}\right)  \approx\sum_{L^{\prime}}Y_{L^{\prime}}\left(
\mathbf{\hat{r}}_{R^{\prime}}\right)  \,\sum_{dL}\mathcal{\hat{P}}_{R^{\prime
}L^{\prime}}\left(  r_{R^{\prime}}\right)  \varrho_{dRL}\left(  \mathbf{r}%
\right)  \mathcal{R}_{RL}^{\left(  d\right)  } \label{onsite}%
\end{equation}
This one-center expansion (\ref{onsite}) is less useful for open than for
close-packed structures because it is strictly valid only for $a_{R^{\prime}%
}\leq r_{R^{\prime}}\leq\min_{R^{\prime\prime}}\left(  d_{R^{\prime\prime
}R^{\prime}}-a_{R^{\prime}}\right)  ,$ where the latter is the distance to the
nearest-neighbor sphere. For $\min_{R^{\prime\prime}}\left(  d_{R^{\prime
\prime}R^{\prime}}-a_{R^{\prime}}\right)  \leq r_{R^{\prime}}<\min
_{R^{\prime\prime}}d_{R^{\prime\prime}R^{\prime}},$ the expansion converges to
the superposition of Hankel functions. Approximating the v\&d functions by the
cubic-harmonic expansion around the \emph{own} site, i.e. choosing
$\mathbf{R}^{\prime}=\mathbf{R,}$ brings great simplification, but only for
close-packed structures, does the expansion hold throughout the Voronoi cell.
The radial functions, $\mathcal{\hat{P}}_{R^{\prime}L^{\prime}}\left(
r\right)  \varrho_{dRL}\left(  \mathbf{r}\right)  ,$ will be derived in the Appendix.

The alert reader will have noted that $\varrho_{d=0\,s}\left(  \mathbf{r}%
\right)  $ in Figs 2 and 3 does \emph{not} start out flat from the central
sphere, but like $1/r.$ This is because we have chosen to carry the prefactor
$r$ in the boundary condition (\ref{v&d}) for the v\&d functions in order to
simplify the formalism leading to eq.s (\ref{vandd})-(\ref{DD}). So what
starts out flat, is $r$ times the spherical average of $\varrho_{d=0\,s}%
\left(  \mathbf{r}\right)  $. Having found these v\&d functions, we may of
course form those, $\bar{\varrho}_{dRL}\left(  \mathbf{r}\right)  ,$ which
satisfy the boundary conditions without the prefactor $r$:%
\begin{equation}
\mathcal{\hat{P}}_{R^{\prime}L^{\prime}}\left(  r\right)  \bar{\varrho}%
_{dRL}\left(  \mathbf{r}\right)  =\frac{\left(  r-a_{R}\right)  ^{d}}%
{d!}\delta_{R^{\prime}R}\delta_{L^{\prime}L}+o. \label{v&dbar}%
\end{equation}
The result, most easily obtained by using the interpolation formalism
(\ref{Input})-(\ref{interpolation}) with $\rho\left(  \mathbf{r}\right)
=\bar{\varrho}_{dRL}\left(  \mathbf{r}\right)  ,$ is:%
\begin{equation}
\bar{\varrho}_{dRL}\left(  \mathbf{r}\right)  =\varrho_{dRL}\left(
\mathbf{r}\right)  a_{R}+\left(  d+1\right)  \varrho_{\left(  d+1\right)
RL}\left(  \mathbf{r}\right)  , \label{vanddbar}%
\end{equation}
with $\varrho_{d>d_{\max}\,RL}\left(  \mathbf{r}\right)  \equiv0.$

The v\&d functions are independent of the scale $\left(  t\right)  $ of the
structure, provided that spatial derivatives are defined with respect to the
dimensionless variable $\mathbf{r/}t$ and that the energy mesh times $t^{2}$
is kept constant. This follows from the fact that USWs solve the wave equation
(\ref{wave eq}).

The main use of expressions (\ref{vandd})-(\ref{DD}) for the v\&d functions as
multi-centered linear combinations of USWs is for solving Poisson's equation
and for forming integrals, as we shall see in Sect.s \ref{Poissons eq} and
\ref{integrals}.

\bigskip

\begin{figure*}[htbp]
  \vspace{0cm}
  \begin{center}
  \includegraphics[width=0.95\textwidth]{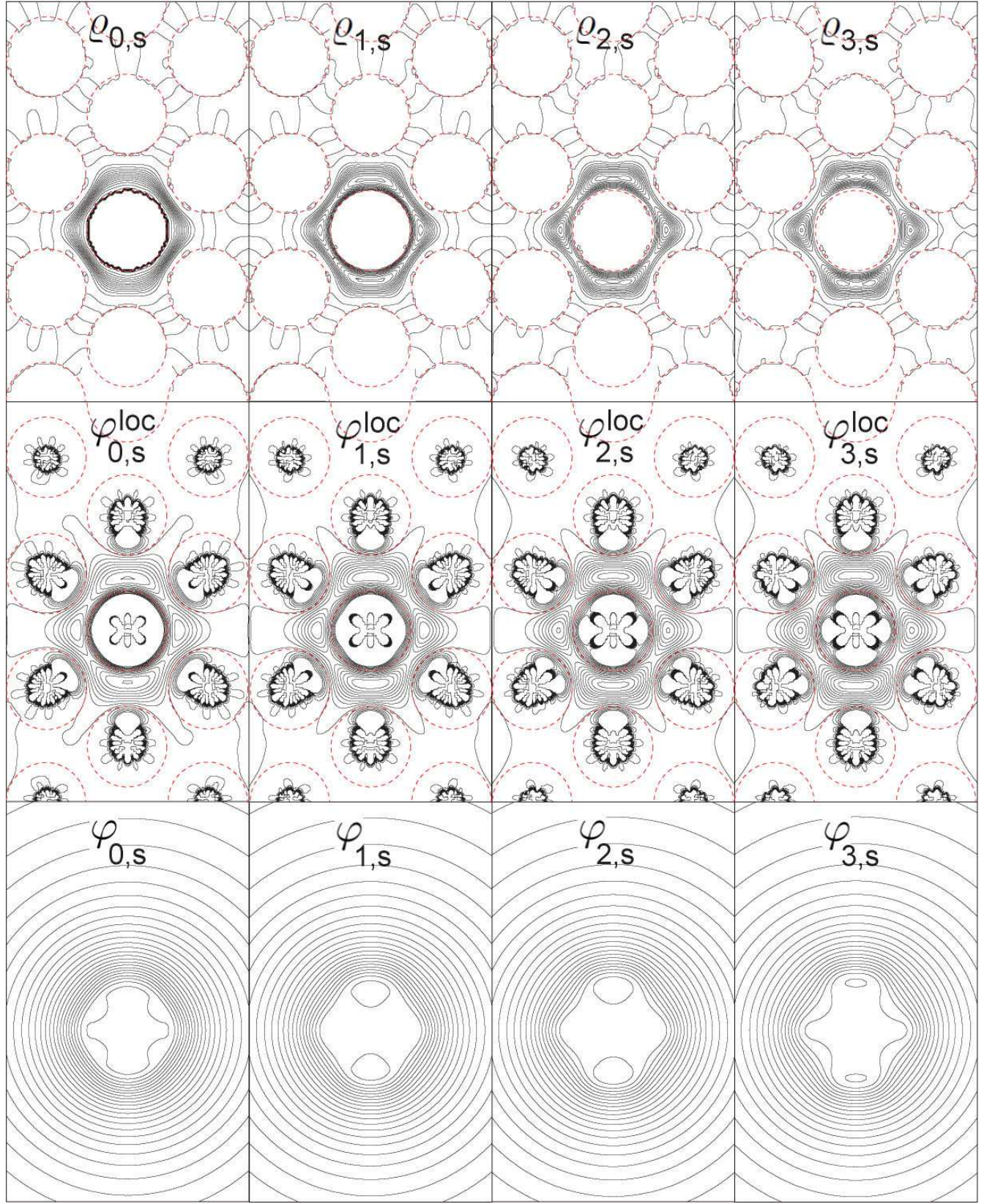}
  \caption{(Color online)
Bcc structure as in the left-hand panel of Fig. 1. \textit{Top row:}
Value-, first-, second-, and third-derivative $s$-like functions,
$\varrho_{ds}\left(  \mathbf{r}\right)  /\left\langle \varrho_{ds}%
\right\rangle $, normalized over the interstitial. See Sect.$\,$%
\ref{v&dfunctions}. \textit{Middle row:} Their localized potentials,
$\varphi_{ds}^{loc}\left(  \mathbf{r}\right)  /\left\langle \varrho
_{ds}\right\rangle .$ \textit{Bottom row:} Their regular potentials,
$\varphi_{ds}\left(  \mathbf{r}\right)  /\left\langle \varrho_{ds}%
\right\rangle .$ See Sect.$\,$\ref{v&dpot}. The localized potentials are
screened by multipoles at the centers of the hard spheres and therefore
diverge there. The regular potentials have these multipoles subtracted out.
For $r$ large, all four regular potentials become that of a point charge,
$-2/r$. The values of the normalization integrals are: $\left\langle
\varrho_{ds}\right\rangle =\,$5.82, 2.05, 0.358, and 0.0256\ (Bohr
radii)$^{d-2}$ for $d\,$=\thinspace0, 1, 2, and 3, respectively. The contours
for the normalized v\&d functions go from $0$ to $0.03$ in steps of $0.002$
(Bohr radii)$^{-3},$ those for the localized potentials from $-0.06$ upwards
in steps of $0.012$ Ry, the zero-potential contour being the one following the
hard spheres most closely, and those for the regular potentials from 0 to 0.99
in steps of 0.03 Ry. The energy mesh is exponential (eq.$\,$(\ref{expmesh}))
with the highest and lowest energies the same as those used in Fig. 1, i.e.
$\epsilon_{1}=-1.54$ and $\epsilon_{4}=-15.7$ Ry.
}
  \end{center}
\end{figure*}

\bigskip

\begin{figure*}[htbp]
  \vspace{0cm}
  \begin{center}
  \includegraphics[width=0.95\textwidth]{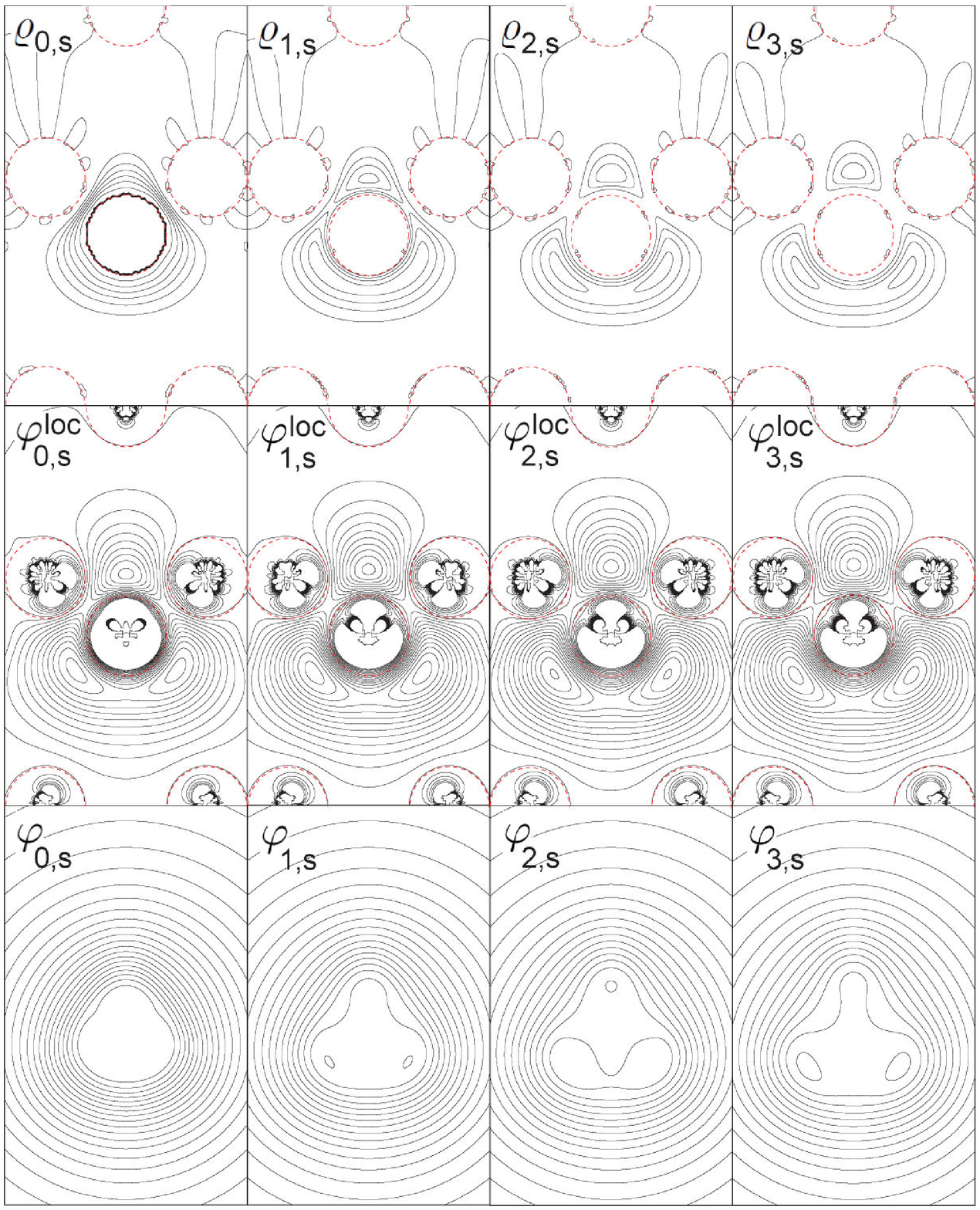}
  \caption{(Color online)
Same as Fig. 2, but for the diamond-structure with the same
nearest-neighbor distance as in Fig.$\,$2. $\left\langle \varrho
_{ds}\right\rangle =$ 11.25, 6.94, 1.89, and 0.188 (Bohr radii)$^{d-2}$ for
$d\,$=\thinspace0, 1, 2, and 3, respectively.
}
  \end{center}
\end{figure*}

\bigskip

\section{Solving Poisson's equation\label{Poissons eq}}

\subsection{Potentials from energy-divided differences of
USWs\label{PotdivdifUSW}}

Poisson's equation (\ref{Poisson}) for a charge density which is a spherical
wave, $\rho\left(  \mathbf{r}\right)  =\psi\left(  \varepsilon,\mathbf{r}%
\right)  ,$ has the particular solution $V\left(  \mathbf{r}\right)  =8\pi
\psi\left(  \varepsilon,\mathbf{r}\right)  /\varepsilon.$ For a charge density
which is the $d$th energy-divided difference (\ref{q0}) of an USW, Poisson's
equation therefore has the solution%
\begin{align}
&  -8\pi\Delta^{-1}\psi_{1..d+1;RL}\left(  \mathbf{r}\right)  =\sum
_{n=1}^{d+1}\frac{8\pi\psi_{nRL}\left(  \mathbf{r}\right)  /\epsilon_{n}}%
{\Pi_{m=1,\neq n}^{d+1}\left(  \epsilon_{n}-\epsilon_{m}\right)  }\nonumber\\
&  =\sum_{n=0}^{d+1}\frac{8\pi\psi_{nRL}\left(  \mathbf{r}\right)  }%
{\Pi_{m=0,\neq n}^{d+1}\left(  \epsilon_{n}-\epsilon_{m}\right)  }-\frac
{8\pi\psi_{RL}\left(  0,\mathbf{r}\right)  }{\Pi_{m=1}^{d+1}\left(
-\epsilon_{m}\right)  }\nonumber\\
&  =8\pi\psi_{0...d+1;RL}\left(  \mathbf{r}\right)  -\frac{8\pi\psi
_{RL}\left(  0,\mathbf{r}\right)  }{\Pi_{m=1}^{d+1}\left(  -\epsilon
_{m}\right)  } \label{Pot}%
\end{align}
in the \emph{interstitial }between the spheres. In the first term on the
right-hand side, we have defined
\begin{equation}
\epsilon_{0}\equiv0, \label{zero}%
\end{equation}
and have used this energy point to take the divided difference for the
potential \emph{one order higher} than for the charge density.

\emph{Inside} the spheres, the solution (\ref{Pot}) is joined smoothly to a
solution of the Laplace equation.

\subsubsection{The localized potential}

The second term, $-8\pi\psi_{RL}\left(  0,\mathbf{r}\right)  \left/  \left(
-\epsilon_{1}\right)  ..\left(  -\epsilon_{d+1}\right)  \right.  ,$ in
expression (\ref{Pot}) satisfies the Laplace equation$.$ We can therefore
choose merely the first term:%
\begin{equation}
8\pi\psi_{0...d+1;RL}\left(  \mathbf{r}\right)  \equiv\phi_{1..d+1;RL}%
^{loc}\left(  \mathbf{r}\right)  , \label{q31}%
\end{equation}
as the particular solution \cite{VPhiphi} of interest in the
\emph{interstitial.} This choice makes the potential \emph{localized} to the
neighborhood of its own sphere because $\psi_{0...d+1;RL}\left(
\mathbf{r}\right)  $ is an energy-divided difference of at least 1st order and
therefore has \emph{vanishing }$L^{\prime}$-averages \emph{at all} spheres for
\emph{all} $l^{\prime}\leq l_{\max}.$ This holds also for the \emph{eigen}%
-projection of $\psi_{0...d+1;RL}\left(  \mathbf{r}\right)  $ because the
eigen-projection of $\psi_{RL}\left(  \varepsilon,\mathbf{r}\right)  $ is
$Y_{L}\left(  \mathbf{\hat{r}}_{R}\right)  $, independently of the energy.
\emph{Near} the own sphere, the localized potential has pure $L$-character$.$

The localized potential may be expanded around any site $\mathbf{R}^{\prime}$
in cubic-harmonics times radial functions:%
\begin{equation}
\phi_{1..d+1;RL}^{loc}\left(  \mathbf{r}\right)  \approx8\pi\sum_{L^{\prime}%
}Y_{L^{\prime}}\left(  \mathbf{\hat{r}}_{R^{\prime}}\right)  \mathcal{\hat{P}%
}_{R^{\prime}L^{\prime}}\left(  r\right)  \psi_{0...d+1;RL}\left(
\mathbf{r}\right)  \label{1cPhiloc}%
\end{equation}
valid in the interstitial \emph{at and outside} the $R^{\prime}$-sphere. Since
$R^{\prime}L^{\prime}$-projection and forming energy-divided differences
commute, we can reverse the order and use the binomial rule (\ref{q1}) to form
the differences of the projections (\ref{r02}) expressed as $\mathcal{\hat{P}%
}\left(  r\right)  \psi\left(  \varepsilon,\mathbf{r}\right)  =u\left(
\varepsilon,r\right)  +g\left(  \varepsilon,r\right)  \mathcal{S}\left(
\varepsilon\right)  .$ The resulting projections are for $d\mathrm{=}0$ to 3:%
\begin{align}
\mathcal{\hat{P}}\left(  r\right)  \psi_{01}\left(  \mathbf{r}\right)   &
=u_{01}\left(  r\right)  +g_{0}\left(  r\right)  S_{01}+g_{01}\left(
r\right)  \mathcal{S}_{1},\nonumber\\
\mathcal{\hat{P}}\left(  r\right)  \psi_{012}\left(  \mathbf{r}\right)   &
=u_{012}\left(  r\right)  +g_{0}\left(  r\right)  S_{012}\nonumber\\
&  +g_{01}\left(  r\right)  S_{12}+g_{012}\left(  r\right)  \mathcal{S}%
_{2},\nonumber\\
\mathcal{\hat{P}}\left(  r\right)  \psi_{0123}\left(  \mathbf{r}\right)   &
=u_{0123}\left(  r\right)  +g_{0}\left(  r\right)  S_{0123}+g_{01}\left(
r\right)  S_{123}\nonumber\\
&  +g_{012}\left(  r\right)  S_{23}+g_{0123}\left(  r\right)  \mathcal{S}%
_{3},\nonumber\\
\mathcal{\hat{P}}\left(  r\right)  \psi_{01234}\left(  \mathbf{r}\right)   &
=u_{01234}\left(  r\right)  +g_{0}\left(  r\right)  S_{01234}\nonumber\\
&  +g_{01}\left(  r\right)  S_{1234}+g_{012}\left(  r\right)  S_{234}%
\nonumber\\
&  +g_{0123}\left(  r\right)  S_{34}+g_{01234}\left(  r\right)  \mathcal{S}%
_{4}. \label{Psi-1}%
\end{align}

\emph{Inside} any sphere, the localized potential is that solution of the
Laplace equation which matches $8\pi\psi_{0...d+1;RL}\left(  \mathbf{r}%
\right)  $ smoothly at the sphere. Of the radial functions in expressions
(\ref{Psi-1}), the only one which does not vanish smoothly at the sphere, and
therefore can provide a slope, is $g_{0}\left(  r\right)  .$ This follows from
eq.s$\,$(\ref{q24})-(\ref{q5}) together with definitions (\ref{UU}) and
(\ref{GG}). According to eq.$\,$(\ref{g}), this slope is $g_{0}^{\prime
}\left(  a\right)  =1/a$. Since $g_{0}\left(  r\right)  \equiv g\left(
\epsilon_{0},r\right)  \equiv g\left(  0,r\right)  $ is a solution of the
radial Laplace equation, the localized potential \emph{inside} the $R^{\prime
}$-sphere is simply:%
\begin{align}
&  \phi_{1..d+1;RL}^{loc}\left(  \mathbf{r}\right)  =\sum_{L^{\prime}%
}Y_{L^{\prime}}\left(  \mathbf{\hat{r}}_{R^{\prime}}\right)  \mathcal{\hat{P}%
}_{R^{\prime}L^{\prime}}\left(  r\right)  \phi_{1..d+1;RL}^{loc}\left(
\mathbf{r}\right) \nonumber\\
&  =8\pi\sum_{L^{\prime}}g_{0;R^{\prime}l^{\prime}}\left(  r_{R^{\prime}%
}\right)  Y_{L^{\prime}}\left(  \mathbf{\hat{r}}_{R^{\prime}}\right)
S_{0...d+1;R^{\prime}L^{\prime},RL}\nonumber\\
&  =8\pi\sum_{L^{\prime}}\left\{  n_{l^{\prime}}\left(  0,r_{R^{\prime}%
}\right)  j_{l^{\prime}}\left(  0,a_{R^{\prime}}\right)  -j_{l^{\prime}%
}\left(  0,r_{R^{\prime}}\right)  n_{l^{\prime}}\left(  0,a_{R^{\prime}%
}\right)  \right\} \nonumber\\
&  \qquad\qquad\qquad\qquad\qquad\times a_{R^{\prime}}Y_{L^{\prime}}\left(
\mathbf{\hat{r}}_{R^{\prime}}\right)  S_{0...d+1;R^{\prime}L^{\prime}%
,RL}\nonumber\\
& \nonumber\\
&  \equiv-\sum_{L^{\prime}}\left(  -\right)  ^{l^{\prime}}\frac{8\pi
}{2l^{\prime}+1}\left\{  1-\left(  \frac{r_{R^{\prime}}}{a_{R^{\prime}}%
}\right)  ^{2l^{\prime}+1}\right\}  \frac{Y_{L^{\prime}}\left(  \mathbf{\hat
{r}}_{R^{\prime}}\right)  }{r_{R^{\prime}}^{l^{\prime}+1}}\nonumber\\
&  \qquad\qquad\qquad\qquad\qquad\times Q_{1..d+1;R^{\prime}L^{\prime},RL},
\label{locinside}%
\end{align}
where we have used eq.s.$\,$(\ref{fginnj}) and (\ref{eps0}). We emphasize that
going from outside to inside a sphere, only the $g_{0}\left(  r\right)
$-terms in the one-centre expansion (\ref{1cPhiloc}) based on projections
(\ref{Psi-1}) survive. Their irregular parts, $\propto-1/r_{R^{\prime}%
}^{l^{\prime}+1},$ clearly seen in the middle rows of Figs 2 and 3, can be
interpreted as due to multipole moments:%
\begin{equation}
Q_{1..d+1;R^{\prime}L^{\prime},RL}=\left(  -\right)  ^{l^{\prime}}%
a_{R^{\prime}}^{l^{\prime}+1}S_{0...d+1;R^{\prime}L^{\prime},RL}, \label{Q}%
\end{equation}
of order $l^{\prime}$ at the sites $R^{\prime}$ which have been
\emph{subtracted} from the interstitial charge density, $\psi_{1..d+1;RL}%
\left(  \mathbf{r}\right)  ,$ in order to make its potential, $\phi
_{1..d+1;RL}^{loc}\left(  \mathbf{r}\right)  ,$ localized. We remark that the
sum over all monopole moments, $\sum_{R^{\prime}}Q_{1..d+1;R^{\prime}0,RL},$
is the total charge, $\left\langle \psi_{1..d+1;RL}\right\rangle ,$ divided by
$\sqrt{4\pi}.$ This follows formally from eq.s (\ref{Q}) and (\ref{int}).

\subsubsection{The regular potential}

The Coulomb potential, $\phi_{1..d+1;RL}\left(  \mathbf{r}\right)  ,$ which is
everywhere \emph{regular} must have the irregular part of the localized
potential inside the spheres (\ref{locinside}) cancelled out. This regular
potential, examples of which are shown in the bottom rows of Fig.s 2 and 3, is
therefore the localized one, \emph{plus} the multipole potential
\emph{extending in all space:}%
\begin{align}
&  \phi_{1..d+1;RL}\left(  \mathbf{r}\right)  =\phi_{1..d+1;RL}^{loc}\left(
\mathbf{r}\right)  -8\pi\label{q32}\\
&  \times\sum_{R^{\prime}L^{\prime}}n_{l^{\prime}}\left(  0,r_{R^{\prime}%
}\right)  Y_{L^{\prime}}\left(  \mathbf{\hat{r}}_{R^{\prime}}\right)
j_{l^{\prime}}\left(  0,a_{R^{\prime}}\right)  a_{R^{\prime}}%
S_{0...d+1;R^{\prime}L^{\prime},RL}.\nonumber
\end{align}

In the \emph{interstitial, }this potential may be expressed entirely in terms
of localized USWs, because the first term is given by eq.$\,$(\ref{q31}) and
the expansion of $n_{l^{\prime}}\left(  0,r_{R^{\prime}}\right)  Y_{L^{\prime
}}\left(  \mathbf{\hat{r}}_{R^{\prime}}\right)  $ in USWs is given by eq.s
(\ref{eps0}) and (\ref{HankinUSW}). As a result:%
\begin{align}
&  \phi_{1..d+1;RL}\left(  \mathbf{r}\right) \nonumber\\
&  =8\pi\psi_{0...d+1;RL}\left(  \mathbf{r}\right)  -8\pi\sum_{R^{\prime
\prime}L^{\prime\prime}}\psi_{R^{\prime\prime}L^{\prime\prime}}\left(
0,\mathbf{r}\right)  j_{l^{\prime\prime}}(0,a_{R^{\prime\prime}})\nonumber\\
&  \qquad\times\sum_{R^{\prime}L^{\prime}}\left(  \frac{n_{l^{\prime}%
}(0,a_{R^{\prime}})}{j_{l^{\prime}}(0,a_{R^{\prime}})}\delta_{R^{\prime\prime
}R^{\prime}}\delta_{L^{\prime\prime}L^{\prime}}+B_{R^{\prime\prime}%
L^{\prime\prime},R^{\prime}L^{\prime}}\left(  0\right)  \right) \nonumber\\
&  \qquad\qquad\times j_{l^{\prime}}\left(  0,a_{R^{\prime}}\right)
a_{R^{\prime}}S_{0...d+1;R^{\prime}L^{\prime},RL}. \label{longint}%
\end{align}
Here, the $R^{\prime}$-sum has long range and may for crystals be computed
with the Ewald method.

\emph{Inside} a sphere, say the one at $\mathbf{R}^{\prime}$, the regular
potential is the regular part of $\phi_{1..d+1;RL}^{loc}\left(  \mathbf{r}%
\right)  $ as given by eq.$\,$(\ref{locinside}), minus the tails from the
multipoles at all other sites, $\mathbf{R}^{\prime\prime}.$ This means that
its cubic-harmonic projection around site $R^{\prime}$ is given by:%
\begin{align}
&  \frac{1}{8\pi}\mathcal{\hat{P}}_{R^{\prime}L^{\prime}}\left(  r\right)
\phi_{1..d+1;RL}\left(  \mathbf{r}\right) \nonumber\\
& \nonumber\\
&  =-j_{l^{\prime}}\left(  0,r\right)  n_{l^{\prime}}\left(  0,a_{R^{\prime}%
}\right)  a_{R^{\prime}}S_{0...d+1;R^{\prime}L^{\prime},RL}\nonumber\\
&  \qquad-\mathcal{\hat{P}}_{R^{\prime}L^{\prime}}\left(  r\right)
\sum_{R^{\prime\prime}\neq R^{\prime}}\sum_{L^{\prime\prime}}n_{l^{\prime
\prime}}\left(  0,r_{R^{\prime\prime}}\right)  Y_{L^{\prime\prime}}\left(
\mathbf{\hat{r}}_{R^{\prime\prime}}\right) \nonumber\\
&  \qquad\qquad\qquad\times j_{l^{\prime\prime}}\left(  0,a_{R^{\prime\prime}%
}\right)  a_{R^{\prime\prime}}S_{0...d+1;R^{\prime\prime}L^{\prime\prime}%
,RL}\nonumber\\
& \nonumber\\
&  =-j_{l^{\prime}}\left(  0,r\right)  \sum_{R^{\prime\prime}L^{\prime\prime}%
}\left(
\begin{array}
[c]{l}%
\frac{n_{l^{\prime}}\left(  0,a_{R^{\prime}}\right)  }{j_{l^{\prime}}\left(
0,a_{R^{\prime}}\right)  }\delta_{R^{\prime}R^{\prime\prime}}\delta
_{L^{\prime}L^{\prime\prime}}\\
\qquad\qquad+B_{R^{\prime}L^{\prime},R^{\prime\prime}L^{\prime\prime}}\left(
0\right)
\end{array}
\right) \nonumber\\
&  \qquad\qquad\qquad\times j_{l^{\prime\prime}}\left(  0,a_{R^{\prime\prime}%
}\right)  a_{R^{\prime\prime}}S_{0...d+1;R^{\prime\prime}L^{\prime\prime},RL},
\label{longinside}%
\end{align}
where we have used the projection (\ref{Ph}), and that $B_{RL^{\prime}%
,RL}\left(  0\right)  =0$ according to eq.$\,$(\ref{bareonsite}). The
long-ranged sum over $R^{\prime\prime}$ is the same as the one over
$R^{\prime}$ in expression (\ref{longint}).

\subsection{Potentials from value-and-derivative
functions\textbf{\label{v&dpot}}}

A v\&d function, $\varrho_{dRL}\left(  \mathbf{r}\right)  ,$ is the
$nR^{\prime}L^{\prime}$-superposition of energy-divided differences of USWs,
$\psi_{1..n;R^{\prime}L^{\prime}}\left(  \mathbf{r}\right)  ,$ given by
(\ref{vandd})-(\ref{DD}). The localized and regular potentials, $\varphi
_{dRL}^{loc}\left(  \mathbf{r}\right)  $ and $\varphi_{dRL}\left(
\mathbf{r}\right)  ,$ from the v\&d function, $\varrho_{dRL}\left(
\mathbf{r}\right)  ,$ are therefore the same superposition of the potentials
$\phi_{1..n;R^{\prime}L^{\prime}}^{loc}\left(  \mathbf{r}\right)  $ and
$\phi_{1..n;R^{\prime}L^{\prime}}\left(  \mathbf{r}\right)  $ from the
energy-divided differences of USWs, $\psi_{0..n;R^{\prime}L^{\prime}}\left(
\mathbf{r}\right)  ,$ given by respectively eq.s (\ref{q31}) and (\ref{q32}).
In the \emph{interstitial, }that is:%
\begin{align}
\varphi_{dRL}^{loc}\left(  \mathbf{r}\right)   &  =\sum_{nR^{\prime}L^{\prime
}}\phi_{1..n;R^{\prime}L^{\prime}}^{loc}\left(  \mathbf{r}\right)
D_{nR^{\prime}L^{\prime},dRL}\nonumber\\
&  =8\pi\sum_{nR^{\prime}L^{\prime}}\psi_{0...n;R^{\prime}L^{\prime}}\left(
\mathbf{r}\right)  D_{nR^{\prime}L^{\prime},dRL} \label{phid}%
\end{align}
for the localized potential, and:%
\begin{align}
\varphi_{dRL}\left(  \mathbf{r}\right)   &  =\varphi_{dRL}^{loc}\left(
\mathbf{r}\right)  -8\pi\label{phiregd}\\
&  \times\sum_{R^{\prime}L^{\prime}}n_{l^{\prime}}\left(  0,r_{R^{\prime}%
}\right)  Y_{L^{\prime}}\left(  \mathbf{\hat{r}}_{R^{\prime}}\right)
j_{l^{\prime}}\left(  0,a_{R^{\prime}}\right)  a_{R^{\prime}}\nonumber\\
&  \qquad\times\sum_{nR^{\prime\prime}L^{\prime\prime}}S_{0...n;R^{\prime
}L^{\prime},R^{\prime\prime}L^{\prime\prime}}D_{nR^{\prime\prime}%
L^{\prime\prime},dRL}.\nonumber
\end{align}
for the regular potential. This means that after right-multiplication by $D$
(including the $n$-sum going from $1$ to $4$) all expressions given in the
previous Sect.$\,$\ref{PotdivdifUSW} for the localized and regular potentials
hold also for the potentials, $\varphi^{loc}\left(  \mathbf{r}\right)  $ and
$\varphi\left(  \mathbf{r}\right)  ,$ from the v\&d functions$.$ Specifically,
the moments of the $R^{\prime}L^{\prime}$-multipoles, which added to the
localized potential (\ref{phid}) make it regular (\ref{phiregd}), are:%
\begin{equation}
Q_{R^{\prime}L^{\prime},dRL}=\sum_{nR^{\prime\prime}L^{\prime\prime}%
}Q_{1..n;R^{\prime}L^{\prime},R^{\prime\prime}L^{\prime\prime}}D_{nR^{\prime
\prime}L^{\prime\prime},dRL}, \label{Qall}%
\end{equation}
where $Q_{1..n;R^{\prime}L^{\prime},R^{\prime\prime}L^{\prime\prime}}$ was
given in eq.$\,$(\ref{Q}).

Usually, we do not want to solve Poisson's equation for merely the
interstitial charge density, but also for the remaining charge density in the
system, such as the one, $\rho_{R}^{rest}\left(  \mathbf{r}\right)  ,$ inside
the sphere at $\mathbf{R.}$ This adds to $Q_{R^{\prime}L^{\prime}}=\sum
_{dRL}Q_{R^{\prime}L^{\prime},dRL}\mathcal{R}_{RL}^{\left(  d\right)  }$ the
(compensating) multipoles:%
\begin{equation}
Q_{R^{\prime}L^{\prime}}^{rest}\equiv\int\rho_{R^{\prime}}^{rest}\left(
\mathbf{r}\right)  \,r^{l^{\prime}}Y_{L^{\prime}}^{\ast}\left(  \mathbf{\hat
{r}}\right)  d^{3}r. \label{rest}%
\end{equation}

Like the localized potentials (\ref{q31}) from energy-divided differences of
USWs, those (\ref{phid}) from the v\&d functions \emph{vanish} at \emph{all}
hard spheres because they are superpositions of the former. In the middle rows
of Fig.s 2 (bcc) and 3 (dia), we show the localized potentials, $\varphi
_{ds}^{loc}\left(  \mathbf{r}\right)  ,$ from the normalized $s$-like v\&d
functions, $\varrho_{ds}\left(  \mathbf{r}\right)  /\left\langle \varrho
_{ds}\right\rangle ,$ for which $\varrho_{ds}\left(  \mathbf{r}\right)  $ with
$d=0,1,2,3$ are shown directly above, in the top row. The zero-potential
contours are seen to closely follow the hard spheres, which are indicated by
(red) dots. Inside the spheres, $\varphi_{ds}^{loc}\left(  \mathbf{r}\right)
$ becomes a multipole-potential, like $\phi_{ds}^{loc}\left(  \mathbf{r}%
\right)  $ in (\ref{locinside}), whose multipole moments are seen to be
dominated by the negative point charge at the centre. Due to the normalization
with $\left\langle \varrho_{ds}\right\rangle $, the sum over all monopoles is
$-1$. The potentials are cut off below $-0.06$ Ry. In the interstitial,
$\varphi_{ds}^{loc}\left(  \mathbf{r}\right)  $ is positive and localized near
the central sphere, around which it oscillates between maxima and
saddlepoints. For $\varphi_{0s}^{loc}\left(  \mathbf{r}\right)  ,$ this
oscillation is 0.10, 0.04, 0.08 Ry in the bcc interstitial and 0.14, 0.03, and
0.11 Ry in the diamond interstitial. As expected, $\varphi_{0s}^{loc}\left(
\mathbf{r}\right)  $ is more isotropic for the closely-packed bcc- than for
the open diamond interstitial. Moreover, the oscillations increase with $d$.
The regular potentials are dominated by the second term in (\ref{phiregd}),
i.e. the multipole potential extending in all space. And this, itself, is
dominated by the potential $-2/r$ from the total charge placed at its center.

The projections, $\mathcal{\hat{P}}_{R^{\prime}L^{\prime}}\left(  r\right)
\varphi_{dRL}^{loc}\left(  \mathbf{r}\right)  ,$ to be used in the
cubic-harmonic expansion like (\ref{1cPhiloc}) are given in at the end of the
Appendix. Specifically, the spherical averages around all sites form the input
to constructing the potential in the so-called overlapping MT approximation
(OMTA)\cite{98MRS,09OMTA} which is used to define the 3rd generation LMTO and
NMTO basis sets.

Provided that it is considered a function of a dimensionless $\mathbf{r/}t,$
the potential from a v\&d function times $t^{2}$ is invariant to a uniform
scaling of the structure. This follows from Poisson's equation (\ref{Poisson}).

\section{Integrals over the interstitial\label{integrals}}

The integral of the \emph{product of two} USWs over the interstitial region
may be calculated as a surface integral over the spheres and by use of Green's
2nd theorem. The result is simply \cite{94Trieste}:%
\begin{equation}
\left\langle \psi_{R^{\prime}L^{\prime}}\left(  \epsilon_{1}\right)  \mid
\psi_{RL}\left(  \epsilon_{2}\right)  \right\rangle =a_{R^{\prime}}%
\frac{S_{R^{\prime}L^{\prime},RL}\left(  \epsilon_{1}\right)  -S_{R^{\prime
}L^{\prime},RL}\left(  \epsilon_{2}\right)  }{\epsilon_{1}-\epsilon_{2}},
\label{ovl}%
\end{equation}
i.e. the first energy-divided difference of the corresponding element of the
structure matrix (\ref{r13}). In our divided-difference notation, this is:
$\left\langle \psi_{1;R^{\prime}L^{\prime}}\mid\psi_{2;RL}\right\rangle
=aS_{12;R^{\prime}L^{\prime},RL}.$ Expression (\ref{ovl}) actually includes
the integrals over the Bessel functions with $l>l_{\max}$ which survive inside
the spheres, but since $j_{l}\left(  \kappa r\right)  \approx r^{l}$ and
$l_{\max}\gtrsim4,$ this contribution is small. Besides, it is usually
counter-balanced by neglecting the high-$l$ components of the target function
as explained after eq.$\,$(\ref{high}).

The integral of a \emph{single} USW over the MT interstitial is:%
\[
\left\langle \psi_{RL}\left(  \varepsilon\right)  \right\rangle =\frac
{\sqrt{4\pi}}{\varepsilon}\sum\nolimits_{R^{\prime}}a_{R^{\prime}}%
S_{R^{\prime}0,RL}\left(  \varepsilon\right)  ,
\]
as obtained by noting that $\left\langle \psi_{RL}\left(  \varepsilon\right)
\right\rangle =\left\langle 1\mid\psi_{RL}\left(  \varepsilon\right)
\right\rangle $ with $\left\langle 1\right\vert $ being a solution of the
Laplace equation, and by using Green's second theorem. An expression with
better $R^{\prime}$-convergence may be obtained by first expanding $\left\vert
1\right\rangle $ in the set of USWs with $\varepsilon=0,$ as done in the upper
right-hand part of Fig.$\,$1. Since $\mathcal{\hat{P}}_{RL}\left(
a_{R}\right)  \left\vert 1\right\rangle =\sqrt{4\pi}\delta_{L,0}$ for any $R,$
only the $s$-USWs contribute, i.e.:%
\begin{equation}
\left\vert 1\right\rangle =\sqrt{4\pi}\sum_{R}\left\vert \psi_{R0}\left(
0\right)  \right\rangle \equiv\sqrt{4\pi}\sum_{R}\left\vert \psi
_{0;R0}\right\rangle , \label{1=LCUSW}%
\end{equation}
and eq.$\,$(\ref{ovl}) then leads to the result:
\begin{align}
\left\langle \psi_{nRL}\right\rangle  &  =\left\langle 1\mid\psi
_{nRL}\right\rangle =\sqrt{4\pi}\sum_{R^{\prime}}\left\langle \psi
_{0R^{\prime}0}\mid\psi_{nRL}\right\rangle \nonumber\\
&  =\sqrt{4\pi}\sum_{R^{\prime}}a_{R^{\prime}}S_{0n;R^{\prime}0,RL}\,.
\label{q34}%
\end{align}
Compared with the slower converging result, this implies that $\sum
_{R^{\prime}}a_{R^{\prime}}S_{R^{\prime}0,RL}\left(  0\right)  =0,$ which
requires that the inversion (\ref{aS}) of the bare structure matrix is
converged with respect to cluster size (see Sect.$\,$\ref{N}).

The integral of an energy-divided difference of a single USW follows from
expression (\ref{q34}):%
\begin{equation}
\left\langle \psi_{1..n;RL}\right\rangle =\sqrt{4\pi}\sum_{R^{\prime}%
}a_{R^{\prime}}S_{0...n;R^{\prime}0,RL}\,, \label{int}%
\end{equation}
because taking an energy-divided difference (\ref{q0}) is a linear operation.
The integral of a single v\&d function is therefore obtained by inserting
expression (\ref{int}) in (\ref{vandd}), yielding:%
\begin{equation}
\left\langle \varrho_{dRL}\right\rangle =\sqrt{4\pi}\sum_{R^{\prime\prime}%
}a_{R^{\prime\prime}}\sum_{n=1}^{4}\sum_{R^{\prime}L^{\prime}}%
S_{0...n;R^{\prime\prime}0,R^{\prime}L^{\prime}}D_{nR^{\prime}L^{\prime},dRL}.
\label{intvandd}%
\end{equation}
This enables us to find the interstitial charge as $\left\langle \rho\left(
\mathbf{r}\right)  \right\rangle =\sum_{dRL}\,\left\langle \varrho
_{dRL}\right\rangle \mathcal{R}_{RL}^{\left(  d\right)  }\left(  a_{R}\right)
.$ This could of course also have been obtained as the sum over the monopole
moments at all centers from expressions (\ref{Q}) and (\ref{Qall}) for the
general multipole moments.

Since our formalism expresses the interstitial density, $\rho\left(
\mathbf{r}\right)  $, its regular potential, $\phi\left(  \mathbf{r}\right)
$, and the potential from site-centered multipoles in terms of energy-divided
differences of USWs (see eq.s$\,$(\ref{interpolation}), (\ref{v&d}),
(\ref{Potential}), and (\ref{longint})), the \emph{electrostatic energies} of
the interstitial charge density are integrals of products of energy-divided
differences of USWs. Expressions (\ref{ovl}) and (\ref{q0}) thus reduce such
an integral to the double sum:%
\begin{align}
&  \left\langle \psi_{1..m}\mid\psi_{0....n}\right\rangle \nonumber\\
&  =a\sum_{\mu=1}^{m}\sum_{\nu=0}^{n}\frac{S_{\mu\nu}}{\Pi_{\sigma=1,\neq\mu
}^{m}\left(  \epsilon_{\mu}-\epsilon_{\sigma}\right)  \Pi_{\tau=0,\neq\nu}%
^{n}\left(  \epsilon_{\nu}-\epsilon_{\tau}\right)  }\nonumber\\
&  \equiv aS_{0\left[  1..\min\left(  m,n\right)  \right]  ..\max\left(
m,n\right)  }\,, \label{Herm}%
\end{align}
where we have returned to matrix notation, i.e. have dropped the $RL$
subscripts. The common energy points, i.e. those used in \emph{both} divided
differences, $\psi_{1..m}$ and $\psi_{0....n},$ give rise to the terms with
$\mu=\nu$ and are seen to involve $\dot{S}_{\nu},$ the first energy derivative
of $S\left(  \varepsilon\right)  $ at $\epsilon_{\nu}.$ In fact
\cite{00DivDif}, the double sum (\ref{Herm}) is the highest non-vanishing
derivative --the $\left(  2+m+n\right)  $th-- of that polynomial in
$\varepsilon$ which coincides with $S\left(  \varepsilon\right)  $ at
\emph{all} the mesh points, $\epsilon_{0},\,...,\,\epsilon_{\max\left(
m,n\right)  },$ and whose 1st derivative coincides with $\dot{S}\left(
\varepsilon\right)  $ at the \emph{common} mesh points, $\epsilon
_{1},...\epsilon_{\min\left(  m,n\right)  }.$ This is \emph{Hermite}
polynomial approximation and the reduction of (\ref{Herm}) to the usual single
sum in terms of $S\left(  \varepsilon\right)  $ evaluated at all the mesh
points and $\dot{S}\left(  \varepsilon\right)  $ evaluated at the common mesh
points is given in Ref. \cite{00DivDif}. For $\dot{S}\left(  \varepsilon
\right)  ,$ we use the analytical expression derived from expressions
(\ref{aS}) and (\ref{bare}).

The local part of the electrostatic self-energy of the interstitial charge
density is then:%
\begin{align*}
&  \frac{1}{2}\int\rho\left(  \mathbf{r}\right)  \phi\left(  \mathbf{r}%
\right)  d^{3}r\\
&  =\frac{1}{2}\sum_{d^{\prime}R^{\prime}L^{\prime}}\sum_{dRL}\mathcal{R}%
_{R^{\prime}L^{\prime}}^{\left(  d^{\prime}\right)  }\left(  a_{R^{\prime}%
}\right)  \left\langle \varrho_{d^{\prime}R^{\prime}L^{\prime}}\mid
\varphi_{dRL}\right\rangle \mathcal{R}_{RL}^{\left(  d\right)  }\left(
a_{R}\right)  .
\end{align*}
Splitting the potential in the interstitial into localized and long-ranged
parts according to (\ref{longint}), the localized part gives:%
\begin{align}
&  \frac{1}{2}\int\rho\left(  \mathbf{r}\right)  \phi^{loc}\left(
\mathbf{r}\right)  d^{3}r=4\pi\times\label{elstatloc}\\
&  \sum_{n^{\prime}d^{\prime}}\sum_{nd}(D_{n^{\prime}d^{\prime}}%
\mathcal{R}^{\left(  d^{\prime}\right)  })^{\dagger}\,aS_{0\left[
1..\min\left(  n^{\prime},n\right)  \right]  ..\max\left(  n^{\prime
},n\right)  }D_{nd}\mathcal{R}^{\left(  d\right)  },\nonumber
\end{align}
using matrix notation in the $RL$-subscripts. The long-ranged part gives:
\begin{align}
&  \frac{1}{2}\int\rho\left(  \mathbf{r}\right)  \phi^{long}\left(
\mathbf{r}\right)  d^{3}r\nonumber\\
&  =-4\pi\sum_{n^{\prime}d^{\prime}}\sum_{nd}(D_{n^{\prime}d^{\prime}%
}\mathcal{R}^{\left(  d^{\prime}\right)  })^{\dagger}aS_{0...n^{\prime}%
}\,j_{0}(a)\nonumber\\
&  \times\left(  \frac{n_{0}(a)}{j_{0}(a)}+B_{0}\right)  j_{0}\left(
a\right)  aS_{0...n}D_{nd}\mathcal{R}^{\left(  d\right)  }. \label{elstatlong}%
\end{align}
since $\left\langle \psi_{0..n}\mid\psi_{0}\right\rangle =aS_{0...n}.$

The charge densities \emph{inside} the spheres from the \emph{rest} of the
system, $\sum_{R}\rho_{R}^{rest}\left(  \mathbf{r}\right)  $ in eq.$\,$%
(\ref{rest}), produce a multipole potential in the interstitial which like in
(\ref{longint}) may be expanded in USWs:%
\begin{align*}
&  \sum_{RL}\left(  -\right)  ^{l}\frac{8\pi}{2l+1}\frac{Y_{L}\left(
\widehat{\mathbf{r}_{R}}\right)  }{r_{R}^{l+1}}Q_{RL}^{rest}\\
&  =-8\pi\psi\left(  0,\mathbf{r}\right)  \,j\left(  0,a\right)  \left(
\frac{n\left(  0,a\right)  }{j\left(  0,a\right)  }+B\left(  0\right)
\right)  \frac{\left(  -1\right)  ^{l}Q^{rest}}{\left(  2l+1\right)  !!},
\end{align*}
where matrix notation has been used in the last line. The electrostatic
interaction between the charge densities inside, $\rho^{rest},$ and outside,
$\rho,$ the spheres is thus:%
\begin{align}
&  \int\rho\left(  \mathbf{r}^{\prime}\right)  \frac{2\sum_{R}\rho_{R}%
^{rest}\left(  \mathbf{r}\right)  }{\left\vert \mathbf{r}^{\prime}%
-\mathbf{r}\right\vert }d^{3}r\nonumber\\
&  =-8\pi\sum_{nd}(D_{nd}\mathcal{R}^{\left(  d\right)  })^{\dagger}%
aS_{0...n}\,j\left(  0,a\right) \nonumber\\
&  \times\left(  \frac{n\left(  0,a\right)  }{j\left(  0,a\right)  }+B\left(
0\right)  \right)  \frac{\left(  -1\right)  ^{l}Q^{rest}}{\left(  2l+1\right)
!!} \label{ElStatinsout}%
\end{align}

\section{How to set the parameters\label{ParamValues}}

In this section we shall first examine how many sites, $N_{R},$ are needed to
screen the bare spherical waves to USWs using the formalism of Sect.$\,$%
\ref{USW}. Then we shall explain how matrix sizes may be reduced by use of
symmetry, and finally shall discuss how to choose the energy mesh.

\subsection{Number of screening sites, $N_{R}$\label{N}}

As derived in Sect.$\,$\ref{USW}, screening the bare spherical waves
(\ref{USWinHank}) in such a way that their averages for all $l\leq l_{\max}$
vanish at all hard spheres except the own, amounts to inverting the symmetric
matrix in the square parenthesis in eq.s (\ref{M}) and (\ref{aS}). We do this
by letting $\mathbf{R}$ and $\mathbf{R}^{\prime}$ be the sites of a cluster of
size $N_{R},$ centered around one of the sites in question. The inversion is
done for all inequivalent sites in the structure.

As illustrated in Fig. 1, for increasing energy the extent of the USWs
increases and herewith the size of the cluster needed for their generation.
For energies above a certain threshold, $\varepsilon_{\hom}>0,\;$delocalized
USWs exist and this threshold increases with the close-packing of the
interstitial; for instance is $\varepsilon_{\hom}t^{2}\sim2.5$ for the
diamond- and $15$ for the bcc structure. In order to interpolate by means of
localized USWs, we must use $\varepsilon<\varepsilon_{\hom}.$ In the following
we simply take $N_{R}$ as the number of sites needed to screen for
$\varepsilon=0;$ this is also the energy needed to describe the Laplace potentials.

In order to monitor the $N_{R}$-convergence, we expand the function
$\left\vert 1\right\rangle $ in USWs with $\varepsilon=0$. The result
(\ref{1=LCUSW}) is exact, and was illustrated for the diamond structured
interstitial in the upper right-hand part of Fig. 1. As a measure of
convergence we take:
\begin{equation}
\left\langle 1\mid1\right\rangle =4\pi\sum_{R}\sum_{R^{\prime}=1}^{N_{R}}%
a_{R}\dot{S}_{Rs},_{R^{\prime}s}\left(  0\right)  , \label{measure}%
\end{equation}
as follows from eq.$\,$(\ref{ovl}) with $\epsilon_{1}=\epsilon_{2}=0.$ Here,
$R^{\prime}$ runs over all sites in the cluster and $R$ over all sites in the
primitive cell. For increasing $l_{\max}$ and $N_{R},$ this measure tends to
the volume of the interstitial.\cite{87ByHand}

Table I gives the relative error for $l_{\max}=4$ as a function of $N_{R}$ for
the bcc and diamond structures and with two hard-sphere radii: $a=0.9t$ and
$0.8t.$ As usual, $t$ is the radius of touching spheres, i.e. half the
nearest-neighbor distance. The closely packed bcc structure has $\pi\sqrt
{3}/8\approx68\%$ of its volume inside touching spheres whereas the open
diamond structure has only half this amount inside. Conversely, the
hard-sphere interstitial with $a=t,\,0.9t,$ or $0.8t$ covers respectively
$32,$ $50,$ or $65\%$ of the volume in the bcc structure and as much as $66$,
$75,$ or $83\%$ in the diamond structure.

\bigskip

\begin{table}[htbp]
\caption{
Relative error $\times10^{3}$ of the interstitial
volume computed as eq.$\,$(\ref{measure})$.$ $l_{\max}\mathrm{=}4.$
}
\bigskip
\begin{tabular}[c]{rrrrrr}\hline\hline
& $N_{R}$ &  & $a=0.9t$ & $a=0.8t$ & \\\hline
\qquad bcc &  &  & \qquad\quad  Int.vol.=50\% & \qquad\quad  Int.vol.=65\% & \\
& $27$ &  & $-0.45$ & $-5.38$ & \\
& $51$ &  & $2.54$ & $0.14$ & \\
& $89$ &  & $0.11$ & $-0.01$ & \\
& $169$ &  & $0.00$ & $0.00$ & \\\hline
dia &  &  & Int.vol.=75\% & Int.vol.=83\% & \\
& $35$ &  & $-44.80$ & $-86.40$ & \\
& $87$ &  & $-3.27$ & $-9.65$ & \\
& $159$ &  & $-0.25$ & $-1.17$ & \\
& $191$ &  & $-0.13$ & $-0.65$ & \\\hline\hline
\end{tabular}
\end{table}

\bigskip

The table shows that in order to have the interstitial volume computed to
better than $10^{-2}$, it suffices to screen with 27 sites, i.e. the 3 first
shells, in the bcc structure and with 87 sites (7 first shells) in the diamond
structure. Computing the volume to better than 10$^{-3},$ requires 59 sites in
the bcc and 159 sites in the diamond structure. The accuracy is seen to be
somewhat better for the larger hard sphere. However, the $l_{\max}%
$-convergence deteriorates when the spheres nearly touch, and this is the
reason for the anomalously large $2.54\times10^{-3}$ relative error for the
bcc structure with $a=0.9t$ and $N_{R}=51.$ Whereas for $a=0.8t,$ the volumes
are essentially \cite{ripples} converged with $l_{\max}=4,$ as also needed for
the purpose of charge interpolation, those for $a=0.9t$ require $l_{\max}=5$
or $N_{R}>51.$

\subsection{Reduction of matrix sizes by use of symmetry\label{Symmetry}}

\subsubsection{Use of local point symmetry in the screening inversion}

In the previous subsection, we considered the number, $N_{R},$ of screening
sites needed when generating each column,\thinspace$RL,$ of the slope matrix,
$S_{R^{\prime}L^{\prime},RL}\left(  \varepsilon\right)  ,$ by inversion in
real space for a cluster centered on site $R$. This number depends on the
hard-sphere packing, but is \emph{not} influenced by symmetry. The only saving
brought about by translational symmetry is that the matrix needs to be
inverted merely for the translationally \emph{in}equivalent sites.

The number, $N_{L},$ of cubic-harmonics needed in the screening inversion is
$\left(  l_{\max}+1\right)  ^{2}\sim25$ when no use is made of symmetry,
whereby the linear dimension, $N_{R}N_{L},$ of the matrix to be inverted is of
order $2500.$ Point symmetry may, however, reduce this significantly. If we
let $N_{L}\left(  R\right)  $ be the number of cubic harmonics with $l\leq
l_{\max}$ in the appropriate irreducible representation of the local point
group at site $R$ of the interstitial function, $\rho\left(  \mathbf{r}%
\right)  ,$ then the matrix dimension is $\sum_{R}^{N_{R}}N_{L}\left(
R\right)  ,$ i.e. $N_{L}$ is now the average of $N_{L}\left(  R\right)  $ over
the $N_{R}$ sites in the cluster. In case $\rho\left(  \mathbf{r}\right)  $ is
the charge density, the appropriate irreducible representation is the identity
representation. Taking as examples the charge density of diamond, Si, or
zincblende-structured binary compounds (see Sect.$\,$\ref{Si}), merely%
\begin{equation}
s,\,3z^{2}-1,\;xyz,\;35z^{4}-30z^{2}+3,\;\mathrm{and\;}x^{4}+y^{4}-6x^{2}%
y^{2}, \label{ch}%
\end{equation}
of the cubic harmonics with $l\leq4$ transform according to the identity
representation of the tetrahedral point group. In this case, use of point
symmetry thus reduces the linear dimension of the matrix to be inverted by a
factor $5,$ i.e. to about $500.$

Said in another way: We only want to construct those USWs (\ref{USWinHank})
which are needed to fit the cubic-harmonic projections (\ref{proj}) of the
target function, and its point symmetries can be used to significantly reduce
the matrix dimension. In the above-mentioned examples it so happens that not
only the projection of the density onto the $25-5=20$ cubic harmonics
\emph{other} than those in (\ref{ch}) \emph{vanish,} but even the projection
onto $3z^{2}-1$ is negligible; so in these cases the linear matrix dimension
is reduced to about $400.$

One might feel that this site-dependent reduction of the number of screening
multipoles will reduce the screening and thus require a larger $N_{R}$ for
convergence. However, as long as the reduction is symmetry dictated, this does
not matter for the relevant USWs and the relevant elements of the slope matrix
\emph{after} they are symmetrized with respect to the space-group symmetry as
explained below, because only the symmetry-allowed $L$-channels will survive
the symmetrization process. Even without this symmetrization, use of the
possibly unconverged quantities in eq.s.$\,$(\ref{vandd}) and (\ref{odd}%
)-(\ref{DD}) for the v\&d functions needed for the interpolation
(\ref{interpolation}), i.e. those with non-vanishing $\mathcal{R}_{RL}%
$-coefficients, will be correct. This means that the reduction due to
point-group symmetry is valid even without space-group symmetrization.

\subsubsection{Use of translational or space-group symmetry}

In order to form the v\&d functions in case the interstitial has translational
symmetry and $\rho\left(  \mathbf{r}\right)  $ has Bloch symmetry, we merely
need Bloch summed USWs:%
\[
\psi_{RL}^{\mathbf{k}}\left(  \varepsilon,\mathbf{r}\right)  =\sum
\nolimits_{T}\psi_{RL}\left(  \varepsilon,\mathbf{r-T}\right)
e^{i\mathbf{k\cdot T}},
\]
and the corresponding slope matrix,
\[
S_{R^{\prime}L^{\prime},RL}^{\mathbf{k}}\left(  \varepsilon\right)
=\sum\nolimits_{T}S_{R^{\prime}L^{\prime},\left(  R-T\right)  L}\left(
\varepsilon\right)  e^{i\mathbf{k\cdot T}},
\]
where $R$ and $R^{\prime}$ are now merely in the primitive cell. After short
range has been achieved through screening, the sum over all translations,
$\mathbf{T,}$ converges fast. For periodic functions like charge densities,
$\mathbf{k=0.}$ The slope matrix entering the matrix expressions
(\ref{odd})-(\ref{DD}) for the Bloch symmetrized v\&d functions in terms of
the Bloch symmetrized USWs (\ref{vandd}), has the number of sites, $N_{R},$
reduced from what was used in the screening inversion (Sect.$\,$\ref{N}), to
the number of translationally inequivalent sites.

In the rightmost panel of Fig. 1 we show symmetrized $s$-USWs for the diamond
structure. Here, symmetrization was done by summing not only over the
translationally equivalent sites forming an fcc lattice, but also over the two
equivalent sites per primitive cell which are related by inversion, so that
$N_{R}=1.$ Hence the slope matrix entering eq.s (\ref{odd})-(\ref{DD}) has
linear dimension $4.$ This of course requires inverting the axes of the cubic
harmonics at every other site. Had the slope matrix been symmetrized with
respect to merely the translational symmetry, its linear dimension would have
been $8,$ in which case the proper point symmetry would be installed only
\emph{after} the v\&d functions have been multiplied with the proper input
coefficients to form the interpolated function (\ref{interpolation}). These
$8\times4$ coefficients satisfy: $\mathcal{R}_{1L}^{\left(  d\right)
}=\left(  -\right)  ^{l}\mathcal{R}_{2L}^{\left(  d\right)  },$ which means
that the sign is flipped on every second site for the $xyz$ projections, but
not for the remaining projections in (\ref{ch}).

In order to profit from space-group symmetry in general, the cubic harmonics,
both in eq.s$\,$(\ref{proj}) and (\ref{bare}), must be defined with respect to
a coordinate system which follows the local symmetry.

Had one only been interested in periodic structures with a few sites per cell,
one could, alternatively to screen the bare structure matrix by inversion
(\ref{aS}) of a large real-space matrix, have started by Bloch-summing the
bare structure matrix, $\sum_{T}B_{R^{\prime}L^{\prime},\left(  R-T\right)
L}\left(  \varepsilon\right)  e^{i\mathbf{k\cdot T}},$ using Ewald's method,
and then performed the screening (\ref{aS}) and the formation of v\&d
functions (\ref{odd})-(\ref{DD}) in the Bloch representation. Knowing the
local point symmetries, e.g. from the $\mathcal{R}_{RL}$-input, it is possible
to sort out the relevant $LL^{\prime}$-block of $B_{R^{\prime}L^{\prime
},\left(  R-T\right)  L}\left(  \varepsilon\right)  $ from the onset.

\subsection{Energy mesh\label{Emesh}}

Here we shall study the influence of the energy mesh on the interpolated
function, as a function of the structure exemplified by the bcc and dia
interstitials, and of the target exemplified by the $\left\vert 1\right\rangle
$-function and the valence charge densities in diamond-structured semiconductors.

As said at the end of Sect.$\,$\ref{v&dfunctions}, the v\&d functions are
independent of the scale $\left(  t\right)  $ of the structure, provided that
spatial derivatives are defined with respect to the dimensionless variable
$\mathbf{r/}t$ and that the energy mesh times $t^{2}$ is kept constant.

Our interpolation is 3-dimensional with $RL$ spanning the 2-dimensional
surfaces of the hard-sphere interstitial and the energy mesh spanning the
perpendicular direction. We shall see that using a 4-point energy mesh to
match the value and first three radial derivatives at the surface, suffices to
make the interpolated function insensitive to the choice of mesh if the
structure is closely packed (bcc), but not if it is open (dia).

In the latter case, the highest energies, i.e. the smallest decay constants,
will determine the behavior of the interpolated function deepest in the
interstitial and must therefore be determined by the behavior of the target
function there. Information about this, such as the value of the integral over
(number of electrons in) the interstitial and/or the value of the target
function at one or more points deep in the interstitial, may conveniently be
included as constraints on the interpolation by adding a higher, 5th energy
and using it to form an additional 4-point mesh with its v\&d functions. The
linear combination of the two sets of v\&d-functions which satisfies the
constraints and the three-times differentiable matching at the spheres is then
easily found. This will be the subject of the following Sect.$\,$%
\ref{Constraints}.

\subsubsection{Constant charge density\label{flat}}

As a first test of our v\&d technique for interpolation across an
interstitial, we return to the function $\left\vert 1\right\rangle $
considered in Sect.$\,$\ref{N}. This function can be expanded \emph{exactly}
in a \emph{single} set of USWs, provided that this set has zero energy, as was
illustrated in the top right part of Fig. 1. We now use USW sets with
\emph{four} \emph{different} energies and fit to the value and first 3 radial
derivatives at the spheres. Hence, the input (\ref{input}) to the
interpolation is:
\[
\mathcal{R}_{RL}^{\left(  0\right)  }=a\sqrt{4\pi}\delta_{L,0},\;\;\mathcal{R}%
_{RL}^{\left(  1\right)  }=\sqrt{4\pi}\delta_{L,0},\;\;\mathrm{and}%
\;\mathcal{R}_{RL}^{\left(  d>1\right)  }=0.
\]
The result is given by eq.$\,$(\ref{interpolation}) in terms of v\&d functions
like the ones shown in the top rows of Figs. 2 and 3.

Like in Sect.$\,$\ref{N}, we use the examples of bcc and diamond-structured
interstitials, and take $a=0.8t,$ $l_{\max}=4,$ and $N_{R}=51$ for bcc and
$159$ for dia. We trace the interpolated function along the line from the
centre of a sphere to that of the most distant nearest-neighbor void. For a
crystal, this is the line from the centre of the Wigner-Seitz cell to its
farthest corner, i.e. the [210] line for the bcc and the [111] line for the
diamond structure. These paths leading deep into the respective interstitials
are sketched in the insets of Fig. 4, which has bcc in the top- and dia in the
bottom panels.

Fig. 4 now exhibits the interpolated functions for 10 different equidistant
4-point meshes, $\epsilon_{4}<\epsilon_{3}<\epsilon_{2}<\epsilon_{1},$ with
$\epsilon_{1}$ stepping from $0$ to $-4$ Ry ($\epsilon_{1}t^{2}$ from 0 to
$-20)$ and $\epsilon_{4}$ stepping from $-4$ to $-8$ Ry ($\epsilon_{4}t^{2}$
from $-20$ to $-40)$ in the left-hand panel and from $-16$ to $-20$ Ry
($\epsilon_{4}t^{2}$ from $-80$ to $-100)$ in the right-hand panel. We notice,
first of all, that the result is exact in all cases where $\epsilon_{1}=0$
(thin red full lines), as is expected. Secondly, all interpolated functions
posses the required value $\left(  =1\right)  $ and first 3 derivatives
$\left(  =0\right)  $ at the spheres. This is strictly true only when
$\epsilon_{1}=0,$ but when $\epsilon_{1}\lesssim$ $-3$ Ry this does not hold
along the part of the line where the spheres come close (see dia). The reason
is that correct values and derivatives are only ensured for the
angular-momentum \emph{averages} over a sphere with $l\leq l_{\max},$ and not
along a single direction (see also the discussion around eq.$\,$(\ref{high})).

Further into the interstitial, the interpolated functions are seen to deviate
from $1,$ a deviation which increases not only with $\epsilon_{1}$ decreasing
below zero, but also with the other energies decreasing below $\epsilon_{1}.$
As illustrated by comparison of the left- and right hand panels, the role of
$\epsilon_{4}$ (the fastest decay) is to modify the behavior close to the
spheres, given the first-, second-, and third radial derivatives. Further away
from a sphere than $1/\sqrt{-\epsilon_{1}}$ and along the radial line far away
from any other sphere, one might have expected the decay to be like that of
the bare $s$-Hankel function with the longest range, i.e. like $\left(
a/r\right)  \exp\left\{  -\left(  r-a\right)  \sqrt{-\epsilon_{1}}\right\}  .$
However, the decays seen in the figure are more gradual. This is connected
with the fact seen in Fig. 1, that the \emph{screened} $s$-Hankel function,
i.e. the $s$-USW, is a structure-adapted $s$\emph{-like} wave which decays
gradually towards the voids and steeply towards the neighboring spheres.

Most striking is the dramatic increase in the sensitivity to the mesh when
going from the closely packed bcc- to the open diamond structure where the
void and hard-sphere structures are identical and interpolation across the
former is therefore difficult. Whereas interpolation of the function
$\left\vert 1\right\rangle $ can be seen to work for bcc as long as
$-2\lesssim\epsilon_{1}\lesssim0\,$Ry and $-6\,\mathrm{Ry}\lesssim\epsilon
_{4}<\epsilon_{1},$ for diamond the requirement is something like:
$-0.4\lesssim\epsilon_{1}\lesssim0\,$Ry and $-1\lesssim\epsilon_{4}%
<\epsilon_{1};$ unless $\epsilon_{1}=0,$ in which case the interpolation is
exact, regardless of the values of the other energies.

\bigskip

\begin{figure}[htbp]
  \vspace{0cm}
  \begin{center}
  \includegraphics[width=0.45\textwidth]{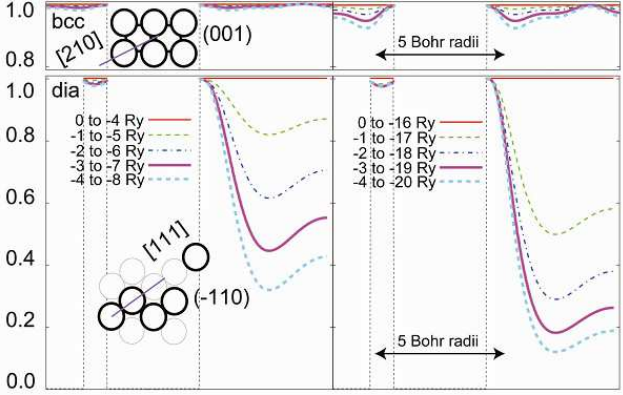}
  \caption{(Color online)
Interpolation of the function $\left\vert 1\right\rangle $ across the
bcc (top, note that the vertical scale starts at 0.8) and dia (bottom)
interstitials using 10 different energy meshes. The results of these v\&d
interpolations are traced along the violet part of the [210] (bcc) and [111]
(dia) lines indicated in the insets. $t=2.22$ Bohr radii, $a=0.8t,$ $l_{\max
}=4,$ $N_{R}=51$ (bcc) and $159$ (dia). See Sect.$\,$\ref{flat}.
}
  \end{center}
\end{figure}

\bigskip

\subsubsection{Charge densities of diamond-structured $sp^{3}$-bonded and
ionic semiconductors from FP NMTO calculations \label{Si}}

Fig.$\,$5 illustrates a realistic example: The valence-electron density in
diamond-structured Si. This charge density, here plotted in the $\left(
110\right)  $ plane through the $sp^{3}$ bonds, is the result of a
density-functional calculation with the full-potential Nth-order MTO method
\cite{12Juelich,15FPNMTO}. In this method, the density has the form:%
\begin{align}
&  \rho\left(  \mathbf{r}\right)  =\label{rhoNMTO}\\
&  \sum_{RL}\sum_{R^{\prime}L^{\prime}}\sum_{nn^{\prime}}\psi_{RL}^{\ast
}\left(  \epsilon_{n},\mathbf{r}_{R}\right)  \Gamma_{nRL,n^{\prime}R^{\prime
}L^{\prime}}\psi_{R^{\prime}L^{\prime}}\left(  \epsilon_{n^{\prime}%
},\mathbf{r}_{R^{\prime}}\right) \nonumber\\
&  +\sum_{R}\sum_{LL^{\prime}}Y_{RL}^{\ast}\left(  \mathbf{\hat{r}}%
_{R}\right)  Y_{RL^{\prime}}\left(  \mathbf{\hat{r}}_{R}\right)
\sum_{nn^{\prime}}\Gamma_{nRL,n^{\prime}R^{\prime}L^{\prime}}\nonumber\\
&  \times\left[  \varphi_{l}\left(  \epsilon_{n},r_{R}\right)  \varphi
_{l^{\prime}}\left(  \epsilon_{n^{\prime}},r_{R}\right)  -\varphi_{l}%
^{o}\left(  \epsilon_{n},r_{R}\right)  \varphi_{l^{\prime}}^{o}\left(
\epsilon_{n^{\prime}},r_{R}\right)  \right] \nonumber
\end{align}
where $\epsilon_{n}$ are the energies chosen for solving Schr\"{o}dinger's
equation; they spread far less that those used for the interpolation and
include positive values. $\Gamma$ is the density matrix and $\psi$ is a
screened spherical wave, basically a USW. $\varphi$ is \emph{not} the
potential from a v\&d function like in Sect.$\,$\ref{v&dpot}, but a solution
of the radial Scr\"{o}dinger equation for the overlapping MT potential
\cite{hardvsMT} which defines the NMTO basis set, and $\varphi^{o}$ is the
solution back-integrated over the MT zero, from the MT sphere to the hard
sphere, inside which it is truncated. Hence, the function $\varphi
\varphi-\varphi^{o}\varphi^{o}$ vanishes smoothly outside the MT sphere and
jumps to $\varphi\varphi$ inside the hard sphere. This discontinuity is
cancelled by the $\psi\Gamma\psi$ term in expression (\ref{rhoNMTO}) which
matches the $\varphi^{o}\varphi^{o}$ term and its first $2\mathrm{N}$ radial
derivatives; here $\mathrm{N}$ is the order of the MTOs ($\mathrm{N=2}$ in the
present calculations). The last term in (\ref{rhoNMTO}) is a single-center sum
over MT densities, each of which may be reduced to the simple warped-ASA
form,\cite{86WASA} $\sum_{L^{\prime\prime}}Y_{L^{\prime\prime}}\left(
\mathbf{\hat{r}}\right)  f_{L^{\prime\prime}}\left(  r\right)  ,$ for which
Poisson's equation is trivially solved. The $\psi\Gamma\psi$-term, however, is
a complicated multi-centre sum occurring in all LCAO-like electronic-structure
methods. This is the one we approximate by interpolation across the
interstitial, matching to the $\varphi^{o}\varphi^{o}$ functions at the hard
spheres when using the NMTO method. The result of this interpolation is shown
in the middle panel of Fig. 5, while the right-hand panel shows the last term
of expression (\ref{rhoNMTO}), the MT part. The sum of the two is shown in the
left-hand panel.

In the top row of Fig. 5, the voids in the diamond structure were filled with
empty spheres such that the structure of the interstitial becomes bcc. The
basis set for the electronic-structure calculation had MTOs on the two silicon
as well as on the two empty spheres in the primitive cell and, hence, the
density-functional calculation was one for 2(SiE).\cite{80Segall} The same
calculation delivered the input for the charge densities shown in the bottom
row, but the E contributions were neglected, i.e. the $R$ sums in expression
(\ref{rhoNMTO}) were over the diamond rather than the bcc lattice. The bottom
(dia) right-hand figure shows the density from the Si MTs, which is the same
as in the top row. The density from the E MTs, present in the top (bcc)
right-hand figure and missing in the bottom (dia) right-hand figure, is
included in the bottom middle figure where it is taken into account by the
interpolation across the dia interstitial matching to the $\varphi^{o}%
\varphi^{o}$ term at the Si hard spheres only. Adding the MT and interpolated
contributions yields the charge densities shown in the left-hand panel. The
ones in the top (bcc) and bottom (dia) rows are almost indistinguishable, but
the densities deep in the diamond interstitial, below the lowest contour used
in the figure (0.01 electrons/(Bohr radius)$^{3}$), are not accurately
interpolated, as we shall see in Fig.$\,7$.

We remark that the purpose of the above-mentioned construction of the Si
charge density without empty spheres is to compare bcc and dia interpolations
for the \emph{same} Si \emph{input}. Of course, we could have performed an
entire selfconsistent FP NMTO calculation for Si without empty spheres
\cite{15FPNMTO}.

\bigskip

\begin{figure}[htbp]
  \vspace{0cm}
  \begin{center}
  \includegraphics[width=0.45\textwidth]{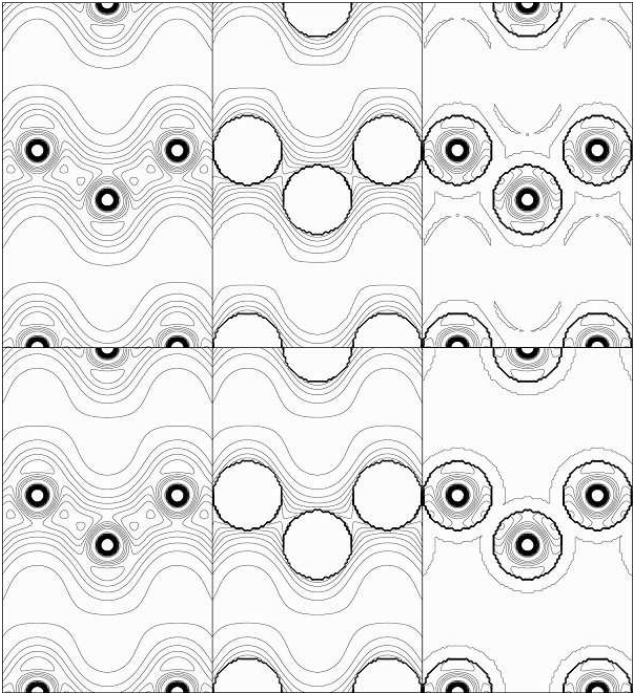}
  \caption{
Valence charge density of diamond-structured Si resulting from a FP
NMTO calculation for 2(SiE), plotted in the (110) plane through the $sp^{3}$
bonds. Like in eq.$\,$(\ref{rhoNMTO}), the right-, middle, and left-hand
panels show respectively the MT density, the $\psi\Gamma\psi$ density
interpolated over the bcc (top) and dia (bottom) interstitials, and the sum of
the two. The contours go from 0 to 0.3 in steps of 0.01 electrons per (Bohr
radii)$^{3}.$ The energy mesh was the same as in Figs 2 and 3, and the
remaining parameters were like in Fig. 4. See Sect.$\,$\ref{Si}.
}
  \end{center}
\end{figure}

\bigskip

Quite a different charge density is that of the zinc-blende structured I-VII
compound CuBr shown in Fig. 6. Rather than being covalent, it is ionic
(Cu$^{+}$Br$^{-}=\,$Cu $3d^{10}$ Br $4p^{8})$ and has a full Cu 3$d$ shell.
Despite this difference, the interstitial $\psi\Gamma\psi$ charge density,
which is the one we interpolate, is not that different from the one in Si,
albeit more concentrated around the atoms. The calculation leading to Fig. 6
was done exactly like the one leading to the top (bcc) row in Fig. 5 for Si.
We shall return to CuBr in Sect.$\,$\ref{Constraints}.

\bigskip

\begin{figure}[htbp]
  \vspace{0cm}
  \begin{center}
  \includegraphics[width=0.45\textwidth]{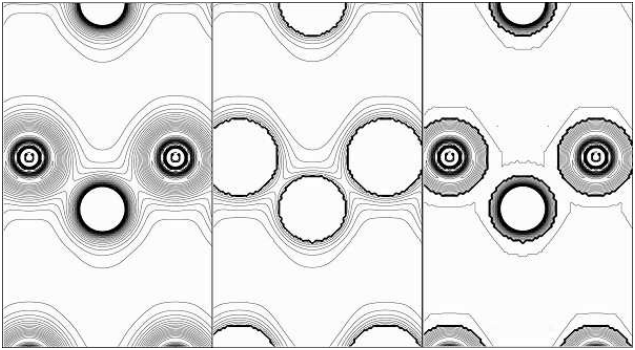}
  \caption{
As in the top (bcc) row of Fig. 5, but for zinc-blende structured CuBr
calculated as CuBrEE' with $a_{\mathrm{Cu}}\mathrm{=}a_{\mathrm{E}}%
\mathrm{=}1.72$ and $a_{\mathrm{Br}}\mathrm{=}a_{\mathrm{E}^{\prime}%
}\mathrm{=}2.00$ Bohr radii, whereby $t\mathrm{=}2.33$Bohr radii. See
Sect.$\,$\ref{Si}.
}
  \end{center}
\end{figure}

\bigskip

Like Fig. 4, but now for silicon, Fig. 7 shows charge densities along the open
[210] and [111] directions in respectively the bcc and dia interstitials using
different energy meshes for the interpolation. The upper figure to the left
shows the $\psi\Gamma\psi$ density interpolated across the bcc interstitial,
from a Si to an E sphere along $\left[  210\right]  ,$ using 6 different
exponential 4-point meshes,
\begin{equation}
\epsilon_{n}=\epsilon_{1}\left(  \epsilon_{4}/\epsilon_{1}\right)  ^{\left(
n-1\right)  /3}, \label{expmesh}%
\end{equation}
with the highest energy: $\epsilon_{1}=-3.0,$ $-2.5,$ $-2.0,$ $-1.5,$ $-1.0,$
or $-0.5$ Ry and the lowest energy: $\epsilon_{4}=-15.7$ Ry. Only near the
local maximum of the density where the $[210]$-line passes closely between a
Si and an E sphere, deviations are detectable. The results for 24 other meshes
with the same values of $\epsilon_{1}$, but with $\epsilon_{4}=-3.3,$ $-5.0,$
$-7.2,$ or $-10.0\,$Ry, deviate even less from each other, and have therefore
not been shown. So for $sp^{3}$-bonded silicon, interpolation across the bcc
interstitial is accurate and robust when using a 4-point mesh with energies
distributed between $-16$ and $-2\,\mathrm{Ry.}$

In the remaining five parts of Fig. 7, the E spheres have been neglected and
the interpolation is across the diamond-structured interstitial using the
above-mentioned 30 different energy meshes. In this case, and like for the
constant density in Fig. 4, the dependence on the energy mesh is strong. In
order to be able to compare with the accurate result (solid black line) of the
SiE calculation, we plot the total rather than the interpolated density. This
we do along the violet [111] line, from slightly before the point inside a
hard Si sphere where the density has fallen to a deep minimum and all
densities are identical, then going into the interstitial, and finally ending
at the midpoint between the two voids along $\left[  111\right]  $ (see also
Fig. 5). Each of the five figures shows (in color) the density obtained using
six different exponential energy meshes with the same $\epsilon_{4},$ and with
$\epsilon_{1}$ running through the above-mentioned values.

We see that all total densities match with value and first 3 derivatives at
the hard Si sphere, as they are designed to. But they deviate as we go deeper
and deeper into the void. It seems that getting the correct 4th derivative
requires $-20\lesssim\epsilon_{4}\lesssim-10$ Ry. In order to prevent the
interpolated function from behaving too wildly deep in the interstitial, we
must take the highest energy $\epsilon_{1}\lesssim-1$ Ry $\left(  \epsilon
_{1}t^{2}\lesssim-5\right)  $. While acceptable interpolation is achieved with
$-3.5\lesssim\epsilon_{1}\lesssim-2.5$ and $-5\lesssim\epsilon_{4}\lesssim-16$
Ry, the best is for the exponential mesh with $\epsilon_{1}=-3.0$ and
$\epsilon_{4}=-15.7$ Ry (thick brown dot-dashed density in the bottom
right-hand panel). This mesh actually reproduces the low densities in the
voids better than does the one with $\epsilon_{1}=-1.54$ Ry and $\epsilon
_{4}=-15.7$ Ry (thin blue dot-dashed density in the bottom right-hand panel)
used for the charge-density contours in the bottom panel of Fig. 5. Of the 4
Si valence electrons, 1.19 are in the Si MT density and 2.81 are in the
interstitial, and this is exactly what interpolation with the best mesh
yields. The mesh with $\epsilon_{1}=-1.54$ Ry and the same $\epsilon_{4},$
giving an electron density along [111] about 0.002 electrons per (Bohr
radius)$^{3}$ higher in the void, yields 2.90 interstitial electrons, which is
barely tolerable.

It is obvious from Fig. 7 that a 4-point mesh exists (the one with
$\epsilon_{1}\sim-3.0$ Ry and $\epsilon_{4}\sim-15.7$) which makes the
interpolation across the open interstitial almost perfect. Moreover, as long
as one starts from a mesh with fixed $-20\lesssim\epsilon_{4}\lesssim-10$ Ry
and $\epsilon_{4}<$ $\epsilon_{3}<$ $\epsilon_{2}\sim-5$ Ry, iterating merely
the value of $\epsilon_{1}$ until the value of some additional constraint like
the density at the centre of the void or the integral over the interstitial
has the correct value, the interpolation will converge to this almost perfect
density. However, iteration of $\epsilon_{1}$ is hardly practical because
computation of the screened structure matrix (Sect.s \ref{N} and
\ref{Symmetry}) is the most expensive part of an interpolation. Moreover, this
method is not a general one for treating additional constraints, so in the
following section we shall devise a different scheme.

\bigskip

\begin{figure}[htbp]
  \vspace{0cm}
  \begin{center}
  \includegraphics[width=0.45\textwidth]{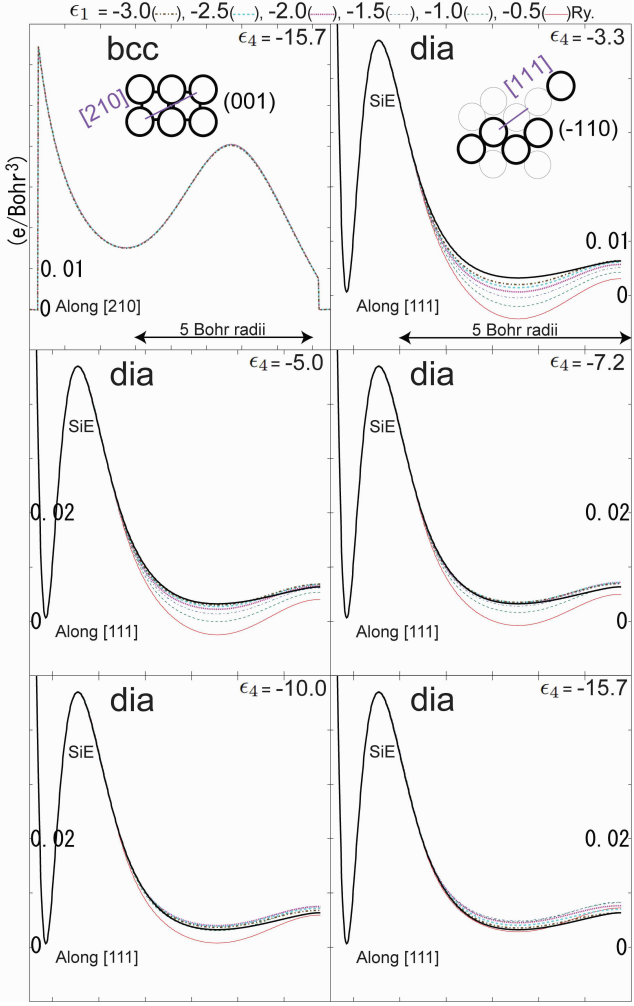}
  \caption{(Color)
As in Fig. 5, but plotted along the [210] line for the bcc- and along
the [111] line for the diamond-structured interstitial, like in Fig. 4. The
first figure shows the SiE $\psi\Gamma\psi$ density interpolated across the
bcc interstitial, while the remaining figures show the total density for SiE
(full black line) and for Si (broken, colored lines) with the $\psi\Gamma\psi$
part interpolated across the dia interstitial using 30 different exponential
energy meshes. See Sect.$\,$\ref{Si} and eq.$\,$(\ref{expmesh}).
}
  \end{center}
\end{figure}

\bigskip

\section{Extra constraints in open structures\label{Constraints}}

Our examples of the closely-packed bcc- and the open diamond structures have
clearly demonstrated (Fig.s 4 and 7) that whereas the energy mesh hardly
matters for the former, whereby the interpolation across the narrow
interstitial is robust, there is a strong dependence for the latter. This
means that in order to interpolate across a bulky interstitial, more
information is needed than the values and first three radial derivatives at
its boundaries.

In density-functional calculations, one basic piece of information is the
total number of valence electrons. It is fixed by the number of occupied bands
(for metals, the occupied part of the Brillouin zone) and the density must
integrate up to this number. Referring now to expression (\ref{rhoNMTO}) and
the corresponding figures 5 and 6 as examples, the MT density is trivial to
integrate accurately, and subtracting this from the number of valence
electrons gives what the integral over the interstitial of the interpolated
density, should be; this is $\left\langle \varrho\right\rangle \mathcal{R}$
from eq.s$\,$(\ref{input}), (\ref{interpolation}), and (\ref{intvandd}). For
Si, this number was 2.81 electrons in the dia interstitial.

An often used option in MTO calculations is to fill the voids with E spheres
whereby the interstitial becomes closely packed \cite{Stuttgart}. This is what
we did in the previous subsections to make the diamond structure bcc. The
additional information provided herewith is the cubic-harmonic projections
(\ref{proj}) at the E-spheres. For some structures, however, it takes numerous
small spheres to fill the voids; melting silicon is one example, solid
C$_{60}$ another.

In such cases, it is more practical to evaluate the density at a few selected
points, $\mathbf{r}_{c},$ deep in the interstitial and then constrain the
interpolated density to those values. In LCAO-type calculations the evaluation
is via the multi-centre expansion $\psi\left(  \mathbf{r}_{c}\right)
\Gamma\psi\left(  \mathbf{r}_{c}\right)  ,$ which is possible for a few
points, but cumbersome for many.

An economic and general implementation of such extra constraints (on top of
those $4N_{R}N_{L}$ constraints given by the matching at the hard spheres)
amounts to computing the structure matrix at merely \emph{one} extra energy
and then with \emph{two} different 4-point meshes generating \emph{two} sets
of v\&d functions, a more localized set, $\varrho^{l}\left(  \mathbf{r}%
\right)  ,$ and a more extended one, $\varrho^{e}\left(  \mathbf{r}\right)  .$
If, for instance, we use expression (\ref{expmesh}) to generate the five
energies: $\epsilon_{5}<\epsilon_{4}<\epsilon_{3}<\epsilon_{2}<\epsilon_{1},$
then $\varrho^{l}\left(  \mathbf{r}\right)  $ and $\varrho^{e}\left(
\mathbf{r}\right)  $ are the sets obtained from respectively points $5$ to $2$
and points $4$ to $1.$ Any of the $4N_{R}N_{L}$ (see Sect.$\,$\ref{Symmetry})
\emph{weighted averages:}%
\begin{equation}
\varrho_{dRL}\left(  \mathbf{r}\right)  \equiv\varrho_{dRL}^{e}\left(
\mathbf{r}\right)  \alpha_{dRL}+\varrho_{dRL}^{l}\left(  \mathbf{r}\right)
\left(  1-\alpha_{dRL}\right)  , \label{IandII}%
\end{equation}
is seen to be a v-or-d function, and we now aim at determining the weights,
$\alpha_{dRL},$ of the extended v\&d functions such that the extra constraints
are satisfied. Note that the number of extra constraints, $N_{c},$ cannot
exceed $N_{R}N_{L}$, because only \emph{one} extra USW set, e.g. $\psi
_{RL}\left(  \epsilon_{1},\mathbf{r}\right)  ,$ has been added in the
expansion (\ref{rho}) of $\rho\left(  \mathbf{r}\right)  .$

Let the extra constraints be $q_{c}\left(  \mathbf{r}\right)  ,$ with $c$
going from 1 to $N_{c},$ and $q_{c}\equiv\left\langle q_{c}\mid\rho
\right\rangle $ be the value of the $c$th constraint. As examples, the
integral of the density in the interstitial is obtained with $q_{c}\left(
\mathbf{r}\right)  \equiv1$ and the value at point $\mathbf{r}_{c}$ is
obtained with $q_{c}\left(  \mathbf{r}\right)  \equiv\delta\left(
\mathbf{r-r}_{c}\right)  .$ The estimate of the $c$th additional constraint is
now:%
\begin{align*}
&  \sum_{dRL}\left\langle q_{c}|\varrho_{dRL}\right\rangle \mathcal{R}%
_{RL}^{\left(  d\right)  }=\\
&  \sum_{dRL}\left\langle q_{c}|\varrho_{dRL}^{e}-\varrho_{dRL}^{l}%
\right\rangle \mathcal{R}_{RL}^{\left(  d\right)  }\alpha_{dRL}+\sum
_{dRL}\left\langle q_{c}|\varrho_{dRL}^{l}\right\rangle \mathcal{R}%
_{RL}^{\left(  d\right)  },
\end{align*}
as obtained by use of the interpolated density (\ref{interpolation}) and the
v\&d functions (\ref{IandII}). Equating this estimate to the true value,
$q_{c},$ of the constraint, leads to the linear equations:%
\begin{equation}
\sum_{dRL}\left(  q_{c,dRL}^{e}-q_{c,dRL}^{l}\right)  \alpha_{dRL}=q_{c}%
-q_{c}^{l}, \label{constr}%
\end{equation}
with\ $c$ going from $1$ to $N_{c},$ for the weights, $\alpha_{dRL}.$ On the
right-hand side,
\[
q_{c}^{l}\equiv\sum_{dRL}q_{c,dRL}^{l}%
\]
is the estimate of the constraint using localized density:%
\[
\rho^{l}\left(  \mathbf{r}\right)  =\sum_{dRL}\varrho_{dRL}^{l}\left(
\mathbf{r}\right)  \mathcal{R}_{RL}^{\left(  d\right)  },
\]
while $q_{c,dRL}^{l}\equiv\left\langle q_{c}|\varrho_{dRL}^{l}\right\rangle
\mathcal{R}_{RL}^{\left(  d\right)  }$ is its $dRL$ component. Similarly for
$q_{c,dRL}^{e}.$ Since the number, $4N_{L}N_{R},$ of unknown weights exceeds
the number, $N_{c},$ of extra constraints, we avoid unphysical solutions of
equations (\ref{constr}) by requiring that the weights, $\alpha_{dRL},$ of the
extended v\&d functions be small. Specifically, we minimize the sum of the
square weights, $\sum_{dRl}\alpha_{dRL}^{2},$ subject to the constraints
(\ref{constr}). This leads to the $4N_{R}N_{L}$ Lagrangian equations:%
\[
\frac{\partial}{\partial\alpha_{d^{\prime}R^{\prime}L^{\prime}}}\sum
_{dRL}\left\{  \alpha_{dRL}^{2}-\sum_{c=1}^{N_{c}}\lambda_{c}\left(
q_{c,dRL}^{e}-q_{c,dRL}^{l}\right)  \alpha_{dRL}\right\}  =0,
\]
for\textbf{\ }$d\mathrm{=}1$ to $4,\;R\mathrm{=}1$ to $N_{R},$%
\ and$\;L\mathrm{=}1$ to $N_{L}\left(  R\right)  ,$ or explicitly:%
\begin{equation}
\alpha_{dRL}=\sum_{c=1}^{N_{c}}\frac{1}{2}\lambda_{c}\left(  q_{c,dRL}%
^{e}-q_{c,dRL}^{l}\right)  , \label{w}%
\end{equation}
to be solved together with eq.s$\,$(\ref{constr}) for the weights,
$\alpha_{dRL},$ and the Langrangian multipliers, $\lambda_{c}.$ Insertion of
(\ref{w}) in eq.$\,$(\ref{constr}) yields the $N_{c}$ linear equations for the
$N_{c}$ Langrangian multipliers:\
\begin{align}
&  \sum_{c^{\prime}}\frac{1}{2}\lambda_{c^{\prime}}\sum_{dRL}\left(
q_{c^{\prime},dRL}^{e}-q_{c^{\prime},dRL}^{l}\right)  \left(  q_{c,dRL}%
^{e}-q_{c,dRL}^{l}\right) \nonumber\\
&  =q_{c}-q_{c}^{l}, \label{lambdas}%
\end{align}
for\ $c\mathrm{=}1$ to $N_{c},$ which may be solved and inserted in
eq.$\,$(\ref{w}) to yield the weights.

The Coulomb potential from the constrained v\&d functions (\ref{IandII}) is of
course given by the same weighted average of the potentials $\varphi_{dRL}%
^{e}\left(  \mathbf{r}\right)  $ and $\varphi_{dRL}^{l}\left(  \mathbf{r}%
\right)  ,$ where the latter are those of the charge densities $\varrho
_{dRL}^{e}\left(  \mathbf{r}\right)  $ and $\varrho_{dRL}^{l}\left(
\mathbf{r}\right)  ,$ obtained as described in Sect.$\,$\ref{v&dpot}.

In Fig. 8 we demonstrate how well this works for the valence electron
densities in diamond-structured Si (top) where $N_{R}N_{L}\mathrm{=}4$, and in
the zinc-blende structured II-VI and I-VII compounds, ZnSe (middle) and CuBr
(bottom) where $N_{R}N_{L}\mathrm{=}8$. For all three materials, we used the
five energies, $\epsilon_{5}$ to $\epsilon_{1},$ obtained with the same
\cite{10percent} exponential mesh: $\epsilon_{n}=-4\left(  4\right)  ^{\left(
n-2\right)  /3}\,\mathrm{Ry.}$ Fig. 8 shows densities along the open [111]
direction, but now all the way across the dia interstitial, because with the A
and B atoms different, the density is not symmetric around any of their
midpoints as in Fig.$\,7$. The aim is to interpolate the density across the
dia interstitial as in the bottom part of Fig. 5, i.e. without using the v\&d
information computed at the empty sphere(s), but obtaining also the densities
below 0.01 electrons per (Bohr radii)$^{3}$ accurately by installing the
following constraints: The total number of electrons be 8 and 18 electrons per
cell for respectively Si and the compound semiconductors $(N_{c}\mathrm{=}1),$
also the density at the centre(s) of the void(s) be correct ($N_{c}%
\mathrm{=}2$ for Si and $3$ for the compounds), and also the density between
the voids be correct ($N_{c}\mathrm{=}3$ and $4)$.

This scheme is seen to work very well, indeed.

\bigskip

\begin{figure}[htbp]
  \vspace{0cm}
  \begin{center}
  \includegraphics[width=0.45\textwidth]{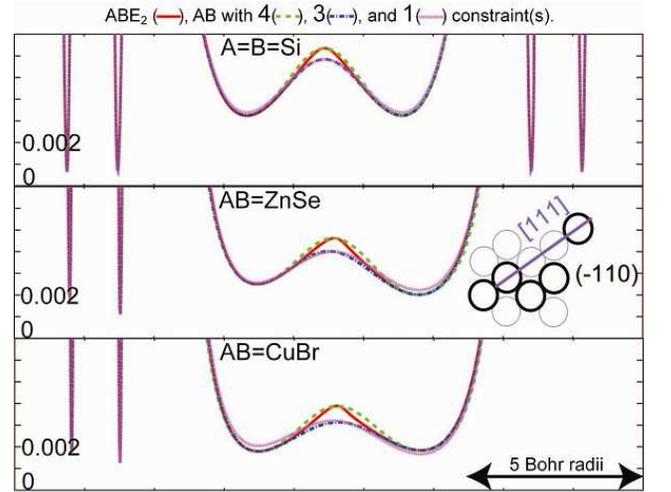}
  \caption{(Color online)
Total densities plotted along the [111] line shown in the inset across
the dia interstitial for Si (top), ZnSe (middle), and CuBr (bottom). This line
crosses the BE'EA spheres from left to right, i.e. the anion is to the left
and the cation to the right. For each material, the full red line shows the
density calculated as ABEE', i.e. the interpolation was merely over the bcc
interstitial, while in the three other curves, the interpolation was across
the dia interstitial and constrained. The dotted magenta curve results from
constraining merely the integral over the interstitial. In the dot-dashed blue
curve also the densities at center(s) of the void(s), i.e. the minima, were
constrained. In the dashed green curve, finally, also the density between the
voids, i.e. near the relative maximum, was constrained. We used the energies:
$\epsilon_{n}=-4\left(  4\right)  ^{\left(  n-2\right)  /3}\,\mathrm{Ry}$ with
$n\mathrm{=}1$ to 5, i.e. $-16,$ $-10.1,$ $-6.3,$ $-4,$ and $-2.5$ Ry. The
screening of the dia structure matrix required $N_{R}\mathrm{=}87$ sites and
$l_{\max}\mathrm{=}4$. See also Figs. 5, 6, and 7, as well as Sect.$\,$%
\ref{Constraints}.
}
  \end{center}
\end{figure}

\bigskip

\section{Conclusions\label{Concl}}

We have carried through the program laid out in the Introduction and have
derived a formalism for numerical 3D interpolation across a hard-sphere
interstitial from the cubic-harmonic projections of the target function,
$\rho\left(  \mathbf{r}\right)  ,$ and its first 3 radial derivatives at the
spheres. Whereas this knowledge suffices for closely-packed structures,
additional information such as the integral of $\rho\left(  \mathbf{r}\right)
$ over the interstitial and/or the values at specific points deep inside the
interstitial is needed for open structures. This was illustrated by
application to a constant function and to the valence charge densities in Si,
ZnSe, and CuBr, interpolated across either the bcc- or the
zinc-blende-structured interstitial, depending on whether or not the voids
were filled with empty spheres (Figs 4-7).

Our interpolation is based on localized, structure-adapted sets of
spherical-waves (USWs), $\psi_{RL}\left(  \epsilon_{n},\mathbf{r}\right)  ,$
with 4 different energies (Fig.$\,$1). These set are combined linearly into
sets of so-called value-and-derivative (v\&d) functions, $\varrho_{dRL}\left(
\mathbf{r}\right)  $ with $d\mathrm{=}0$ to 3, confined essentially to the
Voronoi cells (top row in Fig.s 2 and 3). The formalism is expressed in terms
of energy-divided differences of the USWs and their slope matrix with elements
$S_{RL,R^{\prime}L^{\prime}}\left(  \epsilon_{n}\right)  $ (Sect.s \ref{Prel}
and \ref{v&dfunctions}). For the bcc structure, accurate results were obtained
with an exponential energy mesh spanned by $\epsilon_{4}\mathrm{=}-80/t^{2}$
and $\epsilon_{1}\mathrm{=}-12/t^{2},$ where $t$ is the average radius of
touching spheres and gives the scale of the structure. We expect this mesh to
be satisfactory for all closely-packed structures. The extra constraints
needed for open structures require the use of an extra energy and we
demonstrated that the exponential 5-point mesh with the same limits,
$\epsilon_{5}t^{2}\mathrm{=}-80$ and $\epsilon_{1}t^{2}\mathrm{=}-12,$ gives
excellent results for the zincblende structure (Fig.$\,$8).

Solving Poisson's equation for the interpolated function requires the
solutions $\psi_{RL}\left(  \epsilon_{0}\mathcal{\equiv}0,\mathbf{r}\right)  $
of the Laplace equation as well. The localized potentials (middle row in Fig.s
2 and 3) for the v\&d functions are expressed in terms of energy-divided
differences one order higher than for the v\&d functions. The multipoles
needed in order to make the localized potential regular (bottom row in Fig.s 2
and 3), are given by the same differences of the slope matrix (Sect.
\ref{Poissons eq}). The latter also give the integrals over the interstitial
(Sect.$\,$\ref{integrals}).

The slope matrix is screened through inversion of the analytical, bare
structure matrix for a cluster with $N_{R}$ sites surrounding sites. We found
$N_{R}\mathrm{\sim}60$ for closely packed structures and 2-3 times larger for
open 3D structures like zincblende (Sect.$\,$\ref{N}). For a given site, $R,$
only those $L$-channels must be kept for which the target function does not
vanish due to symmetry. If the local point symmetry is high, the number,
$N_{L}\left(  R\right)  ,$ of $L$-channels is considerably smaller than the
maximum, $\left(  1+l_{\max}\right)  ^{2}\sim25,$ for instance $4-5$ for
tetrahedral symmetry. The dimension of the matrix to be inverted is thus
$\sum_{R}^{N_{R}}N_{L}\left(  R\right)  .$ For interpolating a function
without symmetry this can be large $\left(  10^{3}-10^{4}\right)  ,$ but the
process increases only linearly with the number of sites. Whereas
$N_{R}\mathrm{\sim}100$ is needed in the screening calculation, all subsequent
matrix operations, i.e. those needed to form v\&d functions, potentials, and
integrals, can in case of space-group symmetry and for interpolating the
charge density be performed with a symmetrized slope matrix in which
equivalent sites have been summed over so that $N_{R}$ is merely the number of
inequivalent sites. Hence, $N_{R}\mathrm{=}2$ for the zinc-blende structure
(Sect. \ref{Symmetry})$.$

This method was developed for interpolating charge densities and is currently
used for density-functional FP-NMTO calculations \cite{15FPNMTO,15LiPB}. Since
it is generally applicable, local, and based on cubic harmonics, we expect it
to find numerous uses.

\section{Acknowledgements}

We are grateful to Dr. M. H\"{o}ppner for testing our codes, for providing
valuable feedback, and for help with writing the manual. Prof. H. Takagi is
acknowledged for support and Dr. A. Schnyder for providing many useful examples.

\section{Appendix: One-center cubic-harmonic expansions}

\subsection{$RL$-projections of the v\&d functions}

The expressions derived in Sect.$\,$\ref{v&dfunctions} for the short-ranged
v\&d functions as linear combinations of USWs are particularly useful for
solving Poisson's equation and for computing integrals over the interstitial
as was done in Sect.s \ref{Poissons eq} and \ref{integrals}, respectively. For
other purposes, cubic-harmonic expansions like (\ref{onsite}) around single
centers may be more practical. The coefficients are the projections,
$\mathcal{\hat{P}}_{R^{\prime}L^{\prime}}\left(  r\right)  \varrho
_{dRL}\left(  \mathbf{r}\right)  ,$ given in eq.$\,$(\ref{v&d}) to third order
in the distance from the spheres, and beyond this, by the expresssions derived
below. These higher-order terms are responsible for the sensitivity to the
energy mesh displayed in Fig. 7.

The expressions for $\mathcal{\hat{P}}_{R^{\prime}L^{\prime}}\left(  r\right)
\varrho_{dRL}\left(  \mathbf{r}\right)  $ might be obtained by projecting the
USWs $\psi_{RL}\left(  \epsilon_{n},\mathbf{r}\right)  $ in the first
eq.$\,$(\ref{vandd}) by means of (\ref{r9}), but it is simpler to commute
$RL$-projection with taking $\varepsilon$-divided differences. Like in
(\ref{(r-a)3}), we thus start by taking the $\varepsilon$-divided differences
of the projection using the binomial rule (\ref{q1}). In order that the result
clearly exhibit the parts leading to eq.$\,$(\ref{v&d}), we use $u\left(
\varepsilon,r\right)  \equiv f\left(  \varepsilon,r\right)  -g\left(
\varepsilon,r\right)  $ instead of $f\left(  \varepsilon,r\right)  $ and,
expression (\ref{r02}) divided by $r$ instead of (\ref{r9}). The result is:%
\begin{align}
\mathcal{\hat{P}}\left(  r\right)  \psi_{1}\left(  \mathbf{r}\right)   &
=u_{1}\left(  r\right)  +g_{1}\left(  r\right)  \mathcal{S}_{1},\nonumber\\
\mathcal{\hat{P}}\left(  r\right)  \psi_{12}\left(  \mathbf{r}\right)   &
=u_{12}\left(  r\right)  +g_{1}\left(  r\right)  S_{12}+g_{12}\left(
r\right)  \mathcal{S}_{2},\nonumber\\
\mathcal{\hat{P}}\left(  r\right)  \psi_{123}\left(  \mathbf{r}\right)   &
=u_{123}\left(  r\right)  +g_{1}\left(  r\right)  S_{123}+g_{12}\left(
r\right)  S_{23}\nonumber\\
&  +g_{123}\left(  r\right)  \mathcal{S}_{3},\nonumber\\
\mathcal{\hat{P}}\left(  r\right)  \psi_{1234}\left(  \mathbf{r}\right)   &
=u_{1234}\left(  r\right)  +g_{1}\left(  r\right)  S_{1234}+g_{12}\left(
r\right)  S_{234}\nonumber\\
&  +g_{123}\left(  r\right)  S_{34}+g_{1234}\left(  r\right)  \mathcal{S}_{4}.
\label{PDDUSW}%
\end{align}
The terms present in (\ref{PDDUSW}) and \emph{not} in eq.s (\ref{(r-a)3}) are
of order higher than 3rd in $r-a.$

The projections of the set of 3rd-derivative functions (\ref{DIII}) are then:%
\begin{align}
&  \mathcal{\hat{P}}\left(  r\right)  \varrho_{3}\left(  \mathbf{r}\right)
\nonumber\\
&  =\mathcal{\hat{P}}\left(  r\right)  \psi_{123}\left(  \mathbf{r}\right)
D_{3,3}+\mathcal{\hat{P}}\left(  r\right)  \psi_{1234}\left(  \mathbf{r}%
\right)  D_{4,3}\nonumber\\
&  =-g_{12}\left(  r\right)  +u_{123}\left(  r\right)  D_{3,3}\nonumber\\
&  +g_{123}\left(  r\right)  \left[  \mathcal{S}_{3}D_{3,3}+S_{34}%
D_{4,3}\right] \nonumber\\
&  +u_{1234}\left(  r\right)  D_{4,3}+g_{1234}\left(  r\right)  \mathcal{S}%
_{4}D_{4,3}. \label{PDIII}%
\end{align}
By construction, the coefficient to $g_{1}\left(  r\right)  $ vanishes and the
coefficient to $g_{12}\left(  r\right)  $ is%
\[
S_{23}D_{3,3}+S_{234}D_{4,3}=-1.
\]
With eq.s (\ref{q14})-(\ref{q5}) in mind, we realize that the terms after the
diagonal term, $-g_{12}\left(  r\right)  ,$ in (\ref{PDIII}) are smaller than
$\left(  r-a\right)  ^{3}.$

The projections of the set of 1st-derivative functions (\ref{vandd}%
)-(\ref{odd}) are:%
\begin{align}
&  \mathcal{\hat{P}}\left(  r\right)  \varrho_{1}\left(  \mathbf{r}\right)
\nonumber\\
&  =\mathcal{\hat{P}}\left(  r\right)  \psi_{123}\left(  \mathbf{r}\right)
D_{3,1}+\mathcal{\hat{P}}\left(  r\right)  \psi_{1234}\left(  \mathbf{r}%
\right)  D_{4,1}\nonumber\\
&  =g_{1}\left(  r\right)  -g_{12}\left(  r\right)  \left(  \epsilon
_{1}-w\right)  +u_{123}\left(  r\right)  D_{3,1}\nonumber\\
&  +g_{123}\left(  r\right)  \left[  \mathcal{S}_{3}D_{3,1}+S_{34}%
D_{4,1}\right] \nonumber\\
&  +u_{1234}\left(  r\right)  D_{4,1}+g_{1234}\left(  r\right)  \mathcal{S}%
_{4}D_{4,1}, \label{PDI}%
\end{align}
where the coefficient to $g_{1}\left(  r\right)  $ is:%
\[
S_{123}D_{3,1}+S_{1234}D_{4,1}=1,
\]
and that to $g_{12}\left(  r\right)  $ is:
\[
S_{23}D_{3,1}+S_{234}D_{4,1}=-\left(  \epsilon_{1}-w\right)  .
\]
The first, diagonal $g_{1}\left(  r\right)  $-term gives the 1st derivative,
and its contribution to the 3rd derivative is cancelled by the second term,
$-g_{12}\left(  r\right)  \left(  \epsilon_{1}-w\right)  $.

The projections of the set of 2nd-derivative functions (\ref{vandd}%
)-(\ref{DD}) are:%
\begin{align}
&  \mathcal{\hat{P}}\left(  r\right)  \varrho_{2}\left(  \mathbf{r}\right)
a\nonumber\\
&  =-\mathcal{\hat{P}}\left(  r\right)  \psi_{12}\left(  \mathbf{r}\right)
\nonumber\\
&  +\mathcal{\hat{P}}\left(  r\right)  \psi_{123}\left(  \mathbf{r}\right)
D_{3,2}a+\mathcal{\hat{P}}\left(  r\right)  \psi_{1234}\left(  \mathbf{r}%
\right)  D_{4,2}a\nonumber\\
&  =-u_{12}\left(  r\right)  +u_{123}\left(  r\right)  D_{3,2}a\nonumber\\
&  +g_{123}\left(  r\right)  \left[  S_{34}D_{4,2}+\mathcal{S}_{3}%
D_{3,2}\right]  a\nonumber\\
&  +u_{1234}\left(  r\right)  D_{4,2}a+g_{1234}\left(  r\right)
\mathcal{S}_{4}D_{4,2}a, \label{PDII}%
\end{align}
where the coefficients:%
\[
-S_{12}+S_{123}D_{3,2}a+S_{1234}D_{4,2}a,
\]
and:
\[
-\mathcal{S}_{2}+S_{23}D_{3,2}a+S_{234}D_{4,2}a,
\]
to respectively $g_{1}\left(  r\right)  $ and $g_{12}\left(  r\right)  $
vanish. From eq.$\,$(\ref{q24}) it follows that $-u_{12}\left(  r\right)
=\frac{a}{r}\frac{\left(  r-a\right)  ^{2}}{2!}+\,o,$ which is the behavior of
$\mathcal{\hat{P}}\left(  r\right)  \varrho_{2}\left(  \mathbf{r}\right)  a$
specified by eq.$\,$(\ref{v&d}).

The projections of the set of value functions (\ref{vandd})-(\ref{DD}) are:%
\begin{align}
&  \mathcal{\hat{P}}\left(  r\right)  \rho_{0}\left(  \mathbf{r}\right)
a\nonumber\\
&  =\mathcal{\hat{P}}\left(  r\right)  \psi_{1}\left(  \mathbf{r}\right)
-\mathcal{\hat{P}}\left(  r\right)  \psi_{12}\left(  \mathbf{r}\right)
\left(  \epsilon_{1}-w\right) \nonumber\\
&  +\mathcal{\hat{P}}\left(  r\right)  \psi_{123}\left(  \mathbf{r}\right)
D_{3,0}a+\mathcal{\hat{P}}\left(  r\right)  \psi_{1234}\left(  \mathbf{r}%
\right)  D_{4,0}a\nonumber\\
&  =u_{1}\left(  r\right)  -u_{12}\left(  r\right)  \left(  \epsilon
_{1}-w\right)  +g_{12}\left(  r\right)  aw^{\prime}\nonumber\\
&  +u_{123}\left(  r\right)  D_{3,0}a+g_{123}\left(  r\right)  \left(
\mathcal{S}_{3}D_{3,0}+S_{34}D_{4,0}\right)  a\nonumber\\
&  +u_{1234}\left(  r\right)  D_{4,0}a-g_{1234}\left(  r\right)
\mathcal{S}_{4}D_{4,0}a. \label{PDO}%
\end{align}
Here, the coefficient,%
\[
\mathcal{S}_{1}-S_{12}\left(  \epsilon_{1}-w\right)  +S_{123}D_{3,0}%
a+S_{1234}D_{4,0}a,
\]
to $g_{1}\left(  r\right)  $ has vanished and the coefficient to
$g_{12}\left(  r\right)  $ has worked out to:
\[
-\mathcal{S}_{2}\left(  \epsilon_{1}-w\right)  +S_{23}D_{3,0}a+S_{234}%
D_{4,0}a=aw^{\prime}.
\]
In (\ref{PDO}) then, the $u_{1}\left(  r\right)  $-term gives the value, and
the parts of this term which behave as $\left(  r-a\right)  ^{2}$ and $\left(
r-a\right)  ^{3}$ are cancelled by respectively $-u_{12}\left(  r\right)
\left(  \epsilon_{1}-w\right)  $ and $g_{12}\left(  r\right)  aw^{\prime}.$

The radial functions $u_{Rl}\left(  \varepsilon,r\right)  $ and $g_{Rl}\left(
\varepsilon,r\right)  $ may be generated by numerical integration outwards
from the boundary conditions: $u_{Rl}\left(  \varepsilon,a_{R}\right)  =1,$
$u_{Rl}^{\prime}\left(  \varepsilon,a_{R}\right)  =-1/a_{R},$ and (\ref{g}).
Alternatively, these functions may be expressed in terms of the spherical
Neumann and Bessel functions using eq.s (\ref{fginnj}) and (\ref{nandj}).

As said after eq.$\,$(\ref{onsite}), the one-center cubic-harmonics expansions
with the radial functions $\mathcal{\hat{P}}_{R^{\prime}L^{\prime}}\left(
r\right)  \varrho_{dRL}\left(  \mathbf{r}\right)  $ are valid at and outside
the $R^{\prime}$-sphere and inside the sphere touching the nearest-neighbor
sphere. Inside the $R^{\prime}$-sphere, all v\&d functions vanish.

\subsection{$RL$-projections of the localized potentials from v\&d functions}

Here we shall derive the radial functions in the $L^{\prime}$-expansion
(\ref{one-centre}) around the arbitrary site $R^{\prime}$ of the localized
potential (\ref{phid}) from the v\&d function, $\varrho_{dRL}.$ Examples of
the localized and regular potentials were shown in respectively the middle and
bottom rows of Fig.s 2 and 3, and were discussed in Sect.$\,$\ref{v&dpot}. The
regular potentials look very smooth so that their one-center expansions should
converge well. But this smooth behavior is due to domination by the central
point-charge potential which gives long range, and thus complicates the
summation $\sum_{R}\mathcal{\hat{P}}_{R^{\prime}L^{\prime}}\left(  r\right)
\varphi_{dRL}\left(  \mathbf{r}\right)  \mathcal{R}_{RL}^{\left(  d\right)  }$
for the projection of $V\left(  \mathbf{r}\right)  $, and will in any case be
modified (usually reduced) when adding the potentials from the remaining
charges in the system. It is therefore better, at the end of the $V\left(
\mathbf{r}\right)  $-calculation, to sum up all multipole moments at the
various sites, $R^{\prime\prime},$ and then expand their potentials around the
site, $R^{\prime},$ in question, using the well-known expression
(eq.$\,($\ref{Ph}), $\varepsilon\mathrm{=}0)$ with its large radius of
convergence, $d_{R^{\prime\prime}R^{\prime}}.$ Below, we shall therefore only
consider the localized potential.

The $d$th energy-divided difference, $\psi_{1..d+1;RL}\left(  \mathbf{r}%
\right)  ,$ of a USW gives rise to the localized potential, $\phi
_{1..d+1;RL}^{loc}\left(  \mathbf{r}\right)  ,$ given by eq.$\,$(\ref{q31}).
Its cubic-harmonic expansion (\ref{1cPhiloc}) around the arbitrary site
$\mathbf{R}^{\prime}$ has coefficients, which are $8\pi$ times the projections
given by expressions (\ref{Psi-1}).

Since the v\&d functions are superpositions (\ref{vandd})-(\ref{DD}) of
energy-divided differences of USWs: $\varrho_{d}\left(  \mathbf{r}\right)
=\sum_{n}\psi_{1..n}\left(  \mathbf{r}\right)  D_{n,d},$ the potentials from
$\varrho_{d}\left(  \mathbf{r}\right)  $ are the \emph{same} superpositions of
the potentials, $\phi_{1..n}^{loc}\left(  \mathbf{r}\right)  ,$ from
$\psi_{1..n}\left(  \mathbf{r}\right)  $:
\[
\varphi_{d}^{loc}\left(  \mathbf{r}\right)  =\sum\nolimits_{n}\phi
_{1..n}^{loc}\left(  \mathbf{r}\right)  D_{n,d}=8\pi\sum\nolimits_{n}%
\psi_{0...n}\left(  \mathbf{r}\right)  D_{n,d}\,,
\]
and similarly for the projections of the v\&d functions and of their
potentials:%
\begin{align*}
\mathcal{\hat{P}}\left(  r\right)  \varrho_{d}\left(  \mathbf{r}\right)   &
=\sum_{n=1}^{4}\mathcal{\hat{P}}\left(  r\right)  \psi_{1..n}\left(
\mathbf{r}\right)  D_{n,d}\,,\;\;\mathrm{and}\\
\mathcal{\hat{P}}\left(  r\right)  \varphi_{d}^{loc}\left(  \mathbf{r}\right)
&  =8\pi\sum_{n=1}^{4}\mathcal{\hat{P}}\left(  r\right)  \psi_{0...n}\left(
\mathbf{r}\right)  D_{n,d}\,.
\end{align*}
Comparison of $\mathcal{\hat{P}}\left(  r\right)  \psi_{1..n}\left(
\mathbf{r}\right)  $ in (\ref{PDDUSW}) with $\mathcal{\hat{P}}\left(
r\right)  \psi_{0...n}\left(  \mathbf{r}\right)  $ in (\ref{Psi-1}) now shows
that expressions (\ref{PDIII})-(\ref{PDO}) for $\mathcal{\hat{P}}\left(
r\right)  \varrho_{d}\left(  \mathbf{r}\right)  $ hold also for $\mathcal{\hat
{P}}\left(  r\right)  \varphi_{d}^{loc}\left(  \mathbf{r}\right)  /8\pi,$
provided that (1) in all energy-divided-difference \emph{functions} of
$\mathbf{r}$ and $r$ -- but not in the coefficients -- the subscripts $1..n$
are substituted by $0...n$ and (2) the term $g_{0}\left(  r\right)  \sum
_{n=1}^{4}S_{0...n}D_{n,d}$ is added. (1) is as if Poisson's equation had been
solved by taking the energy-divided differences of only the radial functions,
but not of the slope matrix, and (2) is the Laplace term giving the multipole
potential when continued inside the $R^{\prime}$-sphere (see eq.$\,$%
(\ref{locinside})). With the localized potential from the value function as an
example, we get, starting from eq. (\ref{PDO}):%
\begin{align}
&  \frac{1}{8\pi}\mathcal{\hat{P}}\left(  r\right)  \varphi_{0}^{loc}\left(
\mathbf{r}\right)  a\nonumber\\
&  =g_{0}\left(  r\right)  \sum\nolimits_{n=1}^{4}S_{0...n}D_{n,0}\nonumber\\
&  +u_{01}\left(  r\right)  -u_{012}\left(  r\right)  \left(  \epsilon
_{1}-w\right)  +g_{012}\left(  r\right)  aw^{\prime}\nonumber\\
&  +u_{0123}\left(  r\right)  D_{3,0}a+g_{0123}\left(  r\right)  \left(
\mathcal{S}_{3}D_{3,0}+S_{34}D_{4,0}\right)  a\nonumber\\
&  +u_{01234}\left(  r\right)  D_{4,0}a-g_{01234}\left(  r\right)
\mathcal{S}_{4}D_{4,0}a. \label{PphiO}%
\end{align}

The expansion (\ref{one-centre}) around $\mathbf{R}^{\prime}$ holds also
inside the sphere, i.e. for $0\leq r\leq\min_{R^{\prime\prime}}\left(
d_{R^{\prime\prime}R^{\prime}}-a_{R^{\prime}}\right)  ,$ provided that we keep
only the Laplace term, $g_{0}\left(  r\right)  \sum S_{0...n}D_{n,d},$ there.
Without this term, the radial potential from the value function (\ref{PphiO})
increases smoothly from zero inside the sphere to $u_{01}\left(  r\right)
+o=-\frac{a}{r}\frac{\left(  r-a\right)  ^{2}}{2!}+o$ outside. Here $o$ is
given by eq.$\,$(\ref{o}). The analogous potential from the 1st derivative
function, increases outside as $g_{01}+o=-\frac{1}{r}\frac{\left(  r-a\right)
^{3}}{3!}+\,o,$ and those from the 2nd and 3rd derivative functions as $o.$

These projections with $L^{\prime}\mathrm{=}0$ are used to construct the
overlapping MT potential \cite{98MRS,09OMTA} defining the 3rd generation LMTO
\cite{94Trieste, 98MRS,00Rashmi} and NMTO \cite{00NMTO, 00Odile,
12Juelich,15FPNMTO} basis sets.


\begin{thebibliography}{99}
\bibitem {hardvsMT}In the 3rd generation LMTO \cite{94Trieste, 98MRS} and NMTO
\cite{00NMTO, 00Odile, 00Rashmi, 12Juelich,15FPNMTO} band-structure methods
both hard spheres and MT spheres are used. The former are those of the present
paper and the latter are for presenting the MT part of the potential. The hard
and the MT spheres are concentric with the respective radii $a$ and $s>a.$ A
MT sphere can overlap neighboring MT spheres, but not hard spheres.
\bibitem {Y}We complex conjugate $Y_{L}\left(  \mathbf{\hat{r}}\right)  $ as
appropriate had it been a complex spherical harmonic rather than a real cubic
harmonic because this makes the formalism more general.
\bibitem {kappa}Our bare structure matrix is $\kappa^{l+l^{\prime}}$ times the
one with the usual definition.
\bibitem {VPhiphi}While we use the symbol $V\left(  \mathbf{r}\right)  $ for a
potential in general, $\phi\left(  \mathbf{r}\right)  $ denotes the potential
from an energy-divided difference of USWs as in eq.s (\ref{q31}) and
(\ref{q32}), and $\varphi\left(  \mathbf{r}\right)  $ denotes the potential
from a v\&d function as in eq.s (\ref{phid}) and (\ref{phiregd}), as well as
in the 2nd and 3rd rows of Fig.s 2 and 3.
\bibitem {ripples}No ripples are seen for the symmetrized $s$-USW in the upper
right-hand part of Fig. 1 because it was computed with $l_{\max}=9.$
\bibitem {10percent}Since the lattice constants of ZnSe and CuBr are about 5
\% larger than that of Si, their mesh should have been contracted by 10\% to:
$\epsilon_{n}=-3.6\times\left(  4\right)  ^{\left(  n-2\right)  /5}$ Ry. But
that hardly matters.
\end{thebibliography}
\end{document}